\documentstyle[12pt,epsfig]{article}
\newlength{\dinwidth}
\newlength{\dinmargin}
\newcommand{\ba}{\begin{array}}
\newcommand{\ea}{\end{array}}
\newcommand{\beq}{\begin{equation}}
\newcommand{\eeq}{\end{equation}}
\newcommand{\bea}{\begin{eqnarray}}
\newcommand{\eea}{\end{eqnarray}}

\def\bce{\begin{center}}
\def\ece{\end{center}}

\def\nonu{\nonumber}

\def\pa{\partial}

\def\la{\lambda}
\def\La{\Lambda}

 \def\Si{\Sigma}

\newcommand{\tr}{\mbox{Tr}}

\def\eps6{{\displaystyle \mathop{\epsilon}^{6}}{}}

\def\nab6{{\displaystyle \mathop{\nabla}^{6}}{}}
\def\to{\rightarrow}

\begin{document}

\thispagestyle{empty} \addtocounter{page}{-1}
\begin{flushright}
KIAS-P03016 \\
TIT-HEP-493 \\
{\tt hep-th/0302150}\\
\end{flushright}

\vspace*{1.3cm} \centerline{\Large \bf Phases of} \vskip0.3cm
\centerline{ \Large \bf ${\cal N}=1$ Supersymmetric $SO/Sp$ Gauge
Theories} \vskip0.3cm \centerline{ \Large \bf
 via
Matrix Model}
\vspace*{1.5cm}
\centerline{{\bf Changhyun Ahn}$^1$ and {\bf Yutaka Ookouchi}$^2$}
\vspace*{1.0cm} \centerline{\it $^1$Department of Physics,
Kyungpook National University, Taegu 702-701, Korea}
\vspace*{0.2cm} \centerline{\it $^2$Department of Physics,
 Tokyo Institute of
Technology, Tokyo 152-8511, Japan} \vspace*{0.8cm} \centerline{\tt
ahn@knu.ac.kr, \qquad ookouchi@th.phys.titech.ac.jp} \vskip2cm

\centerline{\bf Abstract} \vspace*{0.5cm} We extend the results of
Cachazo, Seiberg and Witten to ${\cal N}=1$ supersymmetric gauge
theories with gauge groups $SO(2N), SO(2N+1)$ and $Sp(2N)$. By
taking the  superpotential which is an arbitrary polynomial of
adjoint matter $\Phi$ as a small perturbation of ${\cal N}=2$
gauge theories, we examine the singular points preserving ${\cal
N}=1$ supersymmetry in the moduli space where mutually local
monopoles become massless. We derive the matrix model complex
curve for the whole range of the degree of perturbed
superpotential. Then we determine a generalized Konishi anomaly
equation implying the orientifold contribution. We turn to the
multiplication map and the confinement index $K$ and describe both
Coulomb branch and confining branch. In particular, we  construct
a multiplication map from $SO(2N+1)$ to $SO(2KN-K+2)$ where $K$ is
an even integer as well as a multiplication map from $SO(2N)$ to
$SO(2KN-2K+2)$ ($K$ is a positive integer), a map from $SO(2N+1)$
to $SO(2KN-K+2)$ ($K$ is an odd integer) and a map from $Sp(2N)$
to $Sp(2KN+2K-2)$. Finally we analyze some examples which show
some duality: the same moduli space has two different
semiclassical limits corresponding to distinct gauge groups.


\baselineskip=18pt
\newpage
\renewcommand{\theequation}{\arabic{section}\mbox{.}\arabic{equation}}

\section{Introduction}
\setcounter{equation}{0}
The ${\cal N}=1$ supersymmetric gauge theories in four dimensions
are of considerable interest and have a rich physical contents. By
studying the structure of quantum moduli space, one gets some of
very important nonperturbative phenomena such as confinement and
dynamical symmetry breaking because many nonperturbative features
are related to the vacua such as mass generation and fermion
condensates and so on. In particular, in these theories, there
exists a special class of observables that are constrained by
holomorphy, for example, holomorphic effective superpotential. The
holomorphy of the quantum superpotential makes it possible to
determine the light degree of freedom and hence the quantum moduli
space. Although the exact results for many different models have
been obtained, a general recipe for computing the effective
superpotential has remained to be understood fully.

Partially wrapping D-branes over nontrivial cycles of the
geometries yield large classes of interesting gauge theories,
depending on the choice of geometries. Vafa and his collaborators
shed some light on this problem in the viewpoint of gravity dual
\cite{vafa,civ,ckv,cfikv}. They considered four dimensional gauge
theories on the world volume of D5-brane wrapped on ${\bf S}^2$
and claimed, as far as the computation of the effective
superpotential is concerned, that this geometry can be replaced by
a dual geometry where the ${\bf S}^2$ is replaced by the ${\bf
S}^3$ and D5-branes by RR fluxes through these cycles. As
discussed in \cite{gukov1,gukov2, tv}, these RR fluxes generate
the effective superpotential that is equivalent to the effective
superpotential of four dimensional gauge theories on the
worldvolme of D5-brane. The equivalence and validity of this
duality have been tested and proved for several models in many
papers \cite{civ,eot,fo1,cv,feng1,ookouchi,dot1,dot2,ot,dopt}. In
this context, the effective superpotential can be explicitly
expanded in terms of a period integral of the geometry that can be
interpreted as a glueball superfield in four dimensional gauge
theories.

Dijkgraaf and Vafa gave a new recipe to these calculations of the
effective superpotentials \cite{dv1,dv2,dv3,dgkv}. They claimed
that perturbative analysis of effective superpotential was
captured by the matrix model perturbation. Thus what we have  to
do is to sum up Feynman diagram of the matrix model. If one wants
$n$-th order instanton effect, one has only to compute Feynman
diagram up to $(n-1)$ loop. In this context, they also claimed
that the loop equation of matrix model, that plays an important
role in the matrix model, is equivalent to the Riemann surface
that comes from the dual geometry discussed above. After all, this
Riemann surface leads to a fruitful system for studying the
holomorphic information of four dimensional ${\cal N}=1$ gauge
theories. After their works, the correspondence between the
several matrix models  and four dimensional gauge theories
attracted wide attention, and  many papers, which include the
extension to other gauge groups and an adding flavors and so on,
appeared in \cite{cm}-\cite{dv4}. In particular, in
\cite{ferrari1,ferrari2}, Ferrari has discussed the quantum
parameter space of the ${\cal N}=1$ $U(N)$ gauge theory with one
adjoint matter $\Phi$ and a cubic tree level superpotential. These
works are one of the motivations of our paper.

Recently in \cite{cdsw}
they showed that these matrix model analysis could be interpreted
within purely field theoretic point of view. In particular for the
$U(N)$ gauge theory with adjoint matter $\Phi$ and a polynomial
superpotential $W(\Phi)$, a generalized Konishi anomaly equation,
providing  both a connection between the quark condensation and
the gluino condensation in the supersymmetric gauge theories and
possible ways to explore the nonperturbative aspects of the
supersymmetric gauge theories, gives rise to the loop equation of
matrix model. In other words, within gauge theories there is some
aspect that can be interpreted as matrix models.

Later, Cachazo, Seiberg and Witten \cite{csw} have discussed a new
kind of duality. By changing the parameters of $W(\Phi)$, one can
transit several vacua with different broken gauge groups
continuously and holomorphically. There was no restrictions to the
degree of the superpotential $W(\Phi)$ which can be an arbitrary.
They derived the matrix model curve and showed a generalized
Konishi anomaly equation from the strong gauge coupling approach
where the superpotential is considered as a small perturbation of
a strongly coupled ${\cal N}=2$ gauge theory with $W=0$.

In this paper, we extend the analysis given by Cachazo, Seiberg
and Witten to ${\cal N}=1$ $SO/Sp$ gauge theories with arbitrary
tree level superpotentials. We survey the phase structures of the
corresponding gauge theories: Which vacua allow us to smooth
transition? As in $U(N)$ case, are there Coulomb vacua in $SO/Sp$
gauge theories? We answer these questions in this paper.


In section 2.1, we analyze $SO(2N)$ gauge theories in terms of
strong gauge coupling approach. Since we deal with ${\cal N}=2$
theories deformed by tree level superpotential, we have to
constrain ${\cal N}=2$ Coulomb vacua to the special points where
monopoles or dyons become massless. By using ${\cal N}=2$ curve
together with this constraint that will be incorporated to the
effective superpotential and by applying the usual contour
integral formulas heavily, we will derive matrix model curve which
will be used to determine the number of vacua for fixed tree level
superpotential and a generalized Konishi anomaly equation where we
will see the orientifold contribution, peculiar to the gauge
group, which does not vanish even in the semiclassical limit,  for
arbitrary degree of superpotential. In order to produce these, we
are looking for the equations of motion about the Lagrange
multipliers, parameters of  moduli space and the locations of
massless monopoles for given effective superpotential which
depends on these fields, the couplings $g_{2r}$ and the scale
$\La$. Although the discussions of matrix model curve in different
context  have already been studied for special case in
\cite{eot,feng1}, we extend to the analysis for general case.

In section 2.2, we repeat the analysis given in section 2.1 for
$SO(2N+1)$ gauge theories for completeness. We also derive matrix
model curve and a generalized Konishi anomaly equation for
arbitrary degree of superpotential. Remembering the power behavior
of $x$ and $\La$ in ${\cal N}=2$ curve carefully, we also see the
orientifold contribution, with different strength, which does not
vanish even in the semiclassical limit.

In section 2.3, we discuss a confinement index $K$ and
multiplication maps for $SO(N)$ gauge theories. These ideas play
the central role for the study of phase structure. Since we can
construct a multiplication map from $SO(2N+1)$ to $SO(2KN-K+2)$
where $K$ is even  integer, this implies that $SO(2N+1)$ gauge
theory where the number of colors is odd and $SO(2KN-K+2)$ gauge
theory where the number of colors is even are closely related to
each other. There are also a multiplication map from $SO(2N)$ to
$SO(2KN-2K+2)$ where $K$ is a positive integer and a
multiplication map from $SO(2N+1)$ to $SO(2KN-K+2)$ where $K$ is
odd  integer. In checking these multiplication maps, the
properties of first and second kinds of Chebyshev polynomials and
the monopole constraints we will describe are crucial. Moreover it
does not seem to allow one can construct the multiplication map
from $SO(2N)$ to $SO(2M+1)$. Therefore, for $SO(2N)$ and
$SO(2N+1)$ gauge theories, there exist only {\it three} possible
ways to have multiplication maps. The criterion of a confining
phase and a Coulomb phase given in \cite{csw} is extended to
$SO(2N)$ and $SO(2N+1)$ gauge theories. Since the special
combination $(N-2)$ for $SO(N)$ gauge theory, that is dual Coxeter
number, coming from the contribution of unoriented diagram giving
rise to $-2$ term in the effective superpotential provides the
number of vacua of  the gauge theories, we can use the criterion
of $U(N)$ case \cite{csw} straightforwardly and equivalently.

In section 2.4, we analyze several explicit examples illustrating
what we have discussed in section 2.3. We study examples with
gauge group $SO(N)$ for $N=4,5,6,7,8$. In particular, $SO(6)$
gauge theory provides four confining vacua by analyzing the matrix
model curve by turning on a superpotential. According to the
scheme of a multiplication map from $SO(2N)$ to $SO(2KN-2K+2)$ we
have discussed in section 2.3 and by transforming the
characteristic polynomial with confinement index $K=2$ and $N=3$,
we confirm these vacua are really confining phase. Moreover in
$SO(8)$ gauge theory, we can see a smooth transition between
different classical gauge group and multiplication map from
$SO(5)$ by looking at the confining phase, based on a map from
$SO(2N+1)$ to $SO(2KN-K+2)$ described in section 2.3 where $N=2$
and $K=2$.

In section 3.1, as in section 2.1, by restricting ${\cal N}=2$
theories to the special points where monopoles become massless we
discuss ${\cal N}=1$ $Sp(2N)$ gauge theories with arbitrary tree
level superpotential. we will derive matrix model curve which will
be used to determine the number of vacua for fixed tree level
superpotential and a generalized Konishi anomaly equation where we
will see also the orientifold contribution, peculiar to the
$Sp(2N)$ group, which does not vanish even in the semiclassical
limit,  for arbitrary degree of superpotential. Comparing with the
$SO(2N)$ gauge theory we have discussed in section 2.1, there
exists exactly same contribution with an extra minus sign. In
order to produce these, we are studying for the equations of
motion about the Lagrange multipliers, parameters of  moduli space
and the locations of massless monopoles for given effective
superpotential.

In section 3.2, we discuss a confinement index and multiplication
maps for $Sp(2N)$ gauge theories similarly. Since we can construct
a multiplication map from $Sp(2N)$ to $Sp(2KN+2K-2)$ where $K$ is
a positive integer, this implies that $Sp(2N)$ gauge theory  and
$Sp(2KN+2K-2)$ gauge theory  are  related to each other. That is
the number of vacua of $Sp(2KN+2K-2)$ gauge theory is exactly the
confinement index $K$ multiplied by those of $Sp(2N)$ gauge theory
Coulomb phase. In doing this, we will use various properties of
Chebyshev polynomials and monopole constraints frequently.

In section 3.3, we analyze several explicit examples describing
what we have discussed in section 3.2. We study examples with
gauge group $Sp(N)$ for $N=2,4,6$. In particular, the $Sp(6)$
gauge theory provides a smooth transition between different
classical gauge groups, based on a map from $Sp(2N)$ to
$Sp(2KN+2K-2)$ described in section 3.2 where $N=1$ and $K=2$.

Although our work is based on  \cite{csw} completely and there are
not too much originality,  we hope our work is illuminating and
the results we have found newly provide a better understanding of
the structures of vacua of the gauge theories.

\section{$SO(2N)$ and $SO(2N+1)$ gauge theories }
\setcounter{equation}{0}

\subsection{Strong gauge coupling approach:$SO(2N)$ case}

The strong gauge coupling approach was studied in \cite{eot} by
using the method of \cite{aot,ahn98,Tera97}. Now we review it and
extend it to the case where some of $N_i$ (see (\ref{sobreaking})
for the notation) vanish and to allow the more general
superpotential in which the degree of it can be arbitrary without
any restrictions. Let us consider the superpotential regarded as a
small perturbation of an ${\cal N}=2$ $SO(2N)$ gauge theory
\cite{ty,kty,as,aps,hanany,hms,dkp,aot,eot,feng1}
\begin{eqnarray}
W(\Phi)=\sum_{r=1}^{k+1}\frac{g_{2r}}{2r}\mbox{Tr}\Phi^{2r}\equiv
\sum_{r=1}^{k+1} g_{2r} u_{2r}, \qquad u_{2r} \equiv \frac{1}{2r}
\mbox{Tr} \Phi^{2r} \label{treesup}
\end{eqnarray}
where $\Phi$ is an adjoint scalar chiral superfield and we denote
its eigenvalues by  $\pm  \phi_I(I=1,2,\cdots, N)$ that are purely
imaginary values. The degree of the superpotential $W(\Phi)$ is
$2(k+1)$. Since $\Phi$ is an antisymmetric matrix we can transform
to the following simple form, \bea \Phi= \left( {0 \atop -1 }{ 1
\atop 0
 } \right) \otimes \mbox{diag} ( i\phi_1, \cdots,  i\phi_{N}).
\label{Phi} \eea
When we replace $\mbox{Tr} \Phi^{2r} $ with $\left\langle\mbox{Tr}
\Phi^{2r}\right\rangle$, the superpotential becomes the effective
superpotential. We introduce a classical $2N \times 2N$ matrix
$\Phi_{cl}$ such that $\left\langle\mbox{Tr}
\Phi^{2r}\right\rangle = \mbox{Tr} \Phi^{2r}_{cl}$ for
$r=1,2,\cdots, N$. Also $u_{2r} \equiv \frac{1}{2r} \mbox{Tr}
\Phi^{2r}_{cl}$ that are independent. However, for $2r > 2N$, both
$\mbox{Tr} \Phi^{2r} $ and $\left\langle\mbox{Tr}
\Phi^{2r}\right\rangle$ can be written as the $u_r$ of $2r\leq 2N
$. The classical vacua can be obtained by putting all the
eigenvalues of $\Phi$ and $\Phi_{cl}$ equal to the roots of
$W^{\prime}(z)=\sum_{r=1}^{k+1} g_{2r}z^{2r-1}$. We will take the
degree of superpotential to be $2(k+1) \leq 2N$ first in which the
$u_r$ are independent and $\left\langle\mbox{Tr}
\Phi^{2r}\right\rangle = \mbox{Tr} \Phi^{2r}_{cl}$. Then we will
take the degree of superpotential to be arbitrary.

At a generic point on the Coulomb branch of the ${\cal N}=2$
theory, the low energy gauge group is $U(1)^{N}$. We study the
vacua in which the perturbation by $W(\Phi)$ (\ref{treesup})
remains only $U(1)^{n}$ gauge group at low energies, with $2n
\leq 2k$. This happens if the remaining degrees of freedom
become massive for nonzero $W$ due to the condensation of
$(N-n)$ mutually local magnetic monopoles or dyons. This
occurs only at points where at $W^{\prime}=0$ the monopoles are
massless on some particular submanifold $<u_{2r}>$. This can be
done by including the $(N-n)$ monopole hypermultiplets in the
superpotential. Then
 the exact effective superpotential by adding
(\ref{treesup}), near a point with $(N-n)$ massless monopoles,
is given by
\bea W_{eff}=\sqrt{2} \sum_{l=1}^{N-n}
M_{l}(u_{2r}) q_l \widetilde{q}_l + \sum_{r=1}^{k+1} g_{2r}
u_{2r}.
 \nonu \eea Here $q_l$ and $
\widetilde{q}_l$ are the monopole fields and $M_{l}(u_{2r})$ is
the mass of $l$-th monopole as a function of the $u_{2r}$. The
variation of $W_{eff}$ with respect to $q_l$ and $
\widetilde{q}_l$ vanishes. However, the variation of it with
respect to $u_{2r}$ does not lead to the vanishing of $q_l
\widetilde{q}_l$ and the mass of monopoles should vanish for
$l=1,2, \cdots, (N-n)$ in a supersymmetric vacuum. Therefore the
superpotential in this supersymmetric vacuum becomes
$W_{exact}=\sum_{r=1}^{k+1} g_{2r} <u_{2r}>$. The masses $M_i$ are
equal to the periods of some meromorphic one-form over some cycles
of the ${\cal N}=2$ hyperelliptic curve.

It is useful and convenient to consider a singular point in the
moduli space where $(N-n)$ mutually local monopoles are massless.
Then the ${\cal N}=2$ curve of genus $(2N-1)$ degenerates to a
curve of genus $2n$ \cite{eot,feng1} (See also \cite{bl}) 
and it is given by
\begin{eqnarray}
y^2=
P_{2N}^2(x)-4\Lambda^{4N-4}x^4=x^2H_{2N-2n-2}^2(x)F_{2(2n+1)}(x),
\label{curve}
\eea
where
\bea H_{2N-2n-2}(x)=
\prod^{N-n-1}_{i=1}(x^2-p_i^2), \quad F_{2(2n+1)}(x)
=\prod^{2n+1}_{i=1}(x^2-q_i^2)
\nonu
\eea
where
$H_{2N-2n-2}(x)$ is a polynomial in $x$ of degree $(2N-2n-2)$ that
gives $(2N-2n-2)$ double roots and $F_{2(2n+1)}(x)$ is a
polynomial in $x$ of degree $(4n+2)$ that is related to the
deformed superpotential (\ref{sol}). Both functions are even
functions in $x$. That is, a function of $x^2$ which is peculiar
to the gauge group $SO(2N)$. The characteristic function
$P_{2N}(x)$ that is also function of $x^2$ is described in terms
of the eigenvalues of $\Phi$ (\ref{Phi}) as follows:
\begin{eqnarray}
P_{2N}(x)&=&\det (x-\Phi_{cl})=
\prod_{I=1}^N(x^2-\phi^2_I). \nonu
\end{eqnarray}
The degeneracy of the above curve can be checked by computing both
$y^2$ and $\frac{\pa y^2}{\pa x^2}$ at the point $x=\pm p_i $ and
$x=0$ obtaining a zero. The factorization condition (\ref{curve})
can be described and encoded by Lagrange multipliers \cite{csw}.
When the degree $(2k+1)$ of $W^{\prime}(x)$ is equal to $(2n+1)$,
the highest $(2n+2)$ coefficients of $F_{2(2n+1)}(x)$ are given in
terms of $W^{\prime}(x)$ as follows \cite{eot,feng1}: \bea
F_{2(2n+1)}(x) = \frac{1}{g_{2n+2}^2} W^{\prime}_{2n+1}(x)^2
+{\cal O}(x^{2n}), \qquad 2k=2n.
\label{sol} \eea
The
superpotential $W(x)$ is known and one should know the ${\cal
N}=2$ vacua or $P_{2N}(x)$ through (\ref{curve}). The relation
(\ref{sol}) determines $F_{2(2n+1)}(x) $ in terms of $(2n+2)$
undetermined coefficients coming from ${\cal O}(x^{2n})$. Then
these can be obtained by demanding the existence of a polynomial
$H_{2N-2n-2}(x)$ via (\ref{curve}). In this case, there exists a
unique solution. We return to the explicit examples illustrating
the mechanism of this in section 2.4.

$\bullet$ {\bf Superpotential of degree $2(k+1)$ less than $2N$}

Now we study (\ref{sol}) when $2k > 2n$ by introducing the
constraints.
We follow the basic idea of \cite{csw} and repeat
the derivations of (\ref{sol}) and generalize to the arbitrary
degree of superpotential. That is, in the range $2n+2 \le 2k+2\le
2N$, let us consider the superpotential under these constraints
(\ref{curve}),
\begin{eqnarray}
W_{eff} & = & \sum_{r=1}^{k+1}g_{2r}u_{2r} \nonu \\
&+ & \sum_{i=0}^{2N-2n-2} \left[L_i\oint
\frac{P_{2N}(x)-2\epsilon_i x^2
\Lambda^{2N-2}}{(x-p_i)}dx+B_i\oint \frac{P_{2N}(x)-2\epsilon_i
x^2 \Lambda^{2N-2}}{(x-p_i)^2}dx \right]
 \label{eff}
\end{eqnarray}
where $L_i$ and $B_i$ are Lagrange multipliers imposing the
constraints and $\epsilon_i=\pm 1$. The contour integration
encloses all $p_i$'s and the factor $1/2\pi i$ is absorbed in the
symbol of $\oint$ for simplicity. The $p_i$'s where $i=0,1,2,
\cdots, (2N-2n-2)$ are the locations of the double roots of $y^2=
P_{2N}^2(x)-4\Lambda^{4N-4}x^4$. The $P_{2N}(x)$ depends on
$u_{2r}$. Note that massless monopole points appear in pair
$(p_i,-p_i)$ where $i=1,2,\cdots, (N-n-1)$. So we denote half of
the $p_i$'s by $p_{N-n-1+i}=-p_i, i=1,2, \cdots, (N-n-1)$.
Moreover we define $p_0=0$. Since $P_{2N}(x)$ is an even function,
if the following constraints are satisfied at $x=p_i$ where $
i=1,2, \cdots, (N-n-1)$,
\begin{eqnarray}
\left(P_{2N}(x) -2x^2\epsilon_i \Lambda^{2N-2}\right)|_{x=p_i}=0,
\qquad  \frac{\partial }{\partial x}\left(P_{2N}(x) -2x^2\epsilon_i
\Lambda^{2N-2} \right)|_{x=p_i}=0, \nonu
\end{eqnarray}
then they also automatically are satisfied at $x=-p_i$. Then the
numbers of constraint that we should consider are $(N-n-1)$. Thus
we denote the half of the Lagrange multipliers by
$L_{N-n-1+i}=L_i$ and $B_{N-n-1+i}=B_i$ where $i=1,2, \cdots,
(N-n-1)$. Due to the fact that the second derivative of $P_{2N}(x)
-2x^2\epsilon_i \Lambda^{2N-2}$ with respect to $x$ at $x=p_i$
does not vanish, there are no such higher order terms
$(x-p_i)^{-a}, a=3,4,5, \cdots$ in the effective superpotential
(\ref{eff}).

The variation of $W_{eff}$ with respect to $B_i$ leads to, by the
formula of contour integral,
\begin{eqnarray}
0&=&\oint \frac{P_{2N}(x) -2x^2\epsilon_i \Lambda^{2N-2}}
{(x-p_i)^2}dx=\left(P_{2N}(x) -2x^2\epsilon_i \Lambda^{2N-2}
\right)^{\prime} |_{x=p_i} \nonu \\
&=& \left(P_{2N}^{\prime}(x)-\frac{2}{x}2\epsilon_i x^2
\Lambda^{2N-2} \right)|_{x=p_i}  = \left(P_{2N}(x)\sum_{J=1}^N
\frac{2x}{x^2-\phi_J^2}-\frac{2}{x}P_{2N}(x)\right)|_{x=p_i} \nonu
\\
& = & P_{2N}(x) \left(\mbox{Tr} \frac{1}{x-\Phi_{cl}}-\frac{2}{x}
\right)|_{x=p_i} \label{relation7}
\end{eqnarray}
where we used the equation of motion for $L_i$
when we replace $2x^2\epsilon_i \Lambda^{2N-2}$ with $P_{2N}(x)$ at $x=p_i$.
The last equality
comes from the following relation, together with (\ref{Phi}),
\begin{eqnarray}
\mbox{Tr} \frac{1}{x-\Phi_{cl}}&=&\sum_{k=0}^{\infty} x^{-k-1}
\mbox{Tr}\Phi_{cl}^k=\sum_{i=0}^{\infty}x^{-(2i+1)} \mbox{Tr}
\Phi^{2i}_{cl}
 = \sum_{i=0}^{\infty}x  \left(x^2 \right)^{-(i+1)}\sum_{I=1}^N 2
\left(\phi_I^2\right)^i \nonu \\
&=& \sum_{I=1}^N \frac{2x}{x^2-\phi_I^2} \label{relation1}
\end{eqnarray}
where $\Phi_{cl}$ is antisymmetric matrix, the odd power terms are
vanishing. Since $P_{2N}(x=p_i) \neq 0$ for $i=1,2,\cdots ,(N-n-1)$ due to the relation
(\ref{curve}) and $H_{2N-2n-2}(x=p_i)=0$, we arrive at, from
(\ref{relation7}),
 \bea \left(\mbox{Tr}
\frac{1}{x-\Phi_{cl}}-\frac{2}{x} \right)|_{x=p_i}=0,\qquad P_{2N}(x=p_0)=0.
\nonu \eea
Note the presence of $2/x$ term which was not present in the
$U(N)$ case. This implies one can get and solve $p_i$ in terms of
$u_{2r}$.

As we have used the calculation in (\ref{relation7}),
the derivative $P_{2N}(x)$ with respect to $x$ is given by
\begin{eqnarray}
P_{2N}^{\prime}(x)=\left(\prod_{I}^{N}(x^2-\phi_I^2)\right)^{\prime}=2x\sum_{J=1}^N
\prod_{I\ne J}^N(x^2-\phi^2_I)=P_{2N}(x) \sum_{J=1}^N
\frac{2x}{x^2-\phi_J^2}. \nonu
\end{eqnarray}
Taking into account (\ref{relation1}), we can rewrite this result
as,
\begin{eqnarray}
\frac{P_{2N}^{\prime}(x)}{P_{2N}(x)}=
\mbox{Tr}\frac{1}{x-\Phi_{cl}}.
\nonu
\end{eqnarray}
Using (\ref{relation1}), we have
\begin{eqnarray}
\frac{P_{2N}^{\prime}(x)}{P_{2N}(x)}=\sum_{i=0}^{\infty}x^{-(2i+1)}
\mbox{Tr} \Phi^{2i}_{cl} =\sum_{i=0}^{\infty}x^{-(2i+1)} 2i
u_{2i}. \nonu
\end{eqnarray}
After an integration in $x$, we obtain the following result,
\begin{eqnarray}
P_{2N}(x)= \left[x^{2N}\exp
\left(-\sum_{r=1}^{\infty}\frac{u_{2r}}{x^{2r}} \right)
\right]_{+} \label{p2n}
\end{eqnarray}
where we choose an integration constant to zero. This enables us
to write $u_{2r}$ with $2r > 2N$ in terms of $u_{2r}$ with $2r
\leq 2N$ by requiring the vanishing of negative powers of $x$. The
derivative of $P_{2N}(x)$ with respect to $u_{2r}$ is given by
\begin{eqnarray}
\frac{\partial P_{2N}(x)}{\partial
u_{2r}}=-\left[\frac{P_{2N}(x)}{x^{2r}} \right]_{+} \nonu
\end{eqnarray}
where $+$ means the polynomial part of a Laurent series of the
expression inside the bracket. Although  we get
$W_{low}=W_{low}(g_{2r},\La)$ by minimizing
$W_{eff}=W_{eff}(u_{2r},p_i,L_i,B_i;g_{2r},\La)$, we would like to
know the information about $F_{2(2n+1)}(x)$ at the minimum by
looking at the field equations.

Next we consider variations of $W_{eff}$ with respect to $p_j$,
\begin{eqnarray}
0&=&L_j\oint \frac{P_{2N}(x) -2x^2\epsilon_i
\Lambda^{2N-2}}{(x-p_j)^2}dx+2B_j \oint \frac{P_{2N}(x)
-2x^2\epsilon_i \Lambda^{2N-2}}{(x-p_j)^3}dx
\nonu \\
&=&2B_j \oint \frac{P_{2N}(x) -2x^2\epsilon_i
\Lambda^{2N-2}}{(x-p_j)^3}dx. \nonu
\end{eqnarray}
In the last equation we have used the equation of motion for $B_i$
(\ref{relation7}). In general this integral does not vanish, then
we should have $B_i=0$ because the curve does not have more than
cubic roots since $y^2$ contains a polynomial $H^2_{2N-2n-2}(x)$.
We consider variations of $W_{eff}$ with respect to $u_{2r}$,
\begin{eqnarray}
0=g_{2r}-\sum_{i=0}^{2N-2n-2}\oint \left[\frac{P_{2N}}{x^{2r}}
\right]_{+} \frac{L_i}{x-p_i}dx, \nonu
\end{eqnarray}
where we used $B_i=0$ at the level of equation of motion.
Multiplying this with $z^{2r-1}$ and summing over $r$ where $z$ is
inside the contour of integration we can obtain the following
relation,
\begin{eqnarray}
W^{\prime}(z)&=&\sum_{r=1}^{k+1}
g_{2r}z^{2r-1}=\sum_{i=0}^{2N-2n-2}\oint
\sum_{r=1}^{k+1}z^{2r-1}\frac{P_{2N}(x)}{x^{2r}}
\frac{L_i}{(x-p_i)}dx. \label{relation2}
\end{eqnarray}

Let us introduce a new polynomial $Q(x)$ defined as
\begin{eqnarray}
\sum_{i=0}^{2N-2n-2}\frac{xL_i}{(x-p_i)} & = & L_0+
\sum_{i=1}^{N-n-1}xL_i\left(\frac{1}{x-p_i}+\frac{1}{x+p_i}
\right)=L_0+\sum_{i=1}^{N-n-1}\frac{2x^2L_i}{x^2-p_i^2} \nonu \\
& \equiv & \frac{Q(x)}{H_{2N-2n-2}(x)}. \label{relation6}
\end{eqnarray}
This determines the degree of $Q(x)$ is equal to $(2N-2n-2)$ which
is greater than and equal to $(2k-2n)$. Using this new function we
can rewrite (\ref{relation2}) as
\begin{eqnarray}
W^{\prime}(z)=\oint \sum_{r=1}^{k+1}\frac{z^{2r-1}}{x^{2r}}
\frac{Q(x)P_{2N}(x)}{xH_{2N-2n-2}(x)}dx. \nonu
\end{eqnarray}
Since $W^{\prime}(z)$ is a polynomial of degree $(2k+1)$, we found
the order of $Q(x)$ as $(2k-2n)$, so we denote it by
$Q_{2k-2n}(x)$. Thus we have found the order of polynomial $Q(x)$
and therefore the order of integrand in (\ref{relation2}) is like
as ${\cal O}(x^{2k-(2r+1)})$. Thus if $r\ge k+1$ it does not
contribute to the integral because the power of $x$ in this region
implies that the Laurent expansion around the origin vanishes. We
can replace the upper value of summation with the infinity.
\begin{eqnarray}
W^{\prime}(z)=\oint \sum_{r=1}^{\infty}\frac{z^{2r-1}}{x^{2r}}
\frac{Q_{2k-2n}(x)P_{2N}(x)}{xH_{2N-2n-2}(x)}dx=\oint z
\frac{Q_{2k-2n}(x)P_{2N}(x)}{x(x^2-z^2)H_{2N-2n-2}(x)}dx.
\label{prires}
\end{eqnarray}
From (\ref{curve}) we have the relation,
\begin{eqnarray}
P_{2N}(x)=x\sqrt{F_{2(2n+1)}(x)}H_{2N-2n-2}(x)+{\cal
O}(x^{-2N+4}). \nonu
\end{eqnarray}
As we substitute this relation to (\ref{prires}), the ${\cal
O}(x^{-2N+4})$ terms do not contribute the integral because the
contribution of the Laurent expansion around the origin for the
power of $x$ vanishes, so we can drop those terms. Therefore we
have
\begin{eqnarray}
W^{\prime}(z)=\oint z\frac{y_m(x)}{x^2-z^2}dx, \qquad
y_m^2(x)=F_{2(2n+1)}(x)Q_{2k-2n}^2(x) \nonu
\end{eqnarray}
corresponding to the equation of motion for the matrix model and
the matrix model curve respectively.
%
Then we get an expected and generalized result,
\begin{eqnarray}
y_m^2(x)& = & F_{2(2n+1)}(x)Q_{2k-2n}^2(x)
={W^{\prime}}^2_{2k+1}(x)+{\cal O}(x^{2k}) \nonu \\
& = & {W^{\prime}}^2_{2k+1}(x)+f_{2k}(x), \qquad
 2n+2 \le 2k+2\le 2N \nonu
\end{eqnarray}
where both $F_{2(2n+1)}(x)$ and $Q_{2k-2n}(x)$ are functions of
$x^2$, then $f_{2k}(x)$  also a function of $x^2$. We put the
subscript $m$ in the $y_m$ in order to emphasize the fact that
this corresponds to the matrix model curve. When $2k+1=2n+1$, we
reproduce (\ref{sol}) with $Q_0= g_{2n+2}$. The above relation
determines a polynomial $F_{2(2n+1)}(x)$ in terms of $(2n+1)$
unknown parameters by assuming the leading coefficient to be
normalized by 1 by assuming that $W(x)$ is known. These parameters
can be obtained from both $P_{2N}(x)$ and $H_{2N-2n-2}(x)$ through
the factorization condition (\ref{curve}). Now we move on for the
case of large $k$ next section.

$\bullet$ {\bf Superpotential of degree $2(k+1)$ where $k$ is
arbitrary large}

So far we restricted to the case where the degree of
superpotential, $2(k+1)$, is less than $2N$. Let us extend the
study of previous discussion to general case by allowing a
superpotential of any degree. All the $u_i$'s are independent
coordinates on a bigger space with appropriate constraints. For
this range of degree of polynomial, we have to consider different
constraints due to the instanton corrections \cite{csw}. We denote
$\frac{1}{2k}\langle \mbox{Tr} \Phi^{2k} \rangle=U_{2k}$ by
capital letter. The quantum mechanical expression corresponding to
$\mbox{Tr}\frac{1}{x-\Phi} $ is given by
\begin{eqnarray}
\left\langle \mbox{Tr}\frac{1}{x-\Phi} \right\rangle=\frac{2N}{x}+
\sum_{k=1}^{\infty}\frac{2kU_{2k}}{x^{2k+1}}. \nonu
\end{eqnarray}
 Quantum mechanically we do have
\begin{eqnarray}
\left\langle \mbox{Tr}\frac{1}{x-\Phi} \right\rangle=\frac{d}{dx}
\log \left(P_{2N}(x)+\sqrt{P_{2N}^2(x)-4x^4\Lambda^{4N-4}} \right).
\label{10may2}
\end{eqnarray}
By integrating over $x$ and exponentiating, we get
\begin{eqnarray}
Cx^{2N}\exp\left(-\sum_{i=1}^{\infty}\frac{U_{2i}}{x^{2i}} \right)=
P_{2N}(x)+\sqrt{P_{2N}^2(x)-4x^4\Lambda^{4N-4}} \label{relation3}
\end{eqnarray}
where $C$ is an integration constant that can be determined by the
semiclassical limit $\Lambda \to 0$:
\begin{eqnarray}
Cx^{2N}\exp\left(-\sum_{i=1}^{\infty}\frac{U_{2i}}{x^{2i}}
\right)=2P_{2N}(x) \longrightarrow C=2. \nonu
\end{eqnarray}
Solving (\ref{relation3}) with respect to $P_{2N}(x)$,
\begin{eqnarray}
P_{2N}(x)=
x^{2N}\exp\left(-\sum_{i=1}^{\infty}\frac{U_{2i}}{x^{2i}}
\right)+\frac{\Lambda^{4N-4}}{x^{2N-4}}\exp\left(\sum_{i=1}^{\infty}
\frac{U_{2i}}{x^{2i}} \right), \label{relation8}
\end{eqnarray}
corresponding to (\ref{p2n}). Since $P_{2N}(x)$ is a polynomial in
$x$, (\ref{relation8}) can be used to express $U_{2r}$ with
$2r>2N$ in terms of $U_{2r}$ with $2r\le 2N$ by imposing the
vanishing of the negative powers of $x$. Let us introduce a new
polynomial whose coefficients are Lagrange multipliers. The
generalized superpotential with these constraints is described as
\begin{eqnarray}
W_{eff}&=&\sum_{r=1}^{k+1}g_{2r}U_{2r}+\sum_{i=0}^{2N-2n-2}
\left[L_i\oint \frac{P_{2N}(x)-2\epsilon_i x^2
\Lambda^{2N-2}}{(x-p_i)}dx+B_i\oint
\frac{P_{2N}(x)-2\epsilon_i x^2 \Lambda^{2N-2}}{(x-p_i)^2}dx \right] \nonu \\
&+&\oint R_{2k-2N+2}(x) \left[
x^{2N}\exp\left(-\sum_{i=1}^{\infty} \frac{U_{2i}}{x^{2i}}
\right)+\frac{\Lambda^{4N-4}}{x^{2N-4}}\exp
\left(\sum_{i=1}^{\infty}\frac{U_{2i}}{x^{2i}} \right) \right] dx
\nonu
\end{eqnarray}
where $R_{2k-2N+2}(x)$ is a polynomial of degree $(2k-2N+2)$ whose
coefficients are regarded as Lagrange multipliers which impose
constraints $U_{2r}$ with $2r > 2N$ in terms of $U_{2r}$ with $2r
\leq 2N$. This expression is a generalization of (\ref{eff}).

Now we follow the previous method.
The derivative of $W_{eff}$ with respect to $U_{2r}$ leads to
\begin{eqnarray}
0&=&g_{2r}+\oint \frac{R_{2k-2N+2}(x)}{x^{2r}}\left( -x^{2N}
\exp\left(-\sum_{i=1}^{\infty}\frac{U_{2i}}{x^{2i}} \right)+
\frac{\Lambda^{4N-4}}{x^{2N-4}}\exp\left(\sum_{i=1}^{\infty}\frac{U_{2i}}{x^{2i}}
\right) \right)dx
\nonu \\
&+& \oint \sum_{i=0}^{2N-2n-2}\frac{L_i}{(x-p_i)}\frac{\partial
P_{2N}(x)}{\partial U_{2r}}dx. \label{relation4}
\end{eqnarray}
Using (\ref{relation3}) we have the relation,
\begin{eqnarray}
-x^{2N}\exp\left(-\sum_{i=1}^{\infty}\frac{U_{2i}}{x^{2i}}
\right)+\frac{\Lambda^{4N-4}}{x^{2N-4}}\exp\left(\sum_{i=1}^{\infty}
\frac{U_{2i}}{x^{2i}} \right)=-\sqrt{P_{2N}^2(x)-4x^4\Lambda^{4N-4}}
\nonu
\end{eqnarray}
and
\begin{eqnarray}
\frac{\partial P_{2N}(x) }{\partial U_{2r}}=
-\frac{P_{2N}(x)}{x^{2r}} \;\; \mbox{for} \;\; 2r\le 2N, \qquad
\frac{\partial P_{2N}(x) }{\partial U_{2r}}=0 \;\; \mbox{for} \;\;
2r> 2N. \nonu
\end{eqnarray}
Using these relations we can rewrite (\ref{relation4}) and
simplify as follows:
\begin{eqnarray}
0=g_{2r}+\oint \frac{R_{2k-2N+2}(x)}{x^{2r}}\left( -\sqrt{
P_{2N}^2(x)-4x^4\Lambda^{4N-4}}\right)dx- \oint
\sum_{i=0}^{2N-2n-2} \frac{L_i}{(x-p_i)}\frac{P_{2N}(x)}{x^{2r}} dx.
\label{relation10}
\end{eqnarray}
From the massless monopole constraint (\ref{curve}) the
characteristic function has the following form,
\begin{eqnarray}
P_{2N}(x)=x H_{2N-2n-2}(x) \sqrt{F_{2(2n+1)}(x)}+{\cal
O}(x^{-2N+4}). \label{relation9}
\end{eqnarray}
Substituting (\ref{relation9}) and (\ref{curve}) into
(\ref{relation10}), the ${\cal O}(x^{-2N+4})$ terms do not
contribute to integral as we have done before, then one gets the
coupling $g_{2r}$ as follows:
\begin{eqnarray} g_{2r}&=&\oint
\frac{R_{2k-2N+2}(x)}{x^{2r}}xH_{2N-2n-2}(x)\sqrt{F_{2(2n+1)}(x)}dx
\nonu \\ &+& \oint \sum_{i=0}^{2N-2n-2}\frac{L_i}{(x-p_i)}\frac{x
H_{2N-2n-2}(x)\sqrt{F_{2(2n+1)}(x)}}{x^{2r}} dx. \nonu
\end{eqnarray}
As in previous analysis, by multiplying $z^{2r-1}$ and summing
over $r$, we will get
\begin{eqnarray}
W^{\prime}(z)&=&\sum_{r=1}^{k+1}\left(\oint
\frac{R_{2k-2N+2}(x)}{x^{2r}} xH_{2N-2n-2}(x)
\sqrt{F_{2(2n+1)}(x)}dx \right. \nonu \\
&+& \left. \oint \sum_{i=0}^{2N-2n-2}\frac{L_i}{(x-p_i)}
\frac{xH_{2N-2n-2}(x)\sqrt{F_{2(2n+1)}(x)}}{x^{2r}}dx
\right)z^{2r-1}
\nonu \\
&=&\oint \frac{z}{x^2-z^2}\sqrt{F_{2(2n+1)}(x)}
\left[R_{2k-2N+2}(x)H_{2N-2n-2}(x)+ R_{2N-2n-2}(x) \right]dx.
\nonu
\end{eqnarray}
Finally we obtain an expected relation of equations of motion for
the general order tree level superpotential case,
\begin{eqnarray}
W^{\prime}(z)=\oint \frac{zy_m}{x^2-z^2}dx \nonu
\end{eqnarray}
if the matrix model curve is given by
\begin{eqnarray}
y_m^2 &=& F_{2(2n+1)}(x){\widetilde{Q}}_{2k-2n}^2(x) \nonu \\
&\equiv& F_{2(2n+1)}(x)\left(R_{2k-2N+2}(x)H_{2N-2n-2}(x)+
R_{2N-2n-2}(x)\right)^2. \nonu
\end{eqnarray}
When $2n=2N-2$ (no massless monopoles),
${\widetilde{Q}}_{2k-2N+2}(x)=R_{2k-2N+2}(x)$. When the degree of
superpotential is equal to $2N$, in other words, $2k=2N-2$, then
${\widetilde{Q}}_{2k-2n}(x)=R_{0} H_{2N-2n-2}(x)+ R_{2N-2n-2}(x)$.
In particular, for $2n=2N-2$, ${\widetilde{Q}}_{0}$ is a constant
and $y_m^2(x)=F_{4N-2}(x)=x^{-2} (P_{2N}^2(x) -4 \La^{4N-4} x^4)$.
The above $y_m^2$ is known up to a polynomial  $f_{2k}(x)$ of
degree $2k$: $y_m^2= {W^{\prime}}^2_{2k+1}(x)+f_{2k}(x)$.

$\bullet$ {\bf A generalized Konishi anomaly}


Now we are ready to study for a derivation of the generalized
Konishi anomaly equation based on the results of previous section.
As in \cite{csw}, we restrict to the case with $\left\langle
\mbox{Tr} W^{\prime}(\Phi) \right\rangle=\mbox{Tr}
W^{\prime}(\Phi_{cl})$ and assume that the degree of
superpotential $(2k+2)$ is less than $2N$. By substituting
(\ref{relation6}) into (\ref{prires}) we can write the derivative
of  superpotential $W^{\prime}(\phi_I)$
\begin{eqnarray}
W^{\prime}(\phi_I)=\sum_{i=0}^{2N-2n-2} \oint
\phi_I\frac{P_{2N}(x)}{(x^2-\phi^2_I)} \frac{L_i}{(x-p_i)}dx
\label{10may}
\end{eqnarray}
where we varied $W(\phi_I)$ with respect to $\phi_I$ rather than
$u_{2r}$ and used the result of $B_i=0$. Note that $P_{2N}(x)=
\prod_{I=1}^N(x^2-\phi^2_I)$. Using this equation, one obtains the
following relation,
\begin{eqnarray}
\mbox{Tr} \frac{W^{\prime}(\Phi_{cl})}{z-\Phi_{cl}}&=&\mbox{Tr}
\sum_{k=0}^{\infty}z^{-k-1}\Phi_{cl}^kW^{\prime} (\Phi_{cl})  =
\sum_{i=0}^{\infty}z^{-(2i+1)-1}2
\sum_{I=1}^N \phi_I^{2i+1}W^{\prime}(\phi_I) \nonu \\
&=&2\sum_{I=1}^N \phi_I W^{\prime}(\phi_I)\frac{1}{(z^2-\phi_I^2)}
\nonu \\
& = & \sum_{I=1}^N
\frac{2\phi_I^2}{(z^2-\phi_I^2)}\sum_{i=0}^{2N-2n-2} \oint
\frac{P_{2N}(x)}{(x^2-\phi_I^2)}\frac{L_i}{(x-p_i)}dx,
\label{koni}
\end{eqnarray}
where when we change the summation index from $k$ to $i$, the only
odd terms appear because effectively the product of $\Phi_{cl}$
and $W^{\prime}(\Phi_{cl})$ does contribute only under  that
condition. The even terms do not contribute. Here $z$ is outside
the contour of integration.
In order to compute the above expression we recognize that the
following factor can be written as, by simple manipulation between
the property of the trace we have seen before,
\begin{eqnarray}
\sum_{I=1}^N\frac{2\phi_I^2}{(z^2-\phi_I^2)(x^2-\phi_I^2)}=
\frac{1}{(x^2-z^2)}\left(z\mbox{Tr}\frac{1}{z-\Phi_{cl}}-x\mbox{Tr}
\frac{1}{x-\Phi_{cl}} \right).
\nonu
\end{eqnarray}
Thus we can write (\ref{koni}) as
\begin{eqnarray}
\mbox{Tr} \frac{W^{\prime}(\Phi_{cl})}{z-\Phi_{cl}}=\oint
\sum_{i=0}^{2N-2n-2}\frac{P_{2N}(x)L_i}{(x^2-z^2)(x-p_i)}\left(z\mbox{Tr}
\frac{1}{z-\Phi_{cl}}-x\mbox{Tr}\frac{1}{x-\Phi_{cl}} \right)dx.
\label{koni4}
\end{eqnarray}
As in the case of \cite{csw}, we can rewrite outside contour
integral in terms of two parts as follows:
\begin{eqnarray}
\oint_{z_{out}}=
\oint_{z_{in}}-\oint_{C_z+C_{-z}}
\label{contour}
\end{eqnarray}
where $C_z$ and $C_{-z}$ are small contour around $z$ and $-z$
respectively. Thus the first term in (\ref{koni4}) (corresponding to the
second term in $(B.3)$ of \cite{csw}) can be written as
\begin{eqnarray}
\mbox{Tr}\frac{1}{z-\Phi_{cl}}\oint_{z_{out}}
\frac{zQ_{2k-2n}(x)P_{2N}(x)}{x H_{2N-2n-2}(x)(x^2-z^2)}dx \nonu
\eea by using the relation (\ref{relation6}). Let us emphasize
that in this case, we cannot drop the terms of order ${\cal
O}(x^{-2N+4})$. In order to deal with this,  we have to use the
above change of integration, then this is given by
\bea
&&\mbox{Tr}\frac{1}{z-\Phi_{cl}} \left( \oint_{z_{in}}
\frac{zQ_{2k-2n}(x)P_{2N}(x)}{xH_{2N-2n-2}(x)(x^2-z^2)}dx-
\oint_{C_{z}+C_{-z}}
\frac{zQ_{2k-2n}(x)P_{2N}(x)}{xH_{2N-2n-2}(x)(x^2-z^2)}dx \right) \nonu \\
&& = \mbox{Tr}\frac{1}{z-\Phi_{cl}} \left(W^{\prime}(z)- \frac{y_m(z)
P_{2N}(z)}{\sqrt{P_{2N}^2(z)-4z^4\Lambda^{4N-4}}} \right), \nonu
\end{eqnarray}
where the first term was obtained by the method done previously
and in the last equality we used (\ref{prires}) and
\begin{eqnarray}
H_{2N-2n-2}(z)=\frac{\sqrt{P_{2N}^2(z)-4\Lambda^{4N-4}z^4}}
{z\sqrt{F_{2(2n+1)}(z)}}, \qquad y_m^2(z) =
F_{2(2n+1)}(z) Q^2_{2k-2n}(z).
\nonu
\end{eqnarray}
The second term was calculated at the poles.
The crucial difference between $U(N)$ case and $SO(2N)$ case comes
from the second term of (\ref{koni4}) (corresponding to the first
term in $(B.3)$ of \cite{csw}), which vanishes in $U(N)$ case. Let
us write it as, after an integration over $x$,
\begin{eqnarray}
-\sum_{i=0}^{2N-2n-2} \oint \frac{L_iP_{2N}(x)x}
{(x-p_i)(x^2-z^2)}\mbox{Tr}\frac{1}{x-\Phi_{cl}}dx
=-\sum_{i=0}^{2N-2n-2} \frac{L_i p_i P_{2N}(x=p_i)}
{(p_i^2-z^2)}\mbox{Tr}\frac{1}{p_i-\Phi_{cl}}. \nonu \eea Now we
use the result of the equation of motion for $B_i$
(\ref{relation7}) in order to change the trace part and arrive at
the final contribution as follows: \bea -\sum_{i=0}^{2N-2n-2}
2\frac{L_i P_{2N}(x=p_i)}{(p_i^2-z^2)} =-\sum_{i=0}^{2N-2n-2}
\oint \frac{2P_{2N}(x)}{(x^2-z^2)}
\frac{L_i}{(x-p_i)}dx \nonu
\end{eqnarray}
where note that $z$ is {\it outside} the contour of integration in last equation. Therefore taking into account for (\ref{contour}) and (\ref{10may}) we can rewrite the last equation as follows:
\begin{eqnarray}
-2\frac{W^{\prime}(z)}{z}+\sum_{i=0}^{2N-2n-2}\oint_{C_{z}+C_{-z}}\frac{2P_{2N}(x)}{(x^2-z^2)}
\frac{L_i}{(x-p_i)}dx =-2\frac{W^{\prime}(z)}{z}+\frac{2}{z}\frac{y_mP_{2N}(z)}{\sqrt{P_{2N}^2(z)-4z^4\Lambda^{4N-4}}} \nonu
\end{eqnarray}
where once again we substituted (\ref{relation6}) into
(\ref{prires}). Therefore we obtain \bea \mbox{Tr}
\frac{W^{\prime}(\Phi_{cl})}{z-\Phi_{cl}}=\mbox{Tr}\frac{1}{z-\Phi_{cl}}
\left(W^{\prime}(z)- \frac{y_m(z)
P_{2N}(z)}{\sqrt{P_{2N}^2(z)-4z^4\Lambda^{4N-4}}}
\right)-2\frac{W^{\prime}(z)}{z}+\frac{2}{z}\frac{y_mP_{2N}(z)}{\sqrt{P_{2N}^2(z)-4z^4\Lambda^{4N-4}}}. \nonu \eea
Remembering that \bea \mbox{Tr}\frac{1}{z-\Phi_{cl}}=\frac{P_{2N}^{\prime}(z)}{P_{2N}(z)}, \nonu \eea
the second and fourth terms are rewritten as follows;
\begin{eqnarray}
- \mbox{Tr}\frac{1}{z-\Phi_{cl}}\frac{y_m(z)
P_{2N}(z)}{\sqrt{P_{2N}^2(z)-4z^4\Lambda^{4N-4}}}+\frac{2}{z}\frac{y_mP_{2N}(z)}{\sqrt{P_{2N}^2(z)-4z^4\Lambda^{4N-4}}}=\frac{2y_m}{z}-\left\langle \tr \frac{y_m}{z-\Phi} \right\rangle \nonu
\end{eqnarray}
where we used (\ref{10may2}).

Taking into account the relation,
\begin{eqnarray}
\tr \frac{W^{\prime}(\Phi_{cl})-W^{\prime}(z)}{z- \Phi_ {cl}}=
\left\langle \mbox{Tr}
\frac{W^{\prime}(\Phi)-W^{\prime}(z)}{z-\Phi} \right \rangle \nonu
\end{eqnarray}
we can write  as follows:
\begin{eqnarray}
 \left\langle \tr \frac{W^{\prime}(\Phi)}{z-\Phi} \right\rangle &
=& \left( \left\langle \tr \frac{1}{z-\Phi} \right \rangle
-\frac{2}{z} \right) \left[W^{\prime}(z)-y_m(z)\right] \nonu
\end{eqnarray}
which is the generalized Konishi anomaly equation for $SO(2N)$
case. The resolvent of the matrix model $R(z)$ is related to
$W^{\prime}(z)-y_m(z)$.



\subsection{Strong gauge coupling approach:$SO(2N+1)$ case}

As in $SO(2N)$ case, the perturbed superpotential we are
considering  is given by (\ref{treesup}) and the antisymmetric
matrix takes the form
\bea \Phi= \left( {0 \atop -1 }{ 1 \atop 0
 } \right) \otimes \mbox{diag} ( i\phi_1, \cdots,  i\phi_{N},0).
\nonu
\eea Note that 0 element in the above. The corresponding
${\cal N}=2$ curve is characterized by (See also \cite{ds1})
\begin{eqnarray}
y^2=
P_{2N}^2(x)-4\Lambda^{4N-2}x^2=x^2H_{2N-2n-2}^2(x)F_{2(2n+1)}(x),
\label{curveodd} \eea where the polynomials $H_{2N-2n-2}(x)$ and
$F_{2(2n+1)}(x)$ are defined as before. The difference appears in
the left hand side only: the power of $\La$ and $x$. In this case
also, the solutions for $F_{2(2n+1)}(x)$ are given by (\ref{sol})
when the degree of $W^{\prime}$, $(2k+1)$, is equal to $(2n+1)$.
Now one can generalize for different range of the degree of
superpotential.

$\bullet$ {\bf Superpotential of degree $2(k+1)$ less than $2N$}

The superpotential with appropriate constraints (\ref{curveodd})
can be written, by noting the power behavior of $x$ and $\La$, as
follows:
\begin{eqnarray}
W_{eff} & = & \sum_{r=1}^{k+1}g_{2r}u_{2r} \nonu \\
&+ & \sum_{i=0}^{2N-2n-2} \left[L_i\oint
\frac{P_{2N}(x)-2\epsilon_i x \Lambda^{2N-1}}{(x-p_i)}dx+B_i\oint
\frac{P_{2N}(x)-2\epsilon_i x \Lambda^{2N-1}}{(x-p_i)^2}dx
\right].
\nonu
\end{eqnarray}
All the arguments discussed before hold for $SO(2N+1)$ case and
the constraints become
\begin{eqnarray}
\left(P_{2N}(x) -2x\epsilon_i \Lambda^{2N-1}\right)|_{x=p_i}=0,
\qquad  \frac{\partial }{\partial x}\left(P_{2N}(x) -2x\epsilon_i
\Lambda^{2N-1} \right)|_{x=p_i}=0. \nonu
\end{eqnarray}
Note the behavior of $\La$ term which is  linear in $x$.

The variation of $W_{eff}$ with respect to $B_i$ leads to
\begin{eqnarray}
0&=&\oint \frac{P_{2N}(x) -2x\epsilon_i \Lambda^{2N-1}}
{(x-p_i)^2}dx=\left(P_{2N}(x) -2x\epsilon_i \Lambda^{2N-1}
\right)^{\prime} |_{x=p_i} \nonu \\
&=& \left(P_{2N}^{\prime}(x)-\frac{1}{x}2\epsilon_i x
\Lambda^{2N-1} \right)|_{x=p_i}  = \left(P_{2N}(x)\sum_{J=1}^N
\frac{2x}{x^2-\phi_J^2}-\frac{1}{x}P_{2N}(x)\right)|_{x=p_i} \nonu
\\
& = & P_{2N}(x) \left(\mbox{Tr} \frac{1}{x-\Phi}-\frac{1}{x}
\right)|_{x=p_i}
\label{relation7odd}
\end{eqnarray}
where we used the equation of motion for $L_i$ when we replace
$2x\epsilon_i \Lambda^{2N-1}$ with $P_{2N}(x)$ at $x=p_i$. Since
$P_{2N}(x=p_i) \neq 0$ due to the relation (\ref{curveodd}) and
$H_{2N-2n-2}(x=p_i)=0$, we arrive at, from (\ref{relation7odd}),
\bea \left(\mbox{Tr} \frac{1}{x-\Phi}-\frac{1}{x}
\right)|_{x=p_i}=0, \qquad P_{2N}(x=p_0)=0.
\nonu
\eea
Note the
presence of $1/x$ term which was not present in the $U(N)$ case
and also the coefficient 1 is different from 2 in the case of
$SO(2N)$ we have considered before. The properties of the
characteristic function $P_{2N}(x)$ hold. The variation of
$W_{eff}$ with respect to $p_j$ can be obtained and the expression
looks similar except the power of $x$ and $\La$. That is,
\begin{eqnarray}
0=2B_j \oint \frac{P_{2N}(x) -2x\epsilon_i
\Lambda^{2N-1}}{(x-p_j)^3}dx \nonu
\end{eqnarray}
together with the equation of motion for $B_i$
(\ref{relation7odd}). This implies the vanishing of $B_i$ since
the integration over $x$ does not vanish in general, according to
the same reason as the one in $SO(2N)$ case. The variation of
$W_{eff}$ with respect to $u_{2r}$ can be done and from the
expression for $g_{2r}$ we can construct $W^{\prime}(z)$. By
considering the correct change of the upper bound in the summation
we arrive at the same expression (\ref{prires}). The contribution
of the terms ${\cal O}(x^{-2N+2})$ does not appear in the
evaluation of an integration and therefore one obtains
\begin{eqnarray}
W^{\prime}(z)=\oint z\frac{y_m(x)}{x^2-z^2}dx, \qquad
y_m^2(x)=F_{2(2n+1)}(x)Q_{2k-2n}^2(x) \nonu
\end{eqnarray}
corresponding to the equation of motion for the matrix model and
the matrix model curve respectively. Then we get a generalized
result,
\begin{eqnarray}
y_m^2(x)=F_{2(2n+1)}(x)Q_{2k-2n}^2(x)
={W^{\prime}}^2_{2k+1}(x)+{\cal O}(x^{2k})=
{W^{\prime}}^2_{2k+1}(x)+f_{2k}(x) \nonu
\end{eqnarray}
where both $F_{2(2n+1)}(x)$ and $Q_{2k-2n}(x)$ are functions of
$x^2$, then $f_{2k}(x)$  also a function of $x^2$.

$\bullet$ {\bf Superpotential of degree $2(k+1)$ where $k$ is
arbitrary large}


Quantum mechanically we  have
\begin{eqnarray}
\left\langle \mbox{Tr}\frac{1}{x-\Phi} \right\rangle=\frac{d}{dx}
\log \left(P_{2N}(x)+\sqrt{P_{2N}^2(x)-4x^2\Lambda^{4N-2}}
\right). \label{10may2BB}
\end{eqnarray}
Solving this with respect to $P_{2N}(x)$, one gets, with different
power behavior in the second term below,
\begin{eqnarray}
P_{2N}(x)=
x^{2N}\exp\left(-\sum_{i=1}^{\infty}\frac{U_{2i}}{x^{2i}}
\right)+\frac{\Lambda^{4N-2}}{x^{2N-2}}\exp\left(\sum_{i=1}^{\infty}
\frac{U_{2i}}{x^{2i}} \right).
\nonu
\end{eqnarray}
Let us introduce a new polynomial whose coefficients are Lagrange
multipliers as usual. The superpotential with these constraints is
described as
\begin{eqnarray}
W_{eff}&=&\sum_{r=1}^{k+1}g_{2r}U_{2r}+\sum_{i=0}^{2N-2n-2}
\left[L_i\oint \frac{P_{2N}(x)-2\epsilon_i x
\Lambda^{2N-1}}{(x-p_i)}dx+B_i\oint
\frac{P_{2N}(x)-2\epsilon_i x \Lambda^{2N-1}}{(x-p_i)^2}dx \right] \nonu \\
&+&\oint R_{2k-2N+2}(x) \left[
x^{2N}\exp\left(-\sum_{i=1}^{\infty} \frac{U_{2i}}{x^{2i}}
\right)+\frac{\Lambda^{4N-2}}{x^{2N-2}}\exp
\left(\sum_{i=1}^{\infty}\frac{U_{2i}}{x^{2i}} \right) \right] dx
\nonu
\end{eqnarray}
where $R_{2k-2N+2}(x)$ is a polynomial of degree $(2k-2N+2)$ whose
coefficients are regarded as Lagrange multipliers which impose
constraints $U_{2r}$ with $2r > 2N$ in terms of $U_{2r}$ with $2r
\leq 2N$.

The derivative of $W_{eff}$ with respect to $U_{2r}$ leads to
\begin{eqnarray}
0&=&g_{2r}+\oint \frac{R_{2k-2N+2}(x)}{x^{2r}}\left( -x^{2N}
\exp\left(-\sum_{i=1}^{\infty}\frac{U_{2i}}{x^{2i}} \right)+
\frac{\Lambda^{4N-2}}{x^{2N-2}}
\exp\left(\sum_{i=1}^{\infty}\frac{U_{2i}}{x^{2i}}
\right) \right)dx
\nonu \\
&+& \oint \sum_{i=0}^{2N-2n-2}\frac{L_i}{(x-p_i)}\frac{\partial
P_{2N}(x)}{\partial U_{2r}}dx.
\label{relation4odd}
\end{eqnarray}
Therefore, we have the relation,
\begin{eqnarray}
-x^{2N}\exp\left(-\sum_{i=1}^{\infty}\frac{U_{2i}}{x^{2i}}
\right)+\frac{\Lambda^{4N-2}}{x^{2N-2}}\exp\left(\sum_{i=1}^{\infty}
\frac{U_{2i}}{x^{2i}}
\right)=-\sqrt{P_{2N}^2(x)-4x^2\Lambda^{4N-2}} \nonu
\end{eqnarray}
and
\begin{eqnarray}
\frac{\partial P_{2N}(x) }{\partial U_{2r}}=
-\frac{P_{2N}(x)}{x^{2r}} \;\; \mbox{for} \;\; 2r\le 2N, \qquad
\frac{\partial P_{2N}(x) }{\partial U_{2r}}= 0 \;\; \mbox{for}
\;\; 2r> 2N. \nonu
\end{eqnarray}
Using these relations we can rewrite (\ref{relation4odd}) as
follows,
\begin{eqnarray}
0=g_{2r}+\oint \frac{R_{2k-2N+2}(x)}{x^{2r}}\left( -\sqrt{
P_{2N}^2(x)-4x^2\Lambda^{4N-2}}\right)dx- \oint
\sum_{i=0}^{2N-2n-2} \frac{L_i}{(x-p_i)}\frac{P_{2N}(x)}{x^{2r}}
dx.
\nonu
\end{eqnarray}
From the massless monopole constraint (\ref{curveodd}) we have the
relation,
\begin{eqnarray}
P_{2N}(x)=xH_{2N-2n-2}(x) \sqrt{F_{2(2n+1)}(x)}+{\cal
O}(x^{-2N+2}).
\nonu
\end{eqnarray}
Finally we obtain an expected relation of equations of motion for
the general order tree level superpotential case,
\begin{eqnarray}
W^{\prime}(z)=\oint \frac{zy_m}{x^2-z^2}dx \nonu
\end{eqnarray}
if the matrix model curve is given by
\begin{eqnarray}
y_m^2 &=& F_{2(2n+1)}(x){\widetilde{Q}}_{2k-2n}^2(x) \nonu \\
&\equiv& F_{2(2n+1)}(x)\left(R_{2k-2N+2}(x)H_{2N-2n-2}(x)+
R_{2N-2n-2}(x)\right)^2. \nonu
\end{eqnarray}
When $2n=2N-2$ (no massless monopoles),
${\widetilde{Q}}_{2k-2N+2}(x)=R_{2k-2N+2}(x)$. When the degree of
superpotential is equal to $2N$, in other words, $2k=2N-2$, then
${\widetilde{Q}}_{2k-2n}(x)=R_{0} H_{2N-2n-2}(x)+ R_{2N-2n-2}(x)$.
In particular, for $2n=2N-2$, ${\widetilde{Q}}_{0}$ is a constant
and $y_m^2(x)=F_{4N-2}(x)=x^{-2} (P_{2N}^2(x) -4 \La^{4N-2} x^2)$.

$\bullet$ {\bf A generalized Konishi anomaly}


Now we are ready to study for a derivation of the generalized
Konishi anomaly equation. According to the same change of
integration done in previous discussion,  $\mbox{Tr}
\frac{W^{\prime}(\Phi_{cl})}{z-\Phi_{cl}}$ is given by \bea
&&\mbox{Tr}\frac{1}{z-\Phi_{cl}} \left( \oint_{z_{in}}
\frac{zQ_{2k-2n}(x)P_{2N}(x)}{xH_{2N-2n-2}(x)(x^2-z^2)}dx-
\oint_{C_{z}+C_{-z}}
\frac{zQ_{2k-2n}(x)P_{2N}(x)}{xH_{2N-2n-2}(x)(x^2-z^2)}dx \right) \nonu \\
&& = \mbox{Tr}\frac{1}{z-\Phi_{cl}} \left(W^{\prime}(z)-
\frac{y_m(z) P_{2N}(z)}{\sqrt{P_{2N}^2(z)-4z^2\Lambda^{4N-2}}}
\right), \nonu
\end{eqnarray}
where in the last equality we used (\ref{prires}) and
\begin{eqnarray}
H_{2N-2n-2}(z)=\frac{\sqrt{P_{2N}^2(z)-4\Lambda^{4N-2}z^2}}
{z\sqrt{F_{2(2n+1)}(z)}}, \qquad y_m^2(z) = F_{2(2n+1)}(z)
Q^2_{2k-2n}(z). \nonu
\end{eqnarray}
Let us write it as, after an integration over $x$,
\begin{eqnarray}
-\sum_{i=0}^{2N-2n-2} \oint \frac{L_iP_{2N}(x)x}
{(x-p_i)(x^2-z^2)}\mbox{Tr}\frac{1}{x-\Phi_{cl}}dx
=-\sum_{i=0}^{2N-2n-2} \frac{L_i p_i P_{2N}(x=p_i)}
{p_i^2-z^2}\mbox{Tr}\frac{1}{p_i-\Phi_{cl}}. \nonu \eea Now we use
the result of the equation of motion for $B_i$
(\ref{relation7odd}) in order to change the trace part and arrive
at the final contribution as follows: \bea -\sum_{i=0}^{2N-2n-2}
\frac{L_i P_{2N}(x=p_i)}{p_i^2-z^2} =-\sum_{i=0}^{2N-2n-2} \oint
\frac{P_{2N}(x)}{(x^2-z^2)}
\frac{L_i}{(x-p_i)}dx \nonu
\end{eqnarray}
where note that $z$ is {\it outside} the contour of integration in last equation. As in $SO(2N)$ case we can rewrite the last equation as follows:
\begin{eqnarray}
-\frac{W^{\prime}(z)}{z}+\sum_{i=0}^{2N-2n-2}\oint_{C_{z}+C_{-z}}\frac{P_{2N}(x)}{(x^2-z^2)}
\frac{L_i}{(x-p_i)}dx =-\frac{W^{\prime}(z)}{z}+\frac{1}{z}\frac{y_mP_{2N}(z)}{\sqrt{P_{2N}^2(z)-4z^2\Lambda^{4N-2}}} \nonu
\end{eqnarray}
 Therefore we obtain \bea \mbox{Tr}
\frac{W^{\prime}(\Phi_{cl})}{z-\Phi_{cl}}=\mbox{Tr}\frac{1}{z-\Phi_{cl}}
\left(W^{\prime}(z)- \frac{y_m(z)
P_{2N}(z)}{\sqrt{P_{2N}^2(z)-4z^2\Lambda^{4N-2}}}
\right)-\frac{W^{\prime}(z)}{z}+\frac{1}{z}\frac{y_mP_{2N}(z)}{\sqrt{P_{2N}^2(z)-4z^2\Lambda^{4N-2}}}. \nonu \eea
Remembering that \bea \mbox{Tr}\frac{1}{z-\Phi_{cl}}=\frac{P_{2N}^{\prime}(z)}{P_{2N}(z)}, \nonu \eea
the second and fourth terms are rewritten as follows;
\begin{eqnarray}
- \mbox{Tr}\frac{1}{z-\Phi_{cl}}\frac{y_m(z)
P_{2N}(z)}{\sqrt{P_{2N}^2(z)-4z^2\Lambda^{4N-2}}}+\frac{1}{z}\frac{y_mP_{2N}(z)}{\sqrt{P_{2N}^2(z)-4z^2\Lambda^{4N-2}}}=\frac{y_m}{z}-\left\langle \tr \frac{y_m}{z-\Phi} \right\rangle \nonu
\end{eqnarray}
where we used (\ref{10may2BB}).

We can summarize as follows,
\begin{eqnarray}
\left\langle \tr \frac{W^{\prime}(\Phi)}{z-\Phi} \right\rangle &
=& \left( \left\langle \tr \frac{1}{z-\Phi} \right \rangle
-\frac{1}{z} \right) \left[W^{\prime}(z)-y_m(z)\right] \nonu
\end{eqnarray}
This is the generalized Konishi anomaly equation for $SO(2N+1)$
case.

\subsection{The multiplication map and the confinement index}
In this subsection, we discuss some important property of ${\cal
N}=1$ $SO(N)$ gauge theories for surveying phase structure. As
already discussed, this theory is ${\cal N}=2$ theory deformed by
tree level superpotential (\ref{treesup}) with symmetry breaking
pattern,
\begin{eqnarray}
SO(N)\to SO(N_0)\times \prod_{i=1}^n U(N_i).
\label{sobreaking}
\end{eqnarray}

At first let us review some notations and ideas used in
\cite{cdsw}. The matrix model curve $\Sigma$, which have already
been derived from the strong gauge coupling approach, with
$2k=2n$, is characterized by (\ref{sol})
\begin{eqnarray}
y_m^2=W_{2n+1}^{\prime}(x)^2+f_{2n}(x) \label{matrixcurve}
\end{eqnarray}
and  plays an important role for understanding the various phase
structures or the effective superpotential of corresponding gauge
theories. At most, there are $(2n+1)$ branch cuts on the curve
$\Sigma$. Let us define $A_i$ as a cycle surrounding the $i$-th
cut and $B_i$ as a cycle connecting two infinities of the curve
$\Sigma$ through $i$-th cut. These cycles are simplectic pair that
intersect each other with the intersection number,
$(A_i,A_j)=(B_i,B_j)=0, (A_i,B_j)=\delta_{ij}$. Since the
polynomial  $f_{2n}(x)$ is an even function in $x$, this Riemann
surface $\Sigma$ has ${\bf Z}_2$ identification, which is specific
to $SO/Sp$ cases. As in \cite{cdsw}, let us introduce the two
operators,
\begin{eqnarray}
T(x)=\mbox{Tr}\frac{1}{x-\Phi},\qquad
R(x)=-\frac{1}{32\pi^2}\mbox{Tr}
\frac{W_{\alpha}W^{\alpha}}{x-\Phi}. \nonu
\end{eqnarray} The
period integrals of these operators have a very nice
interpretation. At first, the period integrals of $R(x)$ lead to
\begin{eqnarray}
\oint _{A_i}R(x)dx=S_i,\qquad
\oint_{B_i}R(x)dx=\Pi_i=\frac{1}{2\pi i} \frac{\partial {\cal
F}}{\partial S_i} \nonu
\end{eqnarray}
where $S_i$ are glueball superfields of the gauge groups $SO(N_0)$
and $U(N_i)$ in the above, and ${\cal F}$ is the prepotential of
the gauge theory. The others are period integrals of $T(x)$,
\begin{eqnarray}
\oint_{A_i} \overline{T}(x)dx=N_i, \qquad
\oint_{B_i}\overline{T}(x)dx=-\tau_0-b_i \nonu
\end{eqnarray}
where we defined $\tau_0=-\oint_{B_0}\overline{T}(x)dx$ and
introduced $\overline{T}(x)$ and the real number $b_i$ is written
as the sum of a period of $T(x)$ around the other compact cycle
$C_i$ which is related to the noncompact cycle $B_i$, according to
the convention of \cite{csw}.

Next let us review the effective superpotential by using these
quantities. In \cite{ino}, the perturbative analysis of effective
superpotential for $SO/Sp$ gauge theory was discussed from purely
field theory viewpoint. The contribution comes from only ${\bf
S}^2$ and ${\bf RP}^2$ graphs. The explicit representation of the
effective superpotential is given by \cite{ino,ashoketal,jo}
\begin{eqnarray}
W_{eff}(S_i)=(N_0- 2)\frac{\partial {\cal F}}{\partial S_0}+
\sum_{i=1}^n N_i \frac{\partial {\cal F}}{\partial S_i}+2\pi i
\tau_0 \sum_{i=0}^n S_i+2\pi i \sum_{i=1}^n b_iS_i. \label{zz3}
\end{eqnarray}
The first term comes from ${\bf S}^2$ contribution and the second
term comes from ${\bf RP}^2$ contribution. Note that along the
line of \cite{csw}, we introduced a real number $b_i$ and by
including this the theta angle is shifted. For $U(N)$ gauge
theories, by using 't Hooft loop and Wilson loop, the effect of
$b_i$ was discussed. For any given $N$, there are two kinds of
vacua, namely the confining vacua and the Coulomb vacua. The
criterion of these vacua is quite simple. If both $N_i$ and $b_i$
have a common divisor, then the vacua are confining vacua. The
most greatest common divisor is called a confinement index. If
there is no common divisor between them and the vacuum does not
have a confinement, then it is called Coulomb vacua: for example,
those vacua in which any of the $b_j$ is 1.

Since the number of vacua of $SO(N_0)$ gauge theories is $(N_0-2)$
that is the dual Coxeter number of the group, total number of
vacua is given by $(N_0-2)\times \prod_{i=1}^n N_i$. If we
introduce a new notations $\hat{N}_0\equiv N_0-2$ and
$\hat{N}_i=N_i$, we can discuss the counting of vacua in parallel
way as $U(N)$ case.
These low energy vacua are characterized by some integers
$r_i,(i=0, \cdots ,n)$ with $0\le r_i \le \hat{N}_i-1$. We can
easily extend the criterion of $U(N)$ gauge theories to our
$SO(N)$ gauge theories. If both $\hat{N}_i$ and $b_i \equiv
r_i-r_{i+1}$ have a common divisor, the vacua are confining vacua.
If not, those are Coulomb vacua.

In \cite{csw} to probe confinement, 
the center of $SU(N)$, ${\bf Z}_N$, was used. 
The transformation property of some representation 
under the center was important for studying Wilson loop $W$. 
How about $SO/Sp$ gauge theories? Since the centers 
of $SO/Sp$ gauge theories are ${\bf Z}_2$ rather than ${\bf Z}_N$, 
it seems that the 
confinement index of the theories can be smaller. When 
we consider some representation 
${\cal R}$ as a tensor product of $r$ copies of 
the defining representation, for some $r \geq 0$, 
there exist two types of
transformation under the center. When the $r$ is even, it is 
invariant under the ${\bf Z}_2$ while  
it transforms by $-1$ when the $r$ is odd. 
Thus Wilson loop $W$ in ${\cal R}$ 
is defined mod 2 and $W^2$ has no area law (the theory is 
completely unconfining). 
Naively if one proceeds the discussion of \cite{csw}, 
one regards the confinement index $t$ 
in the notation of \cite{csw} as the 
greatest common divisor of three quantities 
2, $N_i$ and $b_i$. When $t=2$, 
there exists  only confining phase, in other words, 
all phases are confining (For $t=1$ which is trivial, 
the theory is
completely unconfining). This result seems to be unnatural because
as we will see later, we explicitly present  examples for
phase that can not be obtained by multiplication map. One can 
interpret this phase as Coulom phase from the 
$U(N)$ results. So the definition of 
confinement index  above is not appropriate and from another 
mechanism \footnote{ For $SO(N_0)$ gauge theory, there exist 
$(N_0-2)$
 vacua which can be labeled by $r_0$ with $0\le r_0 \le (N_0-2)-1$. 
It is natural that one can distinguish each vacuum 
by the type of confinement as in $U(N)$ case. In other 
words in each vacuum  the loop order parameter
$W^{p} H^q$ are 
unconfined where $H$ is 'tHooft loop. 
Which values of $p$ and $q$ are valid for the $r_0$-th 
vacuum? 
When we consider the Wilson loop we have only 
to consider the center because of electric screening explained 
in \cite{csw}. Since the center is ${\bf Z}_2$ 
the Wilson loop $W^2$ seems to have no area law. Thus to 
distinguish the vacua by the type of confinement, it is necessary 
to consider the case with $p=1, 1\le q \le N_0-2$. We conjecture 
that in the $r_0$-th vacuum, the $W_0H_0^{r_0}$ are unconfined and 
from the effect of magnetic screening 
$H^{N_0-2}$ is unconfined. } it allows us to introduce the right definition 
of confinment index. 

Thus we claim that for $SO(N)$ gauge group breaking into 
$SO(N_0)\times \prod_{i=1}^n U(N_i)$, the confinement index 
$t$  
is nothing but the greatest common divisor of $\hat{N}_i$ and 
$b_i$ which is exactly the same as multiplication map index $K$. 
As we will see in the various examples of 
section 2.4, this definition is consistent with 
the results of explicit calculation for given gauge theories and all the 
examples we consider support our claim.

$\bullet$ {\bf $SO(2N)$ case}

For any positive integer $K$, the construction maps vacua of the
$SO(2N)$ theory with a given superpotential $W(x)$ to vacua of the
$SO(2KN-2K+2)$ theory with the {\it same} superpotential. All the
vacua of $SO(2KN-2K+2)$ with confinement index $K$ can be obtained
from the Coulomb vacua of $SO(2N)$ under the multiplication map.

Let us assume that $P_{2N}(x)$ satisfies massless monopole
constraint with some renormalization scale $\La_0$, as we have
seen before (\ref{curve}) and (\ref{sol}),
\begin{eqnarray}
P_{2N}^2(x)-4x^4 \Lambda^{4N-4}_0 & = & x^2
H^2_{2N-2n-2}(x)F_{2(2n+1)}(x), \nonu \\
F_{2(2n+1)}(x) & = & W^{\prime}(x)^2 + f_{2n}(x) \nonu
\end{eqnarray}
where we put $g_{2n+2}=1$ and have the symmetry breaking pattern
$SO(2N)\to SO(2N_0)\times \prod_{i=1}^{n} U(N_i)$ in the
semiclassical limit. The numbers $N_0$ and $N_i$ should satisfy
$2N=2N_0+2\sum_{i=1}^{n} N_i$. By using Chebyshev polynomials
\footnote{These Chebyshev polynomials ${\cal T}_K(x)$ and ${\cal
U}_{K}(x)$ are slightly different from $T_K(x)$ and $U_{K}(x)$
used in \cite{fo1}. These polynomials have the relation,
$\frac{1}{2}T_K(2x)={\cal T}_K(x)$ and $U_{K}(2x)={\cal
U}_{K-1}(x)$. } of the first kind ${\cal T}_K(x)$ of degree $K$
and the second kind ${\cal U}_{K-1}(x)$ of degree $(K-1)$
\cite{ds}, we can construct the solution for the massless monopole
constraint of $SO(2KN-2K+2)$ with the breaking pattern
$SO(2KN_0-2K+2)\times \prod_ {i=1}^{n} U(KN_i)$, motivated by
\cite{fo1}
\begin{eqnarray}
P_{2KN-2K+2}(x)=2\eta^{K} x^2 \Lambda^{2KN-2K}{\cal T}_K
\left(\frac{P_{2N}(x)} {2\eta x^2 \Lambda^{2N-2}} \right),
\label{multisol}
\end{eqnarray}
where $\eta$ is a $2K$-th root of unity, i.e. $\eta^{2K}=1$. One
can check that in the right hand side, the argument of the first
kind Chebyshev polynomial has a degree $K(2N-2)$, by the
definition of the first kind Chebyshev polynomial and the right
hand side has a factor $x^2$, therefore it leads to a polynomial
of degree  $2+K(2N-2)$ totally. It is right to write the left hand
side as $P_{2KN-2K+2}(x)$ which agrees with the number of order in
$x$ in the right hand side. The power of $\La$ for the argument of
Chebychev polynomial was fixed by the power of $x$ which is equal to
$(2N-2)$. The power of $\La$  in front of the Chebyshev polynomial
can be fixed by the dimension consideration of both sides. The
right side should contain $(2KN-2K+2)$ which is equal to
$2+(2KN-2K)$ where the first term come from the power of $x$ and
the second one should be the power of $\La$. Also note that the
same $\eta$ term appears in the denominator of the argument of
Chebyshev polynomial.  Since Chebyshev polynomials have useful
relation, \bea {\cal T}^2_K(x)-1=(x^2-1){\cal U}^2_{K-1}(x), \nonu
\eea we can confirm that (\ref{multisol}) really satisfies the
massless monopole constraint of $SO(2KN-2K+2)$ gauge theories.

For the simplicity of equations, let us define \bea
\widetilde{x}\equiv \frac{P_{2N}(x)}{2\eta x^2
\Lambda^{2N-2}}. \nonu \eea Then one can easily check that
\begin{eqnarray}
P_{2KN-2K+2}^2(x)-4x^4\Lambda^{4KN-4K}&=&4x^4\Lambda^{4KN-4K}
\left[{\cal T}^2_K \left(\widetilde{x} \right)-1 \right] \nonu
\\
&=&4x^4\Lambda^{4KN-4K}
\left(\widetilde{x}^2-1 \right){\cal U}^2_{K-1}\left(\widetilde{x}
\right) \nonu \\
&=&\frac{\Lambda^{4KN-4K}}{\eta^2 \Lambda^{4N-4}}
\left(P_{2N}^2(x)-4x^4\eta^2\Lambda^{4N-4}
\right) {\cal U}^2_{K-1}\left(\widetilde{x} \right) \nonu \\
&=&x^2 \left[H_{2N-2n-2}(x)\eta^{-1}\Lambda^{2(K-1)(N-1)}{\cal
U}_{K-1}(\widetilde{x}) \right]^2F_{2(2n+1)}(x). \nonu
\end{eqnarray}
In the fourth equality we used the identification
$\Lambda_0^{4N-4}=\eta^2 \Lambda^{4N-4}$.
This implies that the following relations
\bea
P_{2KN-2K+2}(x)& = &2\eta^{K} x^2 \Lambda^{2KN-2K}{\cal T}_K \left(\frac{P_{2N}(x)}
{2\eta x^2 \Lambda^{2N-2}} \right), \nonu \\
\widetilde{F}_{2(2n+1)}(x) & = & F_{2(2n+1)}(x),  \nonu \\
H_{(2N-2)K-2n}(x) & = &
H_{2N-2n-2}(x)\eta^{-1}\Lambda^{2(K-1)(N-1)}{\cal
U}_{K-1}\left(\frac{P_{2N}(x)} {2\eta x^2 \Lambda^{2N-2}} \right)
\nonu \eea satisfy the solution of \bea P_{2KN-2K+2}^2(x) -4x^4
\Lambda^{4NK-4K} = x^2
 H_{(2N-2)K-2n}^2(x)  \widetilde{F}_{2(2n+1)}(x).
\nonu \eea Since $\widetilde{F}_{2(2n+1)}(x)=F_{2(2n+1)}(x)$, the
vacua constructed this way for the $SO(2KN-2K+2)$ theory have the
{\it same} superpotential as the vacua of the $SO(2N)$ theory. For
a superpotential $W(x)$, the $SO(2N)$ theory has a finite number
of vacua with given $n$. For $K$ different values of the $SO(2N)$
parameter $\La_0^{4N-4}$, that is, \bea \Lambda_0^{4N-4}=\eta^2
\Lambda^{4N-4} \nonu \eea there exists $SO(2KN-2K+2)$ vacua.


Next we consider the multiplication map of $T(x)$ we introduced,
\begin{eqnarray}
T(x)&=&\frac{d}{d x}\log
\left(P_{2N}(x)+\sqrt{P_{2N}^2(x)-4x^4\Lambda^{4N-4}} \right)
\nonu
\\
&=&\frac{P_{2N}^{\prime}(x)}{\sqrt{P_{2N}^2(x)-4x^4\Lambda^{4N-4}}}-
\frac{2P_{2N}(x)}{x\sqrt{P_{2N}^2(x)-4x^4\Lambda^{4N-4}}}+\frac{2}{x}.
\label{TT}
\end{eqnarray}
Remember that the Seiberg-Witten differential has the form of \bea
d\la_{SW}= \frac{x d x }{\sqrt{P_{2N}^2(x)-4x^4\Lambda^{4N-4}}}
\left(P_{2N}^{\prime}(x) - \frac{2}{x} P_{2N}(x) \right). \nonu
\eea By using (\ref{multisol}) we obtain the following relations,
\begin{eqnarray}
T_K(x)& =& \sqrt{P_{2KN-2K+2}^2(x)-4x^4\Lambda^{4KN-4K}} \nonu \\
&=&\sqrt{\eta^{-2} \Lambda^{4(K-1)(N-1)}{\cal
U}_{K-1}^2(\widetilde{x})\left(P_{2N}^2(x)-4x^4 \Lambda^{4N-4}_0
\right)}
\nonu  \\
&=&\eta^{K-1}\Lambda^{2(K-1)(N-1)} {\cal
U}_{K-1}(\widetilde{x})\sqrt{P_{2N}^2(x)-4x^4\Lambda^{4N-4}_0}.
\label{rel1}
\end{eqnarray}
In the last equality we used the fact that $\eta^{2K}=1$. That is,
$\eta^{-2}=\eta^{2K-2}$. By using the property between the first
kind Chebyshev and the second kind of Chebyshev and changing the
derivative of the former with respect to the argument into the
confinement index $K$ multiplied by the latter, the derivative of
$P_{2KN-2K+2}(x)$ with respect to $x$ is given by
\begin{eqnarray}
P_{2KN-2K+2}^{\prime}(x)&=&2\eta^K \Lambda^{2KN-2K} \left[2x {\cal
T}_{K} (\widetilde{x})+x^2\left(\frac{P_{2N}(x)}{2\eta x^2
\Lambda^{2N-2}}
\right)^{\prime}K {\cal U}_{K-1}(\widetilde{x}) \right] \nonu \\
&=&2\eta^K \Lambda^{2KN-2K} \left[2x {\cal T}_{K}(\widetilde{x})+
\frac{P_{2N}^{\prime}(x)}{2\eta \Lambda^{2N-2}}K{\cal
U}_{K-1}(\widetilde{x})-\frac{P_{2N}(x)}
{\eta x \Lambda^{2N-2}}K{\cal U}_{K-1}(\widetilde{x})\right] \nonu \\
&=&4x\eta^K \Lambda^{2KN-2K}{\cal
T}_{K}(\widetilde{x})+\eta^{K-1}\Lambda^{2(K-1)(N-1)}
P_{2N}^{\prime}(x)K{\cal U}_{K-1}(\widetilde{x}) \nonumber \\
&-& \frac{2}{x}\eta^{K-1} \Lambda^{2(K-1)(N-1)}P_{2N}(x)K{\cal
U}_{K-1}(\widetilde{x}).
\label{rel2}
\end{eqnarray}
From these two relations (\ref{rel1}) and (\ref{rel2}),  we
compute the first term of (\ref{TT}),
\begin{eqnarray}
\frac{P_{2KN-2K+2}^{\prime}(x)}{\sqrt{P_{2KN-2K+2}^2(x)-4x^4\Lambda^{4KN-4K}}}
& = &
\frac{KP_{2N}^{\prime}(x)}{\sqrt{P_{2N}^2(x)-4x^4\Lambda^{4N-4}_0}}-\frac{2}{x}
\frac{KP_{2N}(x)}{\sqrt{P_{2N}^2(x)-4x^4\Lambda^{4N-4}_0}} \nonumber \\
& + & 4x\eta \Lambda^{2N-2}\frac{{\cal T}_K(\widetilde{x})}{{\cal
U}_{K-1}(\widetilde{x})\sqrt{P_{2N}^2(x)-4x^4 \Lambda^{4N-4}_0}}.
\label{zz1}
\end{eqnarray}
The second and third terms of (\ref{TT}) become
\begin{eqnarray}
\frac{2}{x}-\frac{2}{x}\frac{P_{2KN-2K+2}(x)}{\sqrt{P_{2KN-2K+2}^2(x)-4x^4
\Lambda^{4KN-4K}}}=\frac{2}{x}-\frac{4\eta \Lambda^{2N-2} x{\cal
T}_{K}(\widetilde{x})}{{\cal
U}_{K-1}(\widetilde{x})\sqrt{P_{2N}^2(x)-4x^4\Lambda^{4N-4}_0}}.
\label{zz2}
\end{eqnarray}

Thus combining (\ref{zz1}) with (\ref{zz2}), we obtain the
multiplication map of $T(x)$, $T_K(x)$,
\begin{eqnarray}
T_K(x)&=&K\left( \frac{P_{2N}^{\prime}(x)}{\sqrt{P_{2N}^2(x)-
4x^4\Lambda^{4N-4}_0}}-\frac{2}{x}\frac{P_{2N}(x)}{\sqrt{P_{2N}^2(x)-4x^4
\Lambda^{4N-4}_0}}+\frac{2}{x} \right)-K\frac{2}{x}+\frac{2}{x}
\nonumber \\
&=&K T(x)-K\frac{2}{x}+\frac{2}{x}. \nonu
\end{eqnarray} What does this
mean? For the physical meaning, we consider the integral of $T(x)$
around origin,
\begin{eqnarray}
\oint_{x=0}T_Kdx=K\oint_{x=0}Tdx-2K+2 \iff
2N_0^{\prime}-2=K(2N_0-2). \nonu
\end{eqnarray}
This equation implies that under the multiplication map $2N_0$
does not simply multiply by $K$. The special combination
$(2N_0-2)$ have simple multiplication by $K$. This combination is
natural for $SO(2N)$ gauge theory because in the effective
superpotential (\ref{zz3}) unoriented diagram gives $-2$ factor.

If we define a new operator $h(x)$ that corresponds to flux one
form on Riemann surface $\Si$ \cite{feng1,ookouchi} as
$h(x)=T(x)-\frac{2}{x}=\frac{1}{x} \frac{d\la_{SW}}{dx}$, we have
more simple relation,
\begin{eqnarray}
h_K(x)=Kh(x). \nonu
\end{eqnarray} After all we can understand that
the multiplication map multiplies both $(2N_0-2)$ and $N_i$ by a
common  $K$. Since these quantities have common devisor $K$ this
operation generates only confining vacua. Let us denote adjoint
chiral superfield as $\Phi_0$ in the $SO(2N)$ gauge theory. Then
through the multiplication map we constructed the following
quantum operator in the $SO(2KN-2K+2)$ gauge theory can be
obtained by multiplying the confinement index $K$ by the
corresponding operator in the $SO(2N)$ gauge theory as follows:
 \bea \left\langle \mbox{Tr} \frac{1}{x-\Phi}
\right\rangle -\frac{2}{x} = K \left(\left\langle \mbox{Tr}
\frac{1}{x-\Phi_0} \right\rangle -\frac{2}{x} \right) \nonu \eea
As in $U(N)$ case, all the vacua in $SO(2KN-2K+2)$ with
confinement index $K$ arise in this way from the $SO(2N)$ Coulomb
vacua. This can be shown by counting the number of vacua. The
multiplication map has $\eta$ that is $2K$-th root of unity. So
confining vacua constructed by this map have $K$ times as many
vacua as Coulomb vacua. Note that $\Lambda_0^{4N-4}=\eta^2
\Lambda^{4N-4}$.

Since these vacua have a confinement index $K$, the three
quantities  $K(2N_0-2)$, $KN_i$ and
$\widetilde{b}_i=\widetilde{r}_i-\widetilde{r}_{i+1}$ have
greatest common divisor $K$. As in \cite{csw}, we can represent
such $\widetilde{r}_i$ with $0 \le u\le K-1$ as
$\widetilde{r}_i=u+Kr_i$. Since there exist $K$ choices of $u$,
the number of $SO(2KN-2K+2)$ vacua with confinement index $K$ is
indeed $K$ times the corresponding number of $SO(2N)$ Coulomb
vacua. Thus we constructed all confining vacua with confinement
index $K$ from this map explicitly.

$\bullet $ {\bf $SO(2N+1)$ case}

To begin with let us note some property of Chebyshev polynomial
${\cal T}_{K}(x)$,
\begin{eqnarray}
{\cal T}_K(x)=\frac{1}{2}\left[ (x+\sqrt{x^2-1})^K+(x-\sqrt{x^2-1})^K \right].
\nonu
\end{eqnarray} If the number $K$ is odd, then ${\cal
T}_{K}(x)$ is a odd polynomial in $x$. So the function $x{\cal
T}_K(x)$ is an even polynomial in $x$. With this in mind let us
consider the multiplication map of $SO(2N+1)$ gauge theory. By
assumption, we have the characteristic function $P_{2N}(x)$ that
satisfies the massless monopole constraint (\ref{curveodd}),
\begin{eqnarray}
P_{2N}^2(x)-4x^2 \Lambda^{4N-2}=x^2H_{2N-2n-2}^2(x)F_{2(2n+1)}(x)
\nonu
\end{eqnarray}
and have the breaking pattern $SO(2N_0+1)\times \prod_{i=1}^{n}
U(N_i)$, where $2N+1=2N_0+1+\sum_{i=1}^n 2N_i$. By using this
solution we can construct a new solution of $SO(2KN-K+2)$ with
breaking pattern $SO(2KN_0-K+2)\times \prod_{i=1}^n U(KN_i)$ in
the semiclassical limit,
\begin{eqnarray}
P_{2KN-K+1}(x)=2\eta^K x \Lambda^{2KN-K}{\cal T}_K
\left(\frac{P_{2N}(x)} {2\eta x \Lambda^{2N-1}} \right)
\label{oddsolution}
\end{eqnarray}
where $K$ is an odd number, so $(2KN-K+1)$ is an even number. One
can check that in the right hand side, the argument of the first
kind Chebyshev polynomial has a degree $K(2N-1)$ in $x$, by the
definition of the first kind Chebyshev polynomial and the right
hand side has a factor $x$, therefore it leads to a polynomial of
degree $1+K(2N-1)$ totally. It is right to write the left hand
side as $P_{2KN-K+1}(x)$ which agrees with the number of order in
$x$  in the right hand side. The power of $\La$ for the argument
of Chebychev polynomial was fixed by the power of $x$ which is
equal to $(2N-1)$. The power of $\La$  in front of the Chebyshev
polynomial can be fixed by the dimension consideration of both
sides. The right hand side should contain $(2KN-K+1)$ which is
equal to $1+(2KN-K)$ where the first term comes from the power of
$x$ and the second one should be the power of $\La$. Also note
that the same $\eta$ term appears in the denominator of  the
argument of Chebyshev polynomial.

Let us confirm that this function (\ref{oddsolution}) indeed
satisfies the massless monopole constraint of $SO(2KN-K+2)$. Using
the properties we mentioned, one obtains
\begin{eqnarray}
P_{2KN-K+1}^2(x)-4x^2 \Lambda^{4KN-2K}&=&4x^2 \Lambda^{4KN-2K}
\left({\cal T}^2_{K}(\widetilde{x})-1
\right) \nonu \\
&=&4x^2 \Lambda^{4KN-2K} \left(\widetilde{x}^2-1 \right){\cal U}^2_{K-1}
(\widetilde{x}) \nonu \\
&=&\frac{\Lambda^{4KN-2K}}{\eta^2 \Lambda^{4N-2}}
\left(P_{2N}^2(x)-4\eta^2 x^2\Lambda^{4N-2}
\right){\cal U}^2_{K-1}(\widetilde{x}) \nonu \\
&=&x^2 \left[\eta^{-1}\Lambda^{2(K-1)(2N-1)} H_{2N-2n-2}(x){\cal
U}_{K-1}(\widetilde{x}) \right]^2F_{2(2n+1)}(x) \nonu \eea where
$\eta^2 \La^{4N-2}=\La_0^{4N-2}$ and   this implies that the
following identification
\bea P_{2KN-K+1}(x)&= & 2\eta^K x
\Lambda^{2KN-2K}{\cal T}_K \left(\frac{P_{2N}(x)}
{2\eta x \Lambda^{2N-1}} \right), \nonu \\
\widetilde{F}_{2(2n+1)}(x) & = & F_{2(2n+1)}(x),  \nonu \\
H_{(2N-1)K-2n-1}(x) & = & \eta^{-1}\Lambda^{2(K-1)(2N-1)}
H_{2N-2n-2}(x){\cal U}_{K-1}(\widetilde{x}) \nonu \eea leads to
the solution of \bea P_{2KN-K+1}^2(x)-4x^2 \Lambda^{4KN-2K}&= x^2
H^2_{(2N-1)K-2n-1}(x) F_{2(2n+1)}(x). \nonu \eea Since
$\widetilde{F}_{2(2n+1)}(x)=F_{2(2n+1)}(x)$ the vacua constructed
this way for the $SO(2KN-K+2)$ theory have the {\it same}
superpotential as the vacua of the $SO(2N+1)$ theory.

Let us consider the multiplication map of $T(x)$,
\begin{eqnarray}
T(x)&=&\frac{d}{d x}\log
\left(P_{2N}(x)+\sqrt{P_{2N}^2(x)-4x^2\Lambda^{4N-2}} \right)
\nonu
\\
&=&\frac{P_{2N}^{\prime}(x)}{\sqrt{P_{2N}^2(x)-4x^2\Lambda^{4N-2}}}-
\frac{P_{2N}(x)}{x\sqrt{P_{2N}^2(x)-4x^2\Lambda^{4N-2}}}+\frac{1}{x}.
\nonu
\end{eqnarray} One can easily write $T_K(x)$, from the
equation (\ref{oddsolution}), as follows:
 \bea T_{K}(x) =
\eta^{K-1}\Lambda^{(K-1)(2N-1)} {\cal
U}_{K-1}(\widetilde{x})\sqrt{P_{2N}^2(x)-4x^2\Lambda^{4N-2}_0}.
\nonu \eea
By using the properties of Chebyshev polynomials, the
derivative of $P_{2KN-K+1}(x)$ with respect to $x$ can be
summarized by \bea P^{\prime}_{2KN-K+1}(x) & = & 2\eta^K
\Lambda^{2KN-2K}{\cal
T}_{K}(\widetilde{x})+\eta^{K-1}\Lambda^{(K-1)(2N-1)}
P_{2N}^{\prime}(x)K{\cal U}_{K-1}(\widetilde{x}) \nonumber \\
&-& \frac{1}{x}\eta^{K-1} \Lambda^{(K-1)(2N-1)}P_{2N}(x)K{\cal
U}_{K-1}(\widetilde{x}). \nonu \eea Now it is straightforward to
compute the various quantities in $T_K(x)$ and we arrive at and
obtain the multiplication map of $T(x)$,
\begin{eqnarray}
T_K(x)&=&K\left( \frac{P_{2N}^{\prime}(x)}{\sqrt{P_{2N}^2(x)-
4x^2\Lambda^{4N-2}_0}}-\frac{1}{x}\frac{P_{2N}(x)}{\sqrt{P_{2N}^2(x)-4x^2
\Lambda^{4N-2}_0}}+\frac{1}{x} \right)-K\frac{1}{x}+\frac{1}{x}
\nonumber \\
&=&K T(x)-K\frac{1}{x}+\frac{1}{x}. \nonu
\end{eqnarray}
From this result we obtain
\begin{eqnarray}
\oint_{x=0}T_Kdx=K\oint_{x=0}Tdx-K+1 \iff
2N_0^{\prime}-1=K(2N_0-1). \nonu
\end{eqnarray}
The following quantum operator in the $SO(2KN-2K+2)$ gauge theory
can be obtained by multiplying the confinement index $K$ by the
corresponding operator in the $SO(2N+1)$ gauge theory
 \bea \left\langle \mbox{Tr} \frac{1}{x-\Phi}
\right\rangle -\frac{1}{x} = K \left(\left\langle \mbox{Tr}
\frac{1}{x-\Phi_0} \right\rangle -\frac{1}{x} \right). \nonu \eea

$\bullet$ {\bf From $SO(2N+1)$ to $SO(2M)$ }

Are there any multiplication map from $SO(2N)$ to $SO(2M+1)$?
Conversely from $SO(2N+1)$ to $SO(2M)$ ? We can construct the
multiplication map only from $SO(2N+1)$ to $SO(2M)$ where
$2M=2KN-K+2$. If the number $K$ is even, ${\cal T}_K(x)$ is an
even function. Let us consider
\begin{eqnarray} P_{2KN-K+2}(x)=2\eta^Kx^2 \Lambda^{2KN-K}{\cal
T}_{K}\left(\frac{P_{2N}(x)}{2\eta x\Lambda^{2N-1}}
 \right),\quad K\mbox{\ is even}.
 \label{multisoloddeven}
\end{eqnarray}
One can check that in the right hand side, the argument of the
first kind Chebyshev polynomial has a degree $K(2N-1)$, by the
definition of the first kind Chebyshev polynomial and the right
hand side has a factor $x^2$, therefore it leads to a polynomial
of degree  $2+K(2N-1)$ totally. It is right to write the left hand
side as $P_{2KN-K+2}(x)$ which agrees with the number of order in
$x$ in the right hand side. The power of $\La$ for the argument of
Chebychev polynomial was fixed by the power of $x$ which is equal
to $(2N-1)$. The power of $\La$  in front of the Chebyshev
polynomial can be fixed by the dimension consideration of both
sides. The right hand side should contain $(2KN-K+2)$ which is
equal to $2+(2KN-K)$ where the first term comes from the power of
$x$ and the second one should be the power of $\La$. Also note
that the same $\eta$ term appears in the denominator of  the
argument of Chebyshev polynomial. We can check that this
polynomial satisfies the massless monopole constraint of
$SO(2KN-K+2)$ explicitly. As we did before
\begin{eqnarray}
P_{2KN-K+2}^2(x)-4x^4\Lambda^{4KN-2K}&=&4x^4\Lambda^{4KN-2K}
\left[{\cal T}_K^2(\widetilde{x})-1 \right] \nonumber \\
&=&4x^4\Lambda^{4KN-2K}\left(\widetilde{x}^2-1
\right){\cal U}_{K-1}^2(\widetilde{x}) \nonumber \\
&=&\frac{x^2\Lambda^{4KN-2K}}{\eta^2 \Lambda^{4N-2}}
\left(P_{2N}^2(x)-4x^2\eta^2\Lambda^{4N-2} \right){\cal U}_{K-1}^2(\widetilde{x})
 \nonumber \\
&=&x^2\left[x\eta^{-1}\Lambda^{2KN-K-2N+1}H_{2N-2n-2}(x){\cal U}_{K-1}(\widetilde{x})
 \right]^2F_{2(2n+1)}(x). \nonumber
\end{eqnarray}
This implies the identification
\begin{eqnarray}
P_{2KN-K+2}(x)&=&2\eta^K x^2 \Lambda^{2KN-K}{\cal
T}_{K}\left(\frac{P_{2N}(x)}{2\eta x
\Lambda^{2N-1}} \right), \nonumber \\
\widetilde{F}_{2(2n+1)}(x)&=&F_{2(2n+1)}(x), \nonumber \\
H_{(2N-1)K-2n}(x)&=&x\eta^{-1}
\Lambda^{2KN-K-2N+1}H_{2N-2n-2}(x){\cal U}_{K-1}(\widetilde{x}),
\nonumber
\end{eqnarray}
which lead to the solution of
\begin{eqnarray}
P_{2KN-K+2}^2(x)-4x^4\Lambda^{4KN-2K}=x^2H_{(2N-1)K-2n}^2(x)
F_{2(2n+1)}(x). \nonumber
\end{eqnarray}
As in previous case since
$\widetilde{F}_{2(2n+1)}(x)=F_{2(2n+1)}(x)$ the vacua constructed
this way for the $SO(2KN-K+2)$ theory have the {\it same}
superpotential as the vacua of the $SO(2N+1)$ theory.

%

\subsection{Examples }

In this subsection we will deal with and analyze the explicit
examples which describe the discussions we did, with rank $n=1$,
namely $SO(2N)$ gauge group is broken to $SO(2N_0)\times U(N_1)$.
After the explanation for general $SO(2N)$ and $SO(2N+1)$ gauge
theories we will analyze the examples with gauge group $SO(N)$ for
$N=4,5,6,7,8$. As discussed in multiplication map we can construct
a map from $SO(2N+1)$ to $SO(2M)$. So when we survey the confining
phase we should note that multiplication map. Then we will
consider  both $SO(2N)$ and $SO(2N+1)$ in this section. Through
these examples we will see that there exists a smooth
interpolation between different pairs $(N_0,N_1)$.

{\bf $SO(2N)$ case}

At first we study moduli space for ${\cal N}=1$ $SO(2N)$ gauge
theories that is ${\cal N}=2$ theories deformed by tree level
superpotential. For the simplicity we consider the special case
$k=n=1$, then the superpotential is
\begin{eqnarray}
W(\Phi)=\frac{m}{2}\mbox{Tr}\Phi^2+\frac{g}{4}\mbox{Tr}\Phi^4.
\label{tree4} \nonu
\end{eqnarray} ${\cal N}=2$ theory deformed this potential
only has unbroken supersymmetry on submanifolds of the Coulomb
branch, where there are additional massless monopoles or dyons.
Then we want to find a solution of massless monopole constraint
(\ref{curve}) with $n=1$,
\begin{eqnarray}
P_{2N}^2(x)-4x^4\Lambda^{4N-4}= \left( P_{2N}(x)+2x^2 \La^{2N-2}
\right) \left( P_{2N}(x)-2x^2 \La^{2N-2}\right)=x^2 H^2_{2N-4}(x)
F_{6}(x). \label{exsoconst}
\end{eqnarray}
The ${\cal N}=2$ $SO(2N)$ gauge theory whose Coulomb branch was
described by this hyperelliptic curve and the $2N$-th order
polynomial $P_{2N}(x)$ parametrizes the point in the moduli space.
This point is denoted by the $N$ eigenvalues of the matrix $\Phi$.
To get $n=1$, this ${\cal N}=2$ theory should have $(N-2)$
massless magnetic monopoles therefore the polynomial
(\ref{exsoconst}) should have $2(N-2)$ double roots. Since
$P_{2N}(x)$ depends on the $2N$ complex parameters, the subspace
on which $P_{2N}^2(x)-4x^4\Lambda^{4N-4}$ has $2(N-2)$ double
roots is two-dimensional. There exist $s_{+}$ double roots in the
first factor and $s_{-}$ double roots in the second factor.
Different values of $s_{+}$ and $s_{-}$ correspond to different
branches. The right hand side of (\ref{exsoconst}) has the factor
$x^2$. To have a factor $x^2$ in the left hand side, the constant
term in $P_{2N}(x)$ must be zero, that is,
\begin{eqnarray}
P_{2N}(x)=x^{2N}+s_2x^{2N-2}+s_4x^{2N-4}+\cdots +s_{2N-2}x^2+s_{2N}, \ \
 \mbox{with}\ \ s_{2N}=0.
\nonu
\end{eqnarray}
Thus we obtain the relation $P_{2N}(x)=x^2P_{2N-2}(x)$. Taking
into account this relation, we can rewrite the massless monopole
constraint (\ref{exsoconst}) as
\begin{eqnarray}
P_{2N}(x)+2x^2\Lambda^{2N-2}&=&x^2H_{s_+}^2(x)R_{2N-2-2s_+}(x), \nonu   \\
P_{2N}(x)-2x^2\Lambda^{2N-2}&=&x^2H_{s_-}^2(x)\widetilde{R}_{2N-2-2s_-}(x), \nonu \\
x^2H_{s_+}(x)H_{s_-}(x)=xH_{2N-4}(x),&&
F_6(x)=R_{2N-2-2s_+}(x)\widetilde{R}_{2N-2-2s_-}(x), \label{SS}
\end{eqnarray}
where $s_++s_-+1=2N-4$ and $2N-2-2s_{\pm}\ge 0$. Equivalently it
leads to the following factorization problem
\begin{eqnarray}
P_{2N-2}(x)+2\Lambda^{2N-2}&=&H_{s_+}^2(x) R_{2N-2-2s_+}(x),
\label{soconst1}
\\
P_{2N-2}(x)-2\Lambda^{2N-2}&=&H_{s_-}^2(x)
\widetilde{R}_{2N-2-2s_-}(x) \label{soconst2}
\end{eqnarray} where the subscripts of the polynomials denote
their degrees, respectively. These equations are useful for the
later discussion of  the relation between $SO(2N)$ case with
$Sp(2N)$ case.

From the first two equations (\ref{SS}) we obtain
\begin{eqnarray}
H_{s_+}^2(x)R_{2N-2-2s_+}(x)-4\Lambda^{2N-2}=H_{s_-}^2(x)\widetilde{R}_{2N-2-2s_-}(x).
\label{s3}
\end{eqnarray}
This is the relation we will use below for finding the solutions.
Then we will study the semiclassical limit $\La \rightarrow 0$. We
would like to describe for each branch of the moduli space with
$N_0$ and $N_1$. Then we break the ${\cal N}=2$ supersymmetry to
${\cal N}=1$ supersymmetry by turning on a tree level
superpotential. Since $n=1$, the superpotential must have at least
three critical points and we consider the case of a quartic
superpotential.

As in \cite{csw} we introduce the matrix model curve, which give
the same Riemann surface $\Si$ that appeared in the matrix model
context \cite{dv1},
\begin{eqnarray}
y_m^2=F_6(x)=R_{2N-2-2s_+}(x)\widetilde{R}_{2N-2-2s_-}(x) \nonu
\end{eqnarray}
and one writes this as
\begin{eqnarray}
R_{2N-2-2s_+}(x)\widetilde{R}_{2N-2-2s_-}(x)
={W^{\prime}_3}^2(x)+f_{2}(x). \nonu
\end{eqnarray}
Since $s_++s_-+1=2N-4$, the left hand side is a 6-th order
polynomial and this determines $W_3^{\prime}(x)$ and hence
determines $W(x)$ up to a constant. As already discussed in
\cite{feng1} from the coefficient of $x^2$ in $f_2(x)$ we can find
$S=S_0+S_1$.

{\bf $SO(2N+1)$ case}

We can deal with $SO(2N+1)$ gauge theories in the same way as
$SO(2N)$ gauge theories. We consider the same tree level
superpotential (\ref{tree4}) and massless monopole constraint
(\ref{curveodd}) with $n=1$,
\begin{eqnarray}
P_{2N}^2(x)-4x^2\Lambda^{4N-2}=x^2 H^2_{2N-4}(x)F_{6}(x). \nonu
\end{eqnarray}
The matrix model curve is
determined from
 the solution of this factorization problem as follows:
\begin{eqnarray}
y_m^2(x)=F_6(x). \nonu
\end{eqnarray}
Now we are ready to deal with
the explicit examples.

$\bullet$ {\bf $SO(4)$ case}

\noindent The first example we consider is a $SO(4)$ gauge theory.
We are considering the $n=1$ case and $2N-2n-2=4-2-2=0$. The
characteristic function $P_4(x)$ can be represented as
\begin{eqnarray}
P_4(x)=x^2(x^2-v^2). \nonu
\end{eqnarray} This is a solution for
massless monopole constraint (\ref{exsoconst}). In this case
massless monopole constraint, the factorization problem, is
trivial since we do not need to have double roots at all and by
turning on the quartic superpotential it leads to matrix model
curve, according to (\ref{exsoconst})
\begin{eqnarray}
y_m^2=x^{-2}\left(P_{4}^2(x)-4x^4
\Lambda^{4}\right)=x^2(x^2-v^2)^2-4x^2\Lambda^4. \nonu
\end{eqnarray}
From this curve, we can determine the tree level superpotential
and a deformed function $f_2(x)$ as follow,
\begin{eqnarray}
W_3^{\prime}(x)=x(x^2-v^2),\quad f_2(x)=-4x^2\Lambda^4,\quad
S=\Lambda^4. \nonu
\end{eqnarray}
Then there is only one vacuum for a given $W^{\prime}_3(x)$ that
determines $v$ through the above relation. In the semiclassical
limit $\Lambda \to 0$, the characteristic function is independent
of $\La$ and becomes $P_4(x)\to x^2(x^2-v^2)$. So in this vacuum,
the gauge group $SO(4)$ breaks into $SO(2)\times U(1)$.

$\bullet$ {\bf $SO(5)$ case}

\noindent The next example is a $SO(5)$ gauge theory. As in
previous example, the massless monopole constraint is trivial
since we do not need to consider double roots at all. The
characteristic function $P_4(x)$ is given by
\begin{eqnarray}
P_4(x)=x^2(x^2-l^2).
\nonu
\end{eqnarray} From this, we obtain the
matrix model curve, by turning on the quartic superpotential
\begin{eqnarray}
y_m^2 & = & x^{-2} \left( P_{4}^2(x)-4x^2\Lambda^{6} \right)
=x^2(x^2-l^2)^2-4\Lambda^6, \nonu \\
 W_3^{\prime}(x) & = & x^2(x^2-l^2), \nonu \\
  f_2(x) & = & -4\Lambda^6.
  \nonu
\end{eqnarray}
There is only one vacuum for given $W^{\prime}_3(x)$ that
determines $l$. The different point compared to  $SO(4)$ case is
the fact that the sum of glueball superfield is different. As we
can see in the above, the function $f_2(x)$ does not have $x$
dependence, therefore the expectation value of glueball superfield
is zero, $S=0$. In the semiclassical limit, $P_4(x)\to
x^2(x^2-l^2)$, which shows that the gauge group breaks as
$SO(5)\to SO(3)\times U(1)$.

$\bullet$ {\bf $SO(6)$ case}

\noindent The third example is a $SO(6)$ gauge theory which is
more interesting. The massless monopole constraint for this case
is given by
\begin{eqnarray}
P_{6}^2(x)-4x^4\Lambda^8=x^2H^2_2(x)F_6(x). \label{ss1}
\end{eqnarray}
If we parametrize $P_6(x)=x^2P_4(x)=x^2(x^2-a^2)(x^2-b^2)$, the
equation (\ref{ss1}) becomes
\begin{eqnarray}
x^4\left((x^2-a^2)(x^2-b^2)-2\Lambda^4 \right)
\left((x^2-a^2)(x^2-b^2)+2 \Lambda^4\right)=x^2 H^2_2(x) F_6(x).
\nonu
\end{eqnarray}
From this equation, the subspace with a massless monopole can be
determined by searching for points where $y^2$ has a double roots
and  we find the relation as follows, by looking at the inside of
the bracket in the left hand side and requiring it be a complete
square,
\begin{eqnarray}
a^2-b^2=\epsilon 2\sqrt{2}\Lambda^4 \nonu
\end{eqnarray} where
$\epsilon$ is $4$-th root of unity. Thus we have a solution for
massless monopole constraint,
\begin{eqnarray}
P_6(x)=x^2(x^2-a^2)(x^2-a^2+2\sqrt{2}\epsilon \Lambda^4).
\nonu
\end{eqnarray} In the classical limit $P_6(x)\to
x^2(x^2-a^2)^2$, then the gauge group $SO(6)$ break to
$SO(2)\times U(2)$. The matrix model curve by turning on a
superpotential which makes a system to put at the point $(a,b)$ in
the ${\cal N}=2$ moduli space  is given by
\begin{eqnarray}
y_m^2& = & x^2\left( (x^2-a^2+\epsilon
\sqrt{2}\Lambda^4)^2+4\epsilon^2
\Lambda^8      \right), \nonu \\
 W_3^{\prime}(x) & = & x(x^2-a^2+\epsilon
\sqrt{2}\Lambda^4), \nonu \\
 f_2(x) & = & 4x^2\epsilon^2 \Lambda^8, \nonu \\
 S & = & -\epsilon^2\Lambda^8.
\nonu
\end{eqnarray}
These four vacua are confining phase because we can construct
multiplication map from $SO(4)$ by $K=2$ where we denote as
$P^{K=2}_{SO(4)\to SO(6)}(x)$. According to the analysis of
(\ref{multisol}), $2KN-2K+2$ becomes 6 when $N=2$ and $K=2$ and
the characteristic function  is given by, where ${\cal
T}_2(\widetilde{x})=2 \widetilde{x}^2-1$,
\begin{eqnarray}
P^{K=2}_{SO(4)\to SO(6)}(x)&=&2\eta^2 x^2 \Lambda^4 {\cal T}_2
\left(\frac{P_4(x)}{2\eta x^2 \Lambda^2} \right)=2\eta^2 x^2
\Lambda^4 {\cal T}_2 \left(\frac{x^2(x^2-v^2)}{2\eta x^2
\Lambda^2}
\right) \nonu \\
&=&2\eta^2 x^2 \Lambda^4 \left[2\left(\frac{x^2(x^2-v^2)}{2\eta
x^2 \Lambda^2} \right)^2-1 \right]=x^2(x^2-v^2)^2-2\eta^2x^2
\Lambda^4. \nonu
\end{eqnarray}
Then if we choose $v^2=a^2-\sqrt{2}\epsilon \Lambda^4$, one gets
\begin{eqnarray}
P^{K=2}_{SO(4)\to SO(6)}(x)&=&x^2\left( (x^2-a^2)^2+2\sqrt{2}
\epsilon\Lambda^2(x^2-a^2)+2\Lambda^4(\epsilon^2-\eta^2) \right)
\nonu
\\
&=&x^2(x^2-a^2)(x^2-a^2+2\sqrt{2}\epsilon \Lambda^4)=P_6(x)
\nonu
\end{eqnarray}
which we have found before  and we used the fact that both
$\epsilon$ and $\eta$ are $4$-th roots of unity.

Until now we considered confining vacua only. Are there other
vacua in $SO(6)$ gauge theory? We can easily see Coulomb vacua for
this case as in $SO(4)$, $SO(5)$ gauge theories. If we put
$H_2(x)=x^2$ and $P_6(x)=x^2P_4(x)$, then the massless monopole
constraint becomes
\begin{eqnarray} P_4^2(x)-4\Lambda^8=x^2F_6(x).
\nonu
\end{eqnarray} From this equation we can obtain the solution,
$P_4(x)=x^4+Ax^2+2\eta \Lambda^4$ where $\eta $ is $2$-th root of
unity. In the semiclassical limit, $P_6(x)\to x^4(x^2+A)$, which
shows that the gauge group breaks into $SO(6)\to SO(4)\times
U(1)$. Thus the matrix model curve is given by
\begin{eqnarray}
y_m^2&=&x^2(x^2+A)^2+4\eta \Lambda^4(x^2+A), \nonumber \\
W^{\prime}_3(x)&=& x(x^2+A), \nonumber \\
f_2(x)&=&4\eta \Lambda^4(x^2+A), \nonumber \\
S&=&-\eta \Lambda^4.\nonu
\end{eqnarray}
Then there are two vacua for given superpotential $W(x)$.

$\bullet$ {\bf $SO(7)$ case}

\noindent Next example is a $SO(7)$ gauge theory which is first
example for smooth transition between vacua with different unbroken
gauge groups. The massless monopole constraint for this case
is given by
\begin{eqnarray}
P_{6}^2(x)-4x^2\Lambda^{10}=x^2H^2_2(x)F_6(x).
\label{ss17}
\end{eqnarray}
If we appropriately parametrize $P_6(x)$, $H_2(x)$ and $F_6(x)$,
the equation (\ref{ss17}) becomes
\begin{eqnarray}
x^4(x^2-a)^2(x^2-a-b)^2-4x^2\Lambda^{10}=x^2(x^2-A)^2
\left(x^6+Bx^4+Cx^2+D \right). \nonumber
\end{eqnarray}
The solutions of this equation are given by
\begin{eqnarray}
B&=&-2A+\frac{2\epsilon^2 \Lambda^5}{A^{\frac{3}{2}}},\ \
C=A^2-\frac{6\epsilon^2\Lambda^5}{A^{\frac{1}{2}}}+\frac{\Lambda^{10}}{A^{3}},
\ \ D=-\frac{4\Lambda^{10}}{A^2}, \nonumber \\
b&=&\epsilon \left[\frac{\Lambda^5}{A^{\frac{3}{2}}}\left(8A+\frac{\epsilon^2
\Lambda^5}{A^{\frac{3}{2}}} \right) \right]^{\frac{1}{2}},\ \
a=A+\frac{\epsilon^2 \Lambda^5}{2A^{\frac{3}{2}}}-\frac{b}{2} \label{so7sol}
\end{eqnarray}
where $\epsilon$ is $4$-th root of unity. In this case we can take
two semiclassical limits with $\Lambda \to 0$:

1. Fixed A: Since $b \to 0$ and $a\to A$, the characteristic
function $P_6(x)\to x^2(x^2-A)^2$. Therefore $SO(7)$ is broken to
$SO(3)\times U(2)$. This phase is a Coulomb phase because the
numbers $(2N_0+1)-2=3-2=1$ and $N_1=2$ that come from unbroken
gauge groups have no common divisor.

2. $\Lambda, A\to 0$ with fixed $v\equiv \frac{\epsilon^2
\Lambda^5}{A^{\frac{3}{2}}}$: Since $b\to v $ and $a\to 0$,
$P_6(x)\to x^4(x^2-v)$. Therefore $SO(7)$ is broken to
$SO(5)\times U(1)$. As previous limit, we have no common divisor
of $(2N_0+1)-2=5-2=3$ and $N_1=1$. Therefore this phase is a
Coulomb phase also. After all we have obtained smooth transition
within Coulomb phase.

From the solutions (\ref{so7sol}) we find the corresponding
matrix model curve as follows:
\begin{eqnarray}
y_m^2& = & x^2\left( x^2-A+\frac{\epsilon^2 \Lambda^5}{A^{\frac{3}{2}}}
\right)^2-\frac{8\epsilon^2 \Lambda^5}{A^{\frac{1}{2}}}x^2-\frac{4\Lambda^{10}}{A^2},
\nonumber \\
 W_3^{\prime}(x) & = & x\left( x^2-A+\frac{\epsilon^2
 \Lambda^5}{A^{\frac{3}{2}}} \right), \nonumber \\
 f_2(x) & = &-\frac{8\epsilon^2 \Lambda^5}{A^{\frac{1}{2}}}x^2-
 \frac{4\Lambda^{10}}{A^2} , \nonumber \\
 S & = & \frac{2\epsilon^2 \Lambda^5}{A^{\frac{1}{2}}}.
\nonumber
\end{eqnarray}
Next we count the number of vacua for fixed superpotential,
$W_0^{\prime}=x(x^2+\Delta)$. How many vacua do we have for this
fixed superpotential? From the above result of $W_3^{\prime}(x)$
we can represent $\Delta$ as
\begin{eqnarray}
\Delta=\frac{\epsilon^2\Lambda^5}{A^{\frac{3}{2}}}-A. \nonumber
\end{eqnarray}
We evaluate this equation under the two semiclassical limit $1$ and $2$ discussed above.

1. In this limit since $\Delta=-A$ we have 
two functions $f_2(x)$ for each $\epsilon^2$.
Therefore we have two vacua. 
On the other hand, since the unbroken gauge group
is $SO(3)\times U(2)$ under this limit, 
the number of vacua is $2\cdot 2=4$ \footnote{
For $SO(N)$ gauge theories with $N\ge 5$ the Witten index is $N-2$ 
while for $N\le 4$ the Witten indices are $1,2,4$ for $N=2,3,4$ 
respectively.}.
This number is not equal to the one derived from the potential. 
The reason is as follows: We considered only massless monopole 
constraint (\ref{curveodd}) which have $x^2$ factor. 
If we include the case without $x^2$ factor, namely, 
if the branch cut at the origin of the matrix model curve $y_m$ 
is degenerated, 
we will have the remaining vacua with 
the breaking pattern $SO(7)\to SO(3)\times U(2)$ 
in addition to $SO(7)\to U(3)$ 
\footnote{We would like to thank Bo Feng for the discussion on this point.}.

2. In this limit since $\frac{\epsilon^2}{A^{\frac{1}{2}}}= \left(
\frac{\Delta}{\Lambda^5}\right)^{\frac{1}{3}}$, we have three
functions $f_2(x)$. Thus there are three vacua. On the other hand,
we can count the number of vacua from gauge group. In this limit since the
unbroken gauge group is $SO(5)\times U(1)$ the number 
of vacua is $((2N_0+1)-2)\times N_1=(5-2)\cdot 1=3$,
which is the same number as the one from potential.

$\bullet$ {\bf $SO(8)$ case}

\noindent At last we study the most interesting example $SO(8)$.
We are looking for the subspace of the ${\cal N}=2$ gauge theory
with $N-n=4-1=3$ monopoles. In this gauge theory since
$s_++s_-=3$, we have four branches $(s_+,s_-)=(2,1)$, $(1,2)$ and
$(3,0)$, $(0,3)$. The four double roots of the ${\cal N}=2$ curve
are distributed between the two factors in this region.

{\bf 1. Coulomb branch with $(s_+,s_-)=(2,1)$ or $(1,2)$}

\noindent By introducing $\eta$ we can deal with these two cases
in the same way, that is, simultaneously. Taking into account the
characteristic function of degree 8, $P_8(x)$ is an even function,
we can parametrize points where monopoles are massless as
$H_{s_-}(x)=x$ and $H_{s_+}(x)=x^2-a^2$. The solutions of
equations (\ref{SS}) or (\ref{s3}) are given by
\begin{eqnarray}
P_8(x)+2\eta x^2\Lambda^{6}&=&x^2(x^2-a^2)^2(x^2+\frac{4\eta
\Lambda^6}{a^4}), \nonu
\\P_8(x)-2\eta x^2\Lambda^{6}&=&x^4 \left[(x^2-a^2)^2+\frac{4\eta
\Lambda^6}{a^4}(x^2-2a^2) \right]. \nonu
\end{eqnarray}
Then we find the corresponding matrix model curve as follows:
\begin{eqnarray}
y_m^2&=&(x^2+\frac{4\eta \Lambda^6}{a^4})
\left[(x^2-a^2)^2+\frac{4\eta \Lambda^6}
{a^4}(x^2-2a^2) \right] \nonu \\
&=&x^2\left(x^2+\frac{4\eta \Lambda^6}{a^4}-a^2
\right)^2-\frac{8\eta \Lambda^6}{a^2}x^2+\frac{4\eta
\Lambda^6}{a^2}(a^2-2\frac{4\eta \Lambda^6}{a^4})
\nonu
\end{eqnarray}
from which we can see the tree level superpotential and the
deformation function $f_2(x)$
\begin{eqnarray}
W^{\prime}_3(x)=x\left(x^2+\frac{4\eta \Lambda^6}{a^4}-a^2
\right), \qquad f_2(x)=-\frac{8\eta \Lambda^6}{a^2}x^2+\frac{4\eta
\Lambda^6}{a^2} (a^2-2\frac{4\eta \Lambda^6}{a^4}) . \nonu
\end{eqnarray}
From the coefficient of $x^2$ in $f_2(x)$, we find $S=\frac{2\eta
\Lambda^6}{a^2}$. There are two semiclassical limits with $\Lambda
\to 0$,

1. Fixed $a$: The characteristic function behaves $P_8(x) \to
x^4(x^2-a^2)^2$. Therefore the gauge group $SO(8)$ is broken to
$SO(4)\times U(2)$.

2. $\Lambda,a \to 0$ with fixed $v \equiv \frac{4\eta
\Lambda^6}{a^4}$: The characteristic function becomes $P_8(x) \to
x^6 (x^2+v)$, Therefore $SO(8)$ is broken to $SO(6)\times U(1)$.

\noindent These results represent two distinct semiclassical
limits in the same moduli space corresponding to different gauge
groups. By changing parameter continuously one can freely transit
from $SO(4)\times U(2)$ to $SO(6)\times U(1)$. On the other hand,
the confining $SO(4) \times U(2)$ vacua below can not make such a
transformation because there are no confining $SO(6) \times U(1)$
vacua.


At last we count the number of vacua for fixed tree level
superpotential, $W_3^{\prime}(x)=x(x^2+\Delta)$. From the previous
result of $W_3^{\prime}(x)$ we can represent $\Delta$ as
\begin{eqnarray}
\Delta=\frac{4\eta \Lambda^6}{a^4}-a^2, \nonumber
\end{eqnarray}
in this branch. We evaluate this equation under the
two semiclassical limit $1$ and $2$ discussed above.

1. In this limit since $\Delta=-a^2$ we have two functions
$f_2(x)$ for each $\eta$. Thus we have two vacua. As we will see
below, we have two vacua in confining phase that have the same
gauge group $SO(4)\times U(2)$. Thus all the vacua with this gauge
group are four vacua. On the other hand since the gauge group is
$SO(4)\times U(2)$ under this limit, the number of vacua is
$4 \cdot 2=8$. This number is not equal to the
one derived from the superpotential analysis. 
As discussed in $SO(7)$ case if we include the case without $x^2$ factor, 
namely, if  the branch cut at the origin of the 
matrix model curve $y_m$ is degenerated, we will have the 
remaining vacua with the breaking pattern $SO(8)\to SO(4)\times U(2)$ 
in addition to $SO(8)\to U(4)$.

2. In this limit since
$a^2=\left(\frac{2\Lambda^6}{\Delta}\right)^{\frac{1}{2}}$, taking
into account for $\eta$ we have four functions $f_2(x)$. In other
words we have four vacua for each potential. This number is equal
to the one derived from the gauge group $SO(6)\times U(1)$, i.e.
$(2N_0-2) \times N_1=(6-2)\cdot 1=4$.


{\bf 2. Confining branch $(s_+,s_-)=(0,3)$ or $(3,0)$ }

\noindent Next we study the remaining two cases $(s_+,s_-)=(0,3)$
and $(3,0)$. As in previous examples we introduce $\eta=\pm 1$. We
can easily obtain a solution for massless monopole constraint,
\begin{eqnarray}
P_8(x)+2\eta x^2\Lambda^{6}&=&x^4(x^2-a^2)^2+4\eta x^2 \Lambda^6, \nonu \\
P_8(x)-2\eta x^2\Lambda^{6}&=&x^4(x^2-a^2)^2.
\label{P8}
\end{eqnarray}
Then the matrix model curve is given by
\begin{eqnarray}
y_m^2 & = & x^2(x^2-a^2)^2+4\eta \Lambda^6,\nonu \\
W_3^{\prime}(x) & = & x(x^2-a^2), \nonu \\
f_2(x) & = & 4\eta \Lambda^6. \nonu
\end{eqnarray}
Since the coefficient in front of $x^2$ in $f_2(x)$ is zero, the
expectation value of the sum of glueball superfield is zero.

In the semiclassical limit $\Lambda \to 0$, the characteristic
function becomes $P_8(x)\to x^4(x^2-a^2)^2$, which means that
$SO(8)$ breaks to $SO(4)\times U(2)$. This breaking pattern has
already appeared. However in this case, the phase is a confining
phase, which is different from the former.

This solution we have just described is a confining phase because
we can construct a multiplication map from $SO(5)$ to $SO(8)$ by
$K=2$ where we denote as $P^{K=2}_{SO(5)\to SO(8)}(x)$. According
to the analysis of (\ref{multisoloddeven}), $2KN-K+2=8$ for
$N=K=2$. Plugging these values into (\ref{multisoloddeven}), one
gets
\begin{eqnarray}
P^{K=2}_{SO(5)\to SO(8)}(x)&=&2\eta^2 x^2 \Lambda^6 {\cal T}_2
\left(\frac{P_4(x)}{2\eta x \Lambda^3} \right)=2\eta^2 x^2
\Lambda^6 \left[2\left(\frac{x^2(x^2-l^2)}{2\eta x \Lambda^3}
\right)^2-1
 \right] \nonu \\
&=& x^4(x^2-l^2)^2-2\eta^2 x^2 \Lambda^6 \nonu
\end{eqnarray}
where $\eta$ is a 4-th root of unity. If we identify $a^2$ with
$l^2$ and taking the $\eta^2$ here as the minus of $\eta$ in
(\ref{P8}) this equation becomes exactly the solution of
(\ref{P8}). Thus this vacuum is a confining phase.

So far we have seen the multiplication maps from $SO(4)$ to
$SO(6)$ and from $SO(5)$ to $SO(8)$ where the former is an example
of the vacua  of $SO(2KN-2K+2)$ gauge theory from $SO(2N)$ theory
while the latter is an example of the vacua  of $SO(2KN-K+2)$
gauge theory where $K$ is even from $SO(2N+1)$ theory. Then it is
natural to consider whether there exists the multiplication map
from $SO(2N+1)$ to $SO(2KN-K+2)$ where $K$ is odd. For example,
when $K=3$ and $N=2$, then the vacua of $SO(11)$ gauge theory has
the same superpotential as those of $SO(5)$ theory. Therefore we
expect that there exists a solution that is a confining phase from
the explicit multiplication map from $SO(5)$ to $SO(11)$ by $K=3$
where we denote as $P^{K=3}_{SO(5)\to SO(11)}(x)$. Although we
have considered a couple of examples, it would be interesting to
study the possible confining vacua for each general $N$ and $K$
systematically.

One might ask whether one can construct the multiplication map
from $SO(2N)$ to $SO(2M+1)$. From the experience we have learned
from the section 2.3, one can think of $x  {\cal T}_K
\left(\frac{P_{2N}(x)}{x^2} \right)$ as our characteristic
function $P_{2M}(x)$. The reason for linear behavior of $x$ is
that as we remember the curve for $SO(2M+1)$ gauge theory contains
the $x^2$ in $\La$ term so when we make $P_{2M}(x)$ to be square,
we have to have $x^2$ term. Moreover the $1/x^2$ behavior of the
first kind Chebyshev polynomial is  necessary to require that
$SO(2N)$ gauge theory behave like $x^4$ in the $\La$ term and from
the relation of first kind Chebyshev and second kind Chebyshev.
However, this candidate for the characteristic polynomial has a
degree of $1+K(2N-2)$ which is odd, contrary to the fact that in
our $SO(2M+1)$ theory, the degree of polynomial should be even.
From this naive counting, it does not seem to allow one can
construct the multiplication map from $SO(2N)$ to $SO(2M+1)$.
Therefore, for $SO(2N)$ and $SO(2N+1)$ gauge theories, there exist
only three possible ways to have multiplication maps.

\section{$Sp(N)$ gauge theory}
\setcounter{equation}{0}

\subsection{Strong gauge coupling approach \label{spstrong}}


Let us study the superpotential considered as a small perturbation
of an ${\cal N}=2$ $Sp(2N)$ gauge theory \cite{ahn98}. The
superpotential is given by (\ref{treesup}) and $\Phi$ is an
adjoint scalar chiral superfield and one can transform as follows:
\bea \Phi= \left( {1 \atop 0 }{ 0 \atop -1
 } \right) \otimes \mbox{diag} ( i\phi_1, \cdots,  i\phi_{N}).
\nonu \eea
Note that  the trace of odd power of $\Phi$ is zero. We study the
vacua in which the perturbation by $W(\Phi)$ (\ref{treesup})
remains only $U(1)^{n}$ gauge group at low energies with $2n \leq
2k$. If the remaining degrees of freedom become massive for $W
\neq 0$ due to the condensation of $(N-n)$ mutually local
magnetic monopoles, then the exact effective superpotential can be
written as \bea W_{eff}=\sqrt{2} \sum_{l=1}^{N-n} M_{l}(u_{2r})
q_l \widetilde{q}_l + \sum_{r=1}^{k+1} g_{2r} u_{2r}
 \nonu
\eea similar to the one in $SO(2N)$ case.
The mass of monopoles
should vanish for $l=1,2, \cdots, (N-n)$ in a supersymmetric
vacuum.

We consider a singular point in the moduli space where $(N-n)$
mutually local monopoles are massless. We can represent massless
monopole constraint as,
\begin{eqnarray}
y^2 &=& B^2_{2N+2}(x)-4\Lambda^{4N+4}= \left( x^2P_{2N}(x)+
2\Lambda^{2N+2} \right)^2-4\Lambda^{4N+4} \nonu \\
&= & x^2H_{2N-2n}^2(x)F_{2(2n+1)}(x)
\label{spconstraint}
\end{eqnarray}
where we define
$B_{2N+2}(x) \equiv x^2P_{2N}(x)+2\Lambda^{2N+2} $
and
\bea H_{2N-2n}(x)=
\prod^{N-n}_{i=1}(x^2-p_i^2), \qquad F_{2(2n+1)}(x)
=\prod^{2n+1}_{i=1}(x^2-q_i^2).
\nonu \eea The reason for introducing a polynomial $B_{2N+2}(x)$
rather than $P_{2N}(x)$ is that one can treat a curve
(\ref{spconstraint}) in a symmetric way and one can factorize it.
The function $H_{2N-2n}(x)$ is a polynomial in $x$ of degree
$(2N-2n)$ giving $(2N-2n)$ double roots and the function
$F_{2(2n+1)}(x)$ is a polynomial in $x$ of degree $(4n+2)$. These
are even functions in $x$ and depend on only $x^2$. The
characteristic function $P_{2N}(x)$ becomes
\begin{eqnarray}
P_{2N}(x)=\mbox{det}(x-\Phi)=\prod_{I=1}^N(x^2-\phi^2_I).
\nonu
\end{eqnarray}
When the degree $(2k+1)$ of $W^{\prime}(x)$ is equal to $(2n+1)$,
the highest $(2n+2)$ coefficients of $F_{2(2n+1)}$ are given in
terms of $W^{\prime}(x)$ \cite{feng1} and the expression
(\ref{sol}) is the same.

$\bullet$ {\bf Superpotential of degree $2(k+1)$ less than $2N$}

We describe the superpotential when the degree of $W^{\prime}$,
$(2k+1)$ is greater than $(2n+1)$ with tree level superpotential
(\ref{treesup}), namely $2n+2\leq 2k+2 \leq 2N$ case. Let us
consider the superpotential under these constraints,
\begin{eqnarray}
W_{eff}=
\sum_{r=1}^{k+1}g_{2r}u_{2r}+\sum_{i=0}^{2N-2n}
\left[L_i\oint \frac{B_{2N+2}(x)-2\epsilon_i
\Lambda^{2N+2}}{(x-p_i)}dx+B_i\oint \frac{B_{2N+2}(x)-2\epsilon_i  \Lambda^{2N+2}}
{(x-p_i)^2}dx \right],
\nonu
\end{eqnarray} where $L_i$ and $B_i$ are Lagrange multipliers
and $\epsilon_i=\pm 1 $ as we have considered in $SO(2N)$ case.
The convention of an assignment of $\epsilon_i$ is different from
the one in \cite{feng1}. The $p_i$'s are the locations of the
double roots of $y^2 = B_{2N+2}^2(x)-4 \La^{4N+4}$. The massless
monopoles occur in pair $(p_i, -p_i)$ and also both the expression
$B_{2N+2}(x)-2\epsilon_i \Lambda^{2N+2}$ at $x=\pm p_i$ and its
derivative with respect to $x$ at $x=\pm p_i$ are vanishing since
$B_{2N+2}(x)$ is an even function of $x$. The half of the Lagrange
multipliers  are not independent. Totally the number of constraint
is $(N-n)$.

The variation of $W_{eff}$ with respect to $B_i$ leads to
\begin{eqnarray}
0&=&\oint \frac{B_{2N+2}(x) -2\epsilon_i \Lambda^{2N+2}}
{(x-p_i)^2}dx=B_{2N+2}^{\prime}(x)
 |_{x=p_i}=\left(x^2P_{2N}^{\prime}(x)+2xP_{2N}(x) \right)|_{x=p_i} \nonu \\
&=&x^2P_{2N}(x)\left(\mbox{Tr} \frac{1}{x-\Phi}+\frac{2}{x}
\right)|_{x=p_i} \label{sprelation7}
\end{eqnarray} where we exploited the equation of motion for
$L_i$ and the last equality comes from the following relation,
\begin{eqnarray}
\mbox{Tr} \frac{1}{x-\Phi}
=
\sum_{I=1}^N \frac{2x}{x^2-\phi_I^2}.
\nonu
\end{eqnarray}
Since $P_{2N}(x=p_i) \neq 0$ due to the relation
(\ref{spconstraint}) and $H_{2N-2n}(x=p_i)=0$, we arrive at \bea
\left(\mbox{Tr} \frac{1}{x-\Phi}+\frac{2}{x} \right)|_{x=p_i}=0.
\nonu \eea Notice that the presence of $2/x$ with positive sign is
different from the one $-2/x$ for $SO(2N)$ case and $-1/x$ for
$SO(2N+1)$ case.
All the property of the characteristic function $P_{2N}(x)$ for
the previous cases hold for $Sp(2N)$ case also. We do not write
them explicitly here. Next we consider the variations of $W_{eff}$
with respect to $p_j$,
\begin{eqnarray}
0&=&L_j\oint \frac{B_{2N+2}(x) -2\epsilon_i
\Lambda^{2N+2}}{(x-p_i)^2}dx+2B_j
\oint \frac{B_{2N+2}(x) -2\epsilon_i
\Lambda^{2N+2}}{(x-p_i)^3}dx \nonu \\
&=&2B_j \oint \frac{B_{2N+2}(x) -2\epsilon_i
\Lambda^{2N+2}}{(x-p_i)^3}dx
\nonu
\end{eqnarray}
where in the last equality we used the equation of motion for
$B_i$ (\ref{sprelation7}). In general, this integral does not
vanish, then we have $B_i=0$ based on the same reason about the
structure of the algebraic curve in previous cases. Let us
consider variations with respect to $u_{2r}$,
\begin{eqnarray}
0=g_{2r}-\sum_{i=0}^{2N-2n}\oint \left[\frac{x^2P_{2N}(x)}
{x^{2r}} \right]_{+} \frac{L_i}{x-p_i}dx, \nonu
\end{eqnarray}
where we used $B_i=0$ at the level of equation of motion and we
express the explicit form for $B_{2N+2}(x)$ in terms of
$P_{2N}(x)$. Multiplying this with $z^{2r-1}$ and summing over
$r$, we can obtain the following relation,
\begin{eqnarray}
W^{\prime}(z)=\sum_{r=1}^{k+1}
g_{2r}z^{2r-1}=\sum_{i=0}^{2N-2n}\oint
\sum_{r=1}^{k+1}z^{2r-1}\frac{x^2P_{2N}(x)}{x^{2r}}\frac{L_i}{(x-p_i)}dx.
\label{sprelation2}
\end{eqnarray}

Let us introduce a new polynomial $Q(x)$ defined as
\begin{eqnarray}
\sum_{i=0}^{2N-2n}\frac{xL_i}{(x-p_i)}=L_0+
\sum_{i=1}^{N-n}xL_i\left(\frac{1}{x-p_i}+ \frac{1}{x+p_i}
\right)=L_0+\sum_{i=1}^{N-n}\frac{2x^2L_i}{x^2-p_i^2}
\equiv\frac{Q(x)}{H_{2N-2n}(x)}.
 \label{sprelation6}
\end{eqnarray}
Using this new function we can rewrite (\ref{sprelation2}) as
\begin{eqnarray}
W^{\prime}(z)=\oint
\sum_{r=1}^{k+1}\frac{z^{2r-1}}{x^{2r}}\frac{Q(x)x^2P_{2N}(x)}
{xH_{2N-2n}(x)}dx.
\nonu
\end{eqnarray}
Since $W^{\prime}(z)$ is a polynomial of degree $(2k+1)$,  the
order of $Q(x)$ is $(2k-2n)$, so we denote it by $Q_{2k-2n}$. Thus
we have found the order of polynomial $Q(x)$ and the order of
integrand in (\ref{sprelation2}) behaves like as ${\cal
O}(x^{2k-2r+3})$. Thus if $r\ge k+2$ it does not contribute to the
integral since the Laurent expansion around the origin vanishes.
We can replace the upper value of summation to infinity.
\begin{eqnarray}
W^{\prime}(z)=\oint \sum_{r=1}^{\infty}\frac{z^{2r-1}}{x^{2r}}
\frac{Q_{2k-2n}(x) x^2P_{2N}(x)}{xH_{2N-2n}(x)}dx= \oint z
\frac{Q_{2k-2n}(x)x^2P_{2N}(x)}{x(x^2-z^2)H_{2N-2n}(x)}dx.
\label{spprires}
\end{eqnarray}
From (\ref{spconstraint}) one can write,
\begin{eqnarray}
x^2P_{2N}(x)=x\sqrt{F_{2(2n+1)}(x)}H_{2N-2n}(x)+{\cal
O}(x^{-2N-2}). \nonu
\end{eqnarray}
By substituting this relation to (\ref{spprires}), the ${\cal
O}(x^{-2N-2})$ terms do not contribute the integral, so we can
drop those terms.

Therefore we have
\begin{eqnarray}
W^{\prime}(z)=\oint z\frac{y_m(x)}{x^2-z^2}dx, \qquad y_m^2(x)
=F_{2(2n+1)}(x) Q_{2k-2n}^2(x) \nonu
\end{eqnarray}
corresponding to the equation of motion and the curve in the
matrix model context. Then we get the final result
\begin{eqnarray}
y_m^2(x)=F_{2(2n+1)}(x)Q_{2k-2n}^2(x)
={W^{\prime}}^2_{2k+1}(x)+{\cal O}(x^{2k})=
{W^{\prime}}^2_{2k+1}(x)+f_{2k}(x) \nonu
\end{eqnarray}
where both $F_{2(2n+1)}(x)$ and $Q_{2k-2n}(x)$ are functions of
$x^2$, then $f_{2k}(x)$ is also a function of $x^2$. When $2k=2n$,
we reproduce (\ref{sol}) with $Q_0= g_{2n+2}$. The above relation
determines a polynomial $F_{2(2n+1)}(x)$ in terms of $(2n+1)$
unknown parameters by assuming the leading coefficient to be
normalized by 1 by assuming that $W(x)$ is known.

$\bullet$ {\bf Superpotential of degree $2(k+1)$ where $k$ is
arbitrary large}

Let us extend the study of previous discussion to the general case
without imposing the constraint on the degree of superpotential.
As before, we denote $\frac{1}{2k}\langle \mbox{Tr} \Phi^{2k}
\rangle=U_{2k}$. The corresponding quantum expression can be
written as
\begin{eqnarray}
\left\langle \mbox{Tr}\frac{1}{x-\Phi} \right\rangle=\frac{2N}{x}+
\sum_{k=1}^{\infty}\frac{2kU_{2k}}{x^{2k+1}}
\nonu
\end{eqnarray}
and also  quantum mechanically this can be given by
\begin{eqnarray}
\left\langle \mbox{Tr}\frac{1}{x-\Phi}
\right\rangle=\frac{d}{dx}\left[ \log \left(B_{2N+2}(x)+
\sqrt{B^2_{2N+2}(x)-4\Lambda^{4N+4}} \right) -\log x^2 \right].
\label{10may2C}
\end{eqnarray}
Note in order to make the degree of the expression  inside of the
log as $2N$, we inserted the last term.
By integrating and exponentiating, one gets
\begin{eqnarray}
x^{2N+2}\exp\left(-\sum_{i=1}^{\infty}\frac{U_{2i}}{x^{2i}}
\right)= B_{2N+2}(x)+\sqrt{B^2_{2N+2}(x)-4\Lambda^{4N+4}}
\label{sprelation3}
\end{eqnarray}
where $C$ is an integration constant determined by the
semiclassical limit $\Lambda \to 0$:
\begin{eqnarray}
Cx^{2N}\exp\left(-\sum_{i=1}^{\infty} \frac{U_{2i}}{x^{2i}}
\right)=2P_{2N}(x) \longrightarrow C=2.
\nonu
\end{eqnarray}
Solving (\ref{sprelation3}) with respect to $P_{2N}(x)$, we get
\begin{eqnarray}
P_{2N}(x)=x^{2N}\exp\left(-\sum_{i=1}^{\infty}\frac{U_{2i}}{x^{2i}}
\right)+\frac{\Lambda^{4N+4}}{x^{2N+4}}\exp\left(\sum_{i=1}^{\infty}
\frac{U_{2i}}{x^{2i}} \right). \label{sprelation8}
\end{eqnarray}
This looks similar to the previous expression. The power behavior
of $\La$ and $x$ in the second term  is different.
 Since $P_{2N}(x)$ is a polynomial in $x$, the relation
(\ref{sprelation8}) can be used to express $U_{2r}$ with $2r>2N$
in terms of $U_{2r}$ with $2r\le 2N$ by imposing the vanishing of
the negative powers of $x$.

Let us introduce a new polynomial whose coefficients are Lagrange
multipliers. The superpotential with these constraints is
described as
\begin{eqnarray}
W_{eff}&=&\sum_{r=1}^{k+1}g_{2r}u_{2r}+\sum_{i=0}^{2N-2n}
\left[L_i\oint \frac{B_{2N+2}(x)-2 \epsilon_i
\Lambda^{2N+2}}{(x-p_i)}dx+B_i\oint \frac{B(x)-2\epsilon_i
\Lambda^{2N+2}}{(x-p_i)^2}dx \right] \nonu \\
&+&\oint R_{2k-2N+2}(x) \left[
x^{2N}\exp\left(-\sum_{i=1}^{\infty}\frac{U_{2i}}{x^{2i}}
\right)+\frac{\Lambda^{4N+4}}{x^{2N+4}}\exp
\left(\sum_{i=1}^{\infty}\frac{U_{2i}}{x^{2i}} \right) \right]
dx.
\nonu
\end{eqnarray}
Here $R_{2k-2N+2}(x)$ is a polynomial of degree $(2k-2N+2)$ whose
coefficients are regarded as Lagrange multipliers which impose
constraints $U_{2r}$ with $2r > 2N$ in terms of $U_{2r}$ with $2r
\leq 2N$. The derivative of $W_{eff}$ with respect to $U_{2r}$
leads to
\begin{eqnarray}
0&=&g_{2r}+\oint \frac{R_{2k-2N+2}(x)}{x^{2r}}\left( -x^{2N}
\exp\left(-\sum_{i=1}^{\infty}\frac{U_{2i}}{x^{2i}} \right)+
\frac{\Lambda^{4N+4}}{x^{2N+4}}
\exp\left(\sum_{i=1}^{\infty}\frac{U_{2i}}{x^{2i}}
 \right) \right)dx \nonu \\
&+& \oint \sum_{i=0}^{2N-2n}\frac{x^2L_i}{(x-p_i)} \frac{\partial
P_{2N}(x)}{\partial U_{2r}}dx. \label{sprelation4}
\end{eqnarray}
Using (\ref{sprelation3}) we have the relation,
\begin{eqnarray}
-x^{2N+2}\exp\left(-\sum_{i=1}^{\infty}\frac{U_{2i}}{x^{2i}}
 \right)+\frac{x^2\Lambda^{4N+4}}{x^{2N+4}}
 \exp\left(\sum_{i=1}^{\infty}\frac{U_{2i}}{x^{2i}}
 \right)=-\sqrt{B^2_{2N+2}(x)-4\Lambda^{4N+4}}
\nonu
\end{eqnarray}
and
\begin{eqnarray}
\frac{\partial P_{2N}(x) }{\partial U_{2r}}=
-\frac{P_{2N}(x)}{x^{2r}} \;\;  \mbox{for} \;\; 2r \le 2N , \qquad
\frac{\partial P_{2N}(x) }{\partial U_{2r}}=0 \;\; \mbox{for} \;\;
2r> 2N. \nonu
\end{eqnarray}
Using these relations we can rewrite (\ref{sprelation4}) as
follows:
\begin{eqnarray}
0=g_{2r}+\oint \frac{R_{2k-2N+2}(x)}{x^{2r}} \left(
-\frac{1}{x^2}\sqrt{B^2_{2N+2}(x)-4\Lambda^{4N+4}}\right)dx-
 \oint \sum_{i=0}^{2N-2n}\frac{x^2L_i}{x-p_i}\frac{P_{2N}(x)}{x^{2r}}
 dx.
 \label{sprelation10}
\end{eqnarray}
From the massless monopole constraint (\ref{spconstraint}) we have
the relation,
\begin{eqnarray}
x^2P_{2N}(x)=x H_{2N-2n}(x)\sqrt{F_{2(2n+1)}(x)}+{\cal
O}(x^{-2N-2}).
 \label{sprelation9}
\end{eqnarray}
Substituting (\ref{sprelation9}) and (\ref{spconstraint})
 into (\ref{sprelation10}), the
 ${\cal O}(x^{-2N-2})$ terms do not contribute to integral then
we get an expression for the coupling $g_{2r}$
\begin{eqnarray}
g_{2r}&=&\oint
\frac{R_{2k-2N+2}(x)}{x^{2r}}xH_{2N-2n}(x)\sqrt{F_{2(2n+1)}(x)}dx
\nonu \\
 &+ & \oint
\sum_{i=0}^{2N-2n}
\frac{L_i}{(x-p_i)}\frac{xH_{2N-2n}(x)\sqrt{F_{2(2n+1)}(x)}}{x^{2r}}
 dx. \nonumber
\end{eqnarray}

As in previous analysis, we define a new polynomial $Q(x)$ in
(\ref{sprelation6}) and we see
\begin{eqnarray}
W^{\prime}(z) &=&\sum_{r=1}^{k+1}\left[\oint
\frac{R_{2k-2N+2}(x)}{x^{2r}} xH_{2N-2n}(x)\sqrt{F_{2(2n+1)}(x)}dx
\right. \nonu \\
 &+& \left. \oint \sum_{i=0}^{2N-2n}\frac{L_i}{(x-p_i)}
\frac{xH_{2N-2n}(x)\sqrt{F_{2(2n+1)}(x)}}{x^{2r}}dx \right]z^{2r-1} \nonu \\
&=&\oint \frac{z}{x^2-z^2}\sqrt{F_{2(2n+1)}(x)} \left[R_{
2k-2N+2}(x)H_{2N-2n}(x)+ Q_{2N-2n}(x) \right]dx \nonu
\end{eqnarray}
if the matrix model curve is given by
\begin{eqnarray}
y_m^2 &=& F_{2(2n+1)}(x){\widetilde{Q}}_{2k-2n+2}^2(x) \nonu \\
&\equiv&
F_{2(2n+1)}(x)\left(R_{2k-2N+2}(x)H_{2N-2n}(x)+Q_{2N-2n}(x)\right)^2.
\nonu
\end{eqnarray}
When $2n=2N$ (no massless monopoles),
${\widetilde{Q}}_{2k-2N+2}(x)=R_{2k-2N+2}(x)$. When the degree of
superpotential is equal to $2N$, in other words, $2k=2N-2$, then
${\widetilde{Q}}_{2N-2n}(x)=R_{0} H_{2N-2n}(x)+ Q_{2N-2n}(x)$. In
particular, for $2n=2N$, \bea y^2_m(x)= F_{4N+2}(x) = x^{-2}
\left( B^2_{2N+2}(x)-4\Lambda^{4N+4} \right).
 \nonu \eea

$\bullet$ {\bf A generalized Konishi anomaly}

Now we are ready to study the generalized Konishi anomaly
equation. As in \cite{csw}, we restrict to the case with
$\left\langle \mbox{Tr} W^{\prime}(\Phi) \right\rangle=\mbox{Tr}
W^{\prime}(\Phi_{cl})$ and assume that the degree of
superpotential is less than $2N$. One can write
\begin{eqnarray}
W^{\prime}(\phi_I)=\sum_{i=0}^{2N-2n} \oint
\phi_I\frac{x^2P_{2N}(x)}{(x^2-{\phi_I}^2)} \frac{L_i}{(x-p_i)}dx.
\label{Koni2}
\end{eqnarray}
Using this equation we have following relations,
\begin{eqnarray}
\mbox{Tr} \frac{W^{\prime}(\Phi_{cl})}{x-\Phi_{cl}}&=&\mbox{Tr}
 \sum_{k=0}^{\infty}z^{-k-1}\Phi_{cl}^kW^{\prime}(\Phi_{cl})=
 \sum_{k=0}^{\infty}z^{-(2i+1)-1}2\sum_{I=1}^N
  \phi_I^{2i+1}W^{\prime}(\phi_I) \nonu \\
&=&2\sum_{I=1}^{N }\phi_I
W^{\prime}(\phi_I)\frac{1}{(z^2-\phi_I^2)} \nonu \\ & =
&\sum_{I=1}^{N} \frac{2\phi_I^2}{(z^2-\phi_I^2)}
\sum_{i=0}^{2N-2n} \oint
\frac{x^2P_{2N}(x)}{(x^2-\phi_I^2)}\frac{L_i}{(x-p_i)}dx,
\label{Koni}
\end{eqnarray}
where $z$ is outside the contour of integration.
Using (\ref{relation1}) we can write this factor as
\begin{eqnarray}
\frac{2\phi_I^2}{(z^2-\phi_I^2)(x^2-\phi_I^2)}
=\frac{1}{(x^2-z^2)}\left(z\mbox{Tr}\frac{1}{z-\Phi}-
x\mbox{Tr}\frac{1}{x-\Phi} \right). \nonu
\end{eqnarray}
Thus we can write (\ref{Koni}) as
\begin{eqnarray}
\mbox{Tr} \frac{W^{\prime}(\Phi_{cl})}{x-\Phi_{cl}}=\oint
\sum_{i=0}^{2N-2n}
\frac{x^2P_{2N}L_i}{(x^2-z^2)(x-p_i)}\left(z\mbox{Tr}\frac{1}{z-\Phi}-x
\mbox{Tr}\frac{1}{x-\Phi} \right)dx.
\label{Koni4}
\end{eqnarray}
As in the case \cite{csw} we can rewrite the contour of the
integral,
\begin{eqnarray}
\oint_{z_{out}}=\oint_{z_{in}}-\oint_{C_z+C_{-z}} \nonu
\end{eqnarray}
where $C_z$ and $C_{-z}$ are small contour around $z$ and $-z$
respectively. Thus the first term in (\ref{Koni4}) can be
rewritten as
\begin{eqnarray}
\mbox{Tr}\frac{1}{z-\Phi}\oint_{z_{out}}
\frac{zQ_{2k-2n+2}(x)x^2P_{2N}(x)}{H_{2N-2n}(x)x(x^2-z^2)}dx.
\nonu \eea From the change of an integration, this is given by
\bea && \mbox{Tr}\frac{1}{z-\Phi} \left( \oint_{z_{in}}
\frac{zQ_{2k-2n+2}(x)x^2P_{2N}(x)}{H_{2N-2n}(x)x(x^2-z^2)}dx+
\oint_{C_{z}+C{-z}}
\frac{zQ_{2k-2n+2}(x)x^2P_{2N}(x)}{H_{2N-2n}(x)x(x^2-z^2)}dx
\right)
\nonu \\
&& = \mbox{Tr}\frac{1}{z-\Phi} \left( W^{\prime}(z)- \frac{y_m(z)
z^2P_{2N}(z)}{\sqrt{B^2_{2N+2}(z)-4\Lambda^{4N+4}}} \right), \nonu
\end{eqnarray}
in the last equality we used (\ref{Koni2}) and
\begin{eqnarray}
H_{2N-2n}(z)=\frac{\sqrt{B^2_{2N+2}(z)-4\Lambda^{4N+4}}}{z\sqrt{
F_{2(2n+1)}(z)}}, \qquad y_m^2(z)=F_{2(2n+1)}(z)Q_{2k-2n+2}^2(z).
\nonu
\end{eqnarray}
We now make an integration over $x$
\begin{eqnarray}
-\sum_{i=0}^{2N-2n} \oint
\frac{L_ix^2P_{2N}(x)}{(x-p_i)x(x^2-z^2)}
\mbox{Tr}\frac{x}{x-\Phi}dx  = -\sum_{i=0}^{2N-2n} \frac{L_i
p_i^2P_{2N}(x=p_i)}{p_i^2-z^2} \mbox{Tr}\frac{p_i}{p_i-\Phi}.
\nonu \eea By using the equation of motion for $B_i$ to change the
trace part, one gets \bea \sum_{i=0}^{2N-2n} 2\frac{L_i
p_i^2P_{2N}(x=p_i)}{p_i^2-z^2} = +\sum_{i=0}^{2N-2n}\oint
\frac{2x^2P_{2N}(x)}{x^2-z^2} \frac{L_i}{x-p_i}dx  \nonu
\end{eqnarray}
where note that $z$ is {\it outside} the contour of integration in last equation. Therefore as in $SO$ case we can rewrite the last equation as follows:
\begin{eqnarray}
+2\frac{W^{\prime}(z)}{z}-\sum_{i=0}^{2N-2n}\oint_{C_{z}+C_{-z}}\frac{2x^2P_{2N}(x)}{(x^2-z^2)}
\frac{L_i}{(x-p_i)}dx =+2\frac{W^{\prime}(z)}{z}-\frac{2}{z}\frac{y_mz^2P_{2N}(z)}{\sqrt{B_{2N+2}^2(z)-4\Lambda^{4N+4}}} \nonu
\end{eqnarray}
Therefore we obtain \bea \mbox{Tr}
\frac{W^{\prime}(\Phi_{cl})}{z-\Phi_{cl}}=\mbox{Tr}\frac{1}{z-\Phi_{cl}}
\left(W^{\prime}(z)+ \frac{y_m(z)
z^2P_{2N}(z)}{\sqrt{B_{2N+2}^2(z)-4\Lambda^{4N+4}}}
\right)+2\frac{W^{\prime}(z)}{z}-\frac{2}{z}\frac{y_mz^2P_{2N}(z)}{\sqrt{B_{2N+2}^2(z)-4\Lambda^{4N+4}}}. \nonu \eea
Remembering that \bea \mbox{Tr}\frac{1}{z-\Phi_{cl}}=\frac{P_{2N}^{\prime}(z)}{P_{2N}(z)}, \nonu \eea
the second and fourth terms are rewritten as follows;
\begin{eqnarray}
 \mbox{Tr}\frac{1}{z-\Phi_{cl}}\frac{y_m(z)
z^2P_{2N}(z)}{\sqrt{B_{2N+2}^2(z)-4\Lambda^{4N+4}}}-\frac{2}{z}\frac{y_mz^2P_{2N}(z)}{\sqrt{B_{2N}^2(z)-4\Lambda^{4N+4}}}=-\frac{2y_m}{z}-\left\langle \tr \frac{y_m}{z-\Phi} \right\rangle \nonu
\end{eqnarray}
where we used (\ref{10may2C}).


Taking into account the relation,
\begin{eqnarray}
\tr \frac{W^{\prime}(\Phi_{cl})-W^{\prime}(z)}{z- \Phi_ {cl}}=
\left\langle \mbox{Tr}
\frac{W^{\prime}(\Phi)-W^{\prime}(z)}{z-\Phi} \right \rangle
\nonu
\end{eqnarray}
one can summarize as follows,
\begin{eqnarray}
\left\langle \tr \frac{W^{\prime}(\Phi)}{z-\Phi} \right\rangle &
=& \left( \left\langle \tr \frac{1}{z-\Phi} \right \rangle +
\frac{2}{z} \right) \left[W^{\prime}(z)-y_m(z)\right] \nonu
\end{eqnarray}
This is the generalized Konishi anomaly equation for $Sp(2N)$
case.


\subsection{The multiplication map and the confinement index}
Although in this subsection, we discuss $Sp(2N)$ gauge theories,
the matrix model curve which derived in \ref{spstrong} is the same
as $SO(2N)$ case (\ref{matrixcurve}). Thus we can use the same
notations and ideas introduced in $SO(N)$ case. The symmetry
breaking pattern is given by
\begin{eqnarray}
Sp(2N)\to Sp(2N_0)\times \prod_{i=1}^n U(N_i). \label{spbreaking}
\end{eqnarray}
The  difference between $SO(2N)$ case and $Sp(2N)$ case is the
contribution of unoriented diagram to the effective
superpotential,
\begin{eqnarray}
W_{eff}(S_i)=(N_0+ 2)\frac{\partial {\cal F}}{\partial S_i}+ \sum_{i=1}^n
N_i \frac{\partial {\cal F}}{\partial S_i}+2\pi i
\tau_0 \sum_{i=0}^n S_i+2\pi i \sum_{i=1}^n b_iS_i.
\end{eqnarray}
Thus if we introduce $\hat{N}_0\equiv 2N_0+2$ and $\hat{N}_i=N_i$,
 we can study the phase structure in the same way as $SO(2N)$ case.
 The criterion of
 confinement is that if three objects
 $(2N_0+2)$, $N_i$ and $b_i$ have common
 divisor,
 the vacua are confining vacua, if not, it is called Coulomb vacua.

For $Sp(2N)$ case we can construct multiplication map in the same
way as $SO(2N)$ and $SO(2N+1)$ cases. By assumption we have the
characteristic function $P_{2N}(x)$ that satisfies the following
massless monopole constraint,
\begin{eqnarray}
\left( x^2P_{2N}(x)+2\Lambda_0^{2N+2} \right)^2-4
\Lambda^{4N+4}_0=x^2H_{2N-2n}^2(x)F_{2(2n+1)}(x). \nonu
\end{eqnarray}
By using Chebyshev polynomial, let us consider  a solution of
$Sp(2KN+2K-2)$ gauge theory as
\begin{eqnarray}
P_{2KN+2K-2}(x)=\frac{2\eta^{2K}\Lambda^{2KN+2K}}{x^2}{\cal T}_K
\left(\frac{x^2P_{2N}(x)}{2\eta^2 \Lambda^{2N+2}}+1 \right)-
\frac{2\Lambda^{2KN+2K}}{x^2}
 \label{spmap}
\end{eqnarray}
where we introduce
\bea \widetilde{x} =
\frac{x^2P_{2N}(x)}{2\eta^2 \Lambda^{2N+2}}+1
\nonu
\eea
 and
$\eta^{2K}=1$. One can check that in the right hand side of
(\ref{spmap}), the argument $\widetilde{x}$ of the first kind
Chebyshev polynomial has a degree $(2N+2)$, by the definition of
the first kind Chebyshev polynomial and the right hand side has a
factor $1/x^2$, therefore it leads to a polynomial of degree
$-2+K(2N+2)$ totally. It is right to write the left hand side as
$P_{2KN-2K+2}(x)$ which agrees with the number of order  in $x$ in
the right hand side. The power of $\La$ for the $\widetilde{x}$ of
Chebychev polynomial was fixed by the power of $x$ which is equal
to $(2N+2)$. The power of $\La$  in front of the Chebyshev
polynomial can be fixed by the dimension consideration of both
sides. The right hand side should contain $(2KN+2K-2)$ which is
equal to $-2+(2KN+2K)$ where the first term comes from the power
of $x$ and the second one should be the power of $\La$. Also note
that the same $\eta$ term appears in the denominator of  the
argument of Chebyshev polynomial. Contrary to the previous
$SO(2N)$ or $SO(2N+1)$ cases, the curve contains $\La^{4N+4}$ as
well as $\La^{2N+2}$. In order to get the correct number of vacua
of $Sp(2KN+2K-2)$ gauge theory, it is very important to know the
right behavior of $\eta$ and $\La$ in $\widetilde{x}$. The last
term in (\ref{spmap}) is included since the monopole constraint
for $Sp(2N)$ case has an extra term in the left hand side,
contrary to the $SO(2N)$ or $SO(2N+1)$ cases. Let us check that
(\ref{spmap}) is indeed a solution of the massless monopole
constraint  of $Sp(2KN+2K-2)$ gauge theory. By introducing a new
function $B_{2KN+2K}(x)=x^2P_{2KN+2K-2}(x)+2\Lambda^{2KN+2K}$, we
can see the following relations from the solution (\ref{spmap}).
Then one can construct
\begin{eqnarray}
 B^2_{2KN+2K}(x) - 4 \Lambda^{4KN+4K}& = &
4\Lambda^{4KN+4K}\left({\cal T}^2_{K}(\widetilde{x})-1 \right)
\nonu \\
 &= & 4\Lambda^{4KN+4K}\left(\widetilde{x}^2-1 \right)
{\cal U}^2_{K-1}(\widetilde{x})  \nonu \\
&=&\frac{\Lambda^{4KN+4K}}{\eta^4\Lambda^{4N+4}}
\left[\left(x^2P_{2N}(x)+2\eta^2 \Lambda^{2N+2} \right)^2-4\eta^4
\Lambda^{4N+4}\right]{\cal U}^2_{K-1}(\widetilde{x}) \nonu  \\
&=&x^2 \left[\eta^{-2}\Lambda^{2(K-1)(N+1)} H_{2N-2n}(x){\cal
U}_{K-1}(\widetilde{x}) \right]^2 F_{2(2n+1)}(x). \nonu
\end{eqnarray}
In the fourth equality we used the identification
$\Lambda_0^{2N+2}=\eta^2 \Lambda^{2N+2}$. In other words, the
relations
\bea P_{2KN+2K-2}(x) & = &
\frac{2\eta^{2K}\Lambda^{2KN+2K}}{x^2}{\cal T}_K
\left(\frac{x^2P_{2N}(x)}{2\eta^2 \Lambda^{2N+2}}+1 \right)-
\frac{2\Lambda^{2KN+2K}}{x^2}, \nonu \\
\widetilde{F}_{2(2n+1)}(x) & = & F_{2(2n+1)}(x), \nonu \\
H_{K(2N-2)-2n}(x) & = & \eta^{-2}\Lambda^{2(K-1)(N+1)}
H_{2N-2n}(x){\cal U}_{K-1}\left(\frac{x^2P_{2N}(x)}{2\eta^2
\Lambda^{2N+2}}+1\right)
\nonu \eea
satisfy
\bea \left(
x^2P_{2KN+2K-2}(x)+2\Lambda^{2NK+2K} \right)^2-4
\Lambda^{4NK+4K}=x^2H_{K(2N-2)-2n}^2(x)\widetilde{F}_{2(2n+2)}(x).
\nonu \eea As in $SO(2N)$ or $SO(2N+1)$ case, since
$\widetilde{F}_{2(2n+1)}(x)=F_{2(2n+1)}(x)$, the vacua constructed
this way for the $Sp(2KN+2K-2)$ theory have the {\it same}
superpotential as the vacua of the $Sp(2N)$ theory.


Next we consider the multiplication map of $T(x)$ ,
\begin{eqnarray}
T(x)&=& \frac{d}{dx} \log \left[
\left(B_{2N+2}(x)+\sqrt{B_{2N+2}^2(x)-
4\Lambda^{4N+4}} \right) - \log x^2 \right] \nonu \\
&=&\frac{B_{2N+2}^{\prime}(x)}{\sqrt{B^2_{2N+2}(x)-
4\Lambda^{4N+4}}}-\frac{2}{x}.\nonu
\end{eqnarray}
The Seiberg-Witten differential is given by \bea d
\la_{SW}=\frac{x d x B_{2N+2}^{\prime}(x)}{\sqrt{B^2_{2N+2}(x)-
4\Lambda^{4N+4}}}. \nonu \eea
 By using the equation (\ref{spmap})
we have a function $B_{2KN+2K}(x)$ after the map, which denoted as
$B_{2KN+2K}(x)\equiv (x^2P_{2KN+2K-2}+2\Lambda^{2KN+2K})$. Let us
consider
\begin{eqnarray}
B_{2KN+2K}(x)=2\eta^{2K} \Lambda^{2KN+2K}{\cal T}_K
\left(\frac{x^2P_{2N}(x)}{2\eta^2 \Lambda^{2N+2}}+1 \right). \nonu
\end{eqnarray} The derivative of $B_{2KN+2K}(x)$ with respect
to $x$ leads to
\begin{eqnarray}
B_{2KN+2K}^{\prime}(x)=\eta^{2K} \Lambda^{2KN+2K} K{\cal U}_{K-1}
(\widetilde{x})\left(\frac{x^2P_{2N}(x)}{\eta^2 \Lambda^{2N+2}}
\right)^{\prime}. \nonu
\end{eqnarray}
Also we have the relation
\begin{eqnarray}
\sqrt{B_{2KN+2K}^2(x)-4\Lambda^{4N+4}}=
\frac{\Lambda^{2KN+2K}}{\eta^{2+2K} \Lambda^{2N+2}}{\cal
U}_{K-1}(\widetilde{x})\sqrt{B^2_{2N+2}(x)-4\Lambda^{4N+4}_0}.
\nonu
\end{eqnarray} Combining these two relations we have
\begin{eqnarray}
\frac{B_{2KN+2K}^{\prime}(x)}{\sqrt{B^2_{2KN+2K}(x)-
4\Lambda^{4KN+4K}}}=
K\frac{B_{2N+2}^{\prime}(x)}{\sqrt{B^2_{2N+2}(x)-
4\Lambda^{4N+4}}}. \nonu
\end{eqnarray} Thus finally we obtain multiplication map of
$T(x)$,
\begin{eqnarray}
T_K(x)=KT(x)+K\frac{2}{x}-\frac{2}{x} \iff
2N_0^{\prime}+2=K(2N_0+2), \ N_i^{\prime}=KN_i. \nonu
\end{eqnarray} This means that under the
multiplication map the special combination $(2N_0+2)$ have simple
multiplication by $K$. As in $SO(2N)$ case the number of
$Sp(2KN+2K-2)$ vacua with confinement index $K$ is $K$ times the
one of $Sp(2N)$ Coulomb vacua. All of these confining vacua can be
constructed by this map. Let us denote adjoint chiral superfield
as $\Phi_0$ in the $Sp(2N)$ gauge theory. Then through the
multiplication map we constructed the following quantum operator
in the $Sp(2KN+2K-2)$ gauge theory can be obtained by simply
multiplying the confinement index $K$ by the corresponding
operator in the $Sp(2N)$ gauge theory as follows
\bea \left\langle
\mbox{Tr} \frac{1}{x-\Phi} \right\rangle +\frac{2}{x} = K
\left(\left\langle \mbox{Tr} \frac{1}{x-\Phi_0} \right\rangle
+\frac{2}{x} \right). \nonu \eea

\subsection{Examples }

In this subsection we will analyze some examples of $Sp(2N)$ gauge
theory with rank $n=1$, namely $Sp(2N)$ gauge group is broken to
$Sp(2N_0)\times U(N_1)$. We will deal with ${\cal N}=1$ gauge
theories that are ${\cal N}=2$ theories deformed by tree level
superpotential characterized by
\begin{eqnarray}
W(\Phi)=\frac{m}{2}\mbox{Tr}\Phi^2+\frac{g}{4}\mbox{Tr}\Phi^4.
\nonu
\end{eqnarray}
For  simplicity we consider the special case $k=n=1$, namely the
number of critical point of $W(x)$ equal to the one of gauge
groups after Higgs breaking.

As already discussed, massless monopole constraint with $n=1$ is
given by
\begin{eqnarray}
B^2_{2N+2}(x)-4\Lambda^{4N+4}=\left[xH_{2N-2}(x)\right]^2F_{6}(x).
\label{aa3}
\end{eqnarray}
Equivalently we can factorize this equation as,
\begin{eqnarray}
B_{2N+2}(x)+2\Lambda^{2N+2}&=&H^2_{s_+}(x)R_{2N+2-2s_+}(x),
\label{spconst1} \\
B_{2N+2}(x)-2\Lambda^{2N+2}&=&H^2_{s_-}(x)
\widetilde{R}_{2N+2-2s_-}(x),
\label{spconst2} \\
xH_{2N-2}(x)=H_{s_+}(x)H_{s_-}(x),&&
F_6(x)=R_{2N+2-2s_+}(x)\widetilde{R}_{2N+2-2s_-}(x).
 \nonu
\end{eqnarray}
We want to point out the relation to $SO(2N)$ case. Comparing
(\ref{spconst1})and (\ref{spconst2}) with (\ref{soconst1}) and
(\ref{soconst2}), we find the similarities and correspondences in
$P_{2N-2}(x) \iff B_{2N+2}(x)$. So massless monopole constraint
for $SO(2N+2)$ case is the same as the one for $Sp(2N-2)$ case.
However the subtle difference comes from the function
$B_{2N+2}(x)=x^2P_{2N}(x)+2\Lambda^{2N+2}$. The left hand side in
(\ref{spconst2}) leads to
\begin{eqnarray}
B_{2N+2}(x)-2\Lambda^{2N+2}=x^2P_{2N}(x).
\label{aa4} \nonu
\end{eqnarray}

Thus we must have a factor $x^2$ in the right hand side of
(\ref{spconst2}). Thus we cannot describe two case
$(s_+,s_-)=(a,b)$ and $(b,a)$ in the same way by using $\eta
=\pm1$. Taking into account this constraint, we can rewrite
massless monopole constraint (\ref{spconst1}) and (\ref{spconst2})
as
\begin{eqnarray}
B_{2N+2}(x)+2\Lambda^{2N+2}&=&H_{2s_+}^2(x)R_{2N+2-4s_+}(x),  \nonu \\
B_{2N+2}(x)-2\Lambda^{2N+2}&=&x^2H_{2s_-}^2(x)\widetilde{R}_{2N-4s_-}(x), \nonu \\
H_{2N-2}(x)=H_{2s_+}(x)H_{2s_-}(x),&&
F_6(x)=R_{2N+2-4s_+}(x)\widetilde{R}_{2N-4s_-}(x) \label{AA}
\end{eqnarray}
where $2s_++2s_-=2N-2$ and $2N-4s_-\ge 0$ and $2N+2-4s_+\ge 0$.
From the first two equations (\ref{AA})  we have a constraint that
is useful for analysis below,
\begin{eqnarray}
H_{2s_+}^2(x)R_{2N+2-4s_+}(x)-
4\Lambda^{2N+2}=x^2H_{2s_-}^2(x)\widetilde{R}_{2N-4s_-}(x).
\label{Spconstrelation}
\end{eqnarray}
After solving this equation, we can find matrix model curve,
\begin{eqnarray}
y_m^2(x)=R_{2N+2-4s_+}(x)\widetilde{R}_{2N-4s_-}(x), \nonu
\end{eqnarray}
and
\begin{eqnarray}
R_{2N+2-4s_+}(x)\widetilde{R}_{2N-4s_-}(x)
={W^{\prime}_3}^2(x)+f_{2}(x).
\label{sd1}
\end{eqnarray}
As in $SO(2N)$ or $SO(2N+1)$ case, from the coefficient of $x^2$
in $f_2(x)$ we can read off the expectation value of glueball
superfield, $S=S_0 + S_1$. Let us start with the explicit analysis
of $Sp(2)$, $Sp(4)$ and $Sp(6)$.

$\bullet$ {\bf $Sp(2)$ case}

\noindent The first example is $Sp(2)$ gauge theory. In this case
massless monopole constraint (\ref{aa3}) is trivial since
$2N-2n=2-2=0$. If we parameterize characteristic function as
$P_2(x)=x^2-v^2$, we can rewrite  as, from (\ref{AA}),
\begin{eqnarray}
B_4(x) + 2\La^4 & = & x^2(x^2-v^2)+4\Lambda^4=R_{4}(x), \nonu \\
B_4(x) - 2 \La^4 & = &  x^2(x^2-v^2)= x^2 \widetilde{R}_{2}(x).
\nonu
\end{eqnarray}
Thus the matrix model curve is given by
\begin{eqnarray}
F_6(x)&=&y_m^2=x^2(x^2-v^2)^2+4\Lambda^2(x^2-v^2), \nonumber \\
\quad W_{3}^{\prime}(x)&=&x(x^2-v^2),\nonu \\
f_{2}(x) & = & 4\Lambda^2(x^2-v^2), \nonu \\
 S & = & -\Lambda^2. \nonu
\end{eqnarray} Then there is only one vacuum for given $W(x)$.
In the semiclassical limit $\Lambda \to 0$ the characteristic
function becomes $P_2(x)\to (x^2-v^2)$. So in this vacuum, the
gauge group $Sp(2)$ breaks into $U(1)$.

$\bullet$ {\bf $Sp(4)$ case}

\noindent The second example is a  $Sp(4)$ gauge theory where
$2N-2n=4-2=2$. As already discussed, this case corresponds to
$SO(8)$ case and is somewhat interesting example. The number of
massless monopoles is determined by $2s_++2s_-=2$. Thus we have
two branches $(s_+,s_-)=(1,0)$ and $(0,1)$. At first, we study the
case $(1,0)$. We can parameterize the massless monopole constraint
as
\begin{eqnarray}
B_{6}(x)+2\Lambda^{6}&=&(x^2-a^2)^2(x^2+A), \nonu \\
B_{6}(x)-2\Lambda^{6}&=&x^2 \left(x^4+Bx^2+C \right). \nonu
\end{eqnarray}
These parameters must satisfy (\ref{sd1}). The solution for the
constraint is given by, from (\ref{Spconstrelation}),
\begin{eqnarray}
B_6(x)+2\Lambda^{6}&=&(x^2-a^2)^2 \left(x^2+
\frac{4\Lambda^6}{a^4} \right), \nonu \\
B_6(x)-2\Lambda^{6}&=&x^2
\left[(x^2-a^2)^2+\frac{4\Lambda^6}{a^4}(x^2-2a^2) \right].
\nonu
\end{eqnarray}
We can find the matrix model curve from the
solution as,
\begin{eqnarray}
y_m^2(x)=\left(x^2+\frac{4\Lambda^6}{a^4}
\right)\left[(x^2-a^2)^2+\frac{4\Lambda^6}{a^4}(x^2-2a^2) \right]
\nonu
\end{eqnarray}
and from this one gets
\begin{eqnarray}
W_{3}^{\prime}(x) & = & x\left(x^2+\frac{4 \Lambda^6}{a^4}-a^2
\right),\nonu \\
f_2(x) & = & -\frac{8 \Lambda^6}{a^2}x^2+\frac{4\Lambda^6}{a^2}\left(a^2-2\frac{4
\Lambda^6}{a^4}\right), \nonu \\
 S & = & \frac{2 \Lambda^6}{a^2}. \nonu
\end{eqnarray}
For these vacua we consider the semiclassical limit. In this case
there exist two limits.

1. Fix a: The characteristic function behaves $P_{4}\to
(x^2-a^2)^2$ and the gauge group $Sp(4)$ breaks into $U(2)$.

2. Fix $v\equiv \frac{4\Lambda^6}{a^4}$: The characteristic
function goes to $P_4 \to x^2(x^2+v)$ and  the $Sp(4)$ breaks into
$Sp(2)\times U(1)$.

\noindent Thus we can  transit continuously by changing the
parameters. At last, as in $SO(8)$ case we count the number of
vacua for fixed tree level superpotential,
$W_3^{\prime}(x)=x(x^2+\Delta)$. From the previous result we can
represent $\Delta$ as
\begin{eqnarray}
\Delta=\frac{4 \Lambda^6}{a^4}-a^2. \nonumber
\end{eqnarray}
We evaluate this equation under the two semiclassical limit $1$ and $2$
discussed above.

1. In this limit since $\Delta=-a^2$ we have only one function
$f_2(x)$. Thus we have one vacua. As we will see below we have one
vacua that have the same gauge group. Thus all the vacua with this
gauge group are two vacua. On the other hand, gauge group becomes
$U(2)$ under this limit. So the number of vacua is two. This equal
to the one derived from above.

2. In this limit since
$a^2=\left(\frac{4\Lambda^6}{\Delta}\right)^{\frac{1}{2}}$, we
have two functions $f_2(x)$. In other words we have two vacua for
each potential. This number is equal to the one derived from gauge
group $Sp(2)\times U(1)$, $(N_0+1) \times N_1=(1+1)\cdot 1=2$,
because dual Coxeter number of $Sp(2N_0)$ gauge theories are
$(N_0+1)$.


Next let us consider $(s_+,s_-)=(0,1)$. We can easily solve the
massless monopole constraint as follows:
\begin{eqnarray}
B_6(x)+2\Lambda^{6}&=&x^2(x^2-a^2)^2+4\Lambda^6, \nonu \\
B_6(x)-2\Lambda^{6}&=&x^2(x^2-a^2)^2. \nonu
\end{eqnarray}
From this
solution we can find matrix model curve
\begin{eqnarray}
F_{6}(x) & = & y_m^2(x)=x^2(x^2-a^2)^2+4\Lambda^6, \nonu \\
W_3^{\prime}(x) & = & x(x^2-a^2),\nonu \\
f_2(x) & = & 4\Lambda^6,\nonu \\
 S & = & 0. \nonu
\end{eqnarray}
In the semiclassical limit it is easy to see $P_4(x)\to
(x^2-a^2)^2$, which shows that the gauge group $Sp(4)$ breaks into
$U(2)$. This branch is considerably different as the $SO(8)$
branch $(s_+,s_-)=(1,2)$, though $(2,1)$ is exactly same as the
$Sp(6)$ branch. This agrees with the comment under
(\ref{spconst2}).

$\bullet$ {\bf $Sp(6)$ case}

\noindent The  next example is a $Sp(6)$ gauge theory, which is
the most interesting example. As already discussed, this case
corresponds to $SO(10)$ case. In this case the number of massless
monopoles is given by $2s_++2s_-=4$ and $2\ge s_+, \frac{3}{2}\ge
s_-$. There are two branches $(s_+,s_-)=(2,0)$ and $(1,1)$ in this
theory.

{\bf 1. Confining branch}

At first, let us study $(2,0)$ branch, there exists a solution
\begin{eqnarray}
B_8(x)+2\Lambda^8&=&(x^2-a)^2(x^2-b)^2, \nonu \\
B_8(x)-2\Lambda^8&=&x^2 (x^6+Ax^4+Bx^2+C). \nonu
\end{eqnarray}
From (\ref{Spconstrelation}),  the coefficients $A,B,C,b$ can be
represented in terms of $a$ and they are
\begin{eqnarray} A&=&-2a-4\frac{\eta
\Lambda^4}{a},
\qquad B=a^2+8\eta \Lambda^4+4\frac{\Lambda^8}{a^2}, \nonu \\
C&=&-4a\eta \Lambda^4-\frac{8\Lambda^8}{a},\qquad b=2\frac{\eta
\Lambda^4}{a}, \nonu
\end{eqnarray}
where $\eta $ is $2$-th root of unity. Thus the characteristic function $P_6(x)$ becomes
\begin{eqnarray}
P_6(x)=\frac{1}{x^2}\left[(x^2-a)^2\left(x^2-\frac{2\eta
\Lambda^4}{a}\right)^2-4\Lambda^8\right]. \label{sd3}
\end{eqnarray}
From these
solutions we can find matrix model curve
\begin{eqnarray}
y_m^2&=&x^2\left(x^2+\frac{A}{2}
\right)^2+x^2 \left(B-\frac{A^2}{4} \right)+C  \nonu \\
&=&x^2\left(x^2-a-2\frac{\eta \Lambda^4}{a} \right)^2+4x^2 \eta
\Lambda^4-4a\eta \Lambda^4-\frac{8\Lambda^8}{a}
\nonu
\end{eqnarray}
and moreover we have
\begin{eqnarray}
W_3^{\prime}(x) & = & x \left(x^2-a-2\frac{\eta \Lambda^4}{a}
\right),\nonu \\
 f_2(x) & = & 4x^2 \eta \Lambda^4-4a\eta
\Lambda^4-\frac{8\Lambda^8}{a}, \nonu \\
S & = & -\eta \Lambda^4. \nonu
\end{eqnarray} Next we consider the semiclassical limit. There
are two limits:

1. Fixed a: As $\Lambda \to 0$, the characteristic function has
the relation $P_6(x) \to x^2(x^2-a)^2$. This means
 the gauge group $Sp(6)$ is broken to $Sp(2)\times U(2)$.

2. Fixed $w\equiv \frac{2\eta \Lambda^4}{a}$: The characteristic
function goes to  $P_6(x) \to x^2(x^2-w)^2$. This means the gauge
group $Sp(6)$ is broken to $Sp(2)\times U(2)$. Thus we can see the
continuous transition between two semiclassical limits. But both
limits have the same gauge group. We want to survey whether these
phases are the confining phase or not. Multiplication map for
$Sp(2N)$ gauge theories has already been discussed, i.e.
(\ref{spmap}). If we choose $K=2$, we can construct the
multiplication map from $Sp(2)$ to $Sp(6)$ where we denote as
$P^{K=2}_{Sp(2)\to Sp(6)}(x)$,
\begin{eqnarray}
P^{K=2}_{Sp(2)\to Sp(6)}(x)&=&\frac{2\epsilon^{4} \Lambda^8 }{x^2}
\left[2\left(\frac{x^2P_2(x)}{2\epsilon^2 \Lambda^4}+1 \right)^2-1
\right]-\frac{2\Lambda^8}{x^2} \nonumber \\
&=&\frac{1}{x^2}\left[(x^2(x^2-v^2)+2\epsilon^2
\Lambda^4)^2-4\Lambda^8 \right]. \nonu
\end{eqnarray}
where $\epsilon$ is $4$-th root of unity. If we choose
$v^2=a+\frac{2\epsilon^2 \Lambda^4}{a}$ and identify $\epsilon^2$
with $\eta$, we can reach the solution (\ref{sd3}).
\begin{eqnarray}
P^{K=2}_{Sp(2)\to
Sp(6)}(x)=\frac{1}{x^2}\left[(x^2-a)^2\left(x^2-\frac{2\eta
\Lambda^4}{a}\right)^2-4 \Lambda^8 \right]. \nonu
\end{eqnarray}
Thus this vacua is a confining phase. After all we obtain
continuous transition within the Coulomb phase that does not
change the gauge group.

{\bf 2. Coulomb branch}

Next we consider the other branch $(s_+,s_-)=(1,1)$ and there are
\begin{eqnarray}
B_8(x)+2\Lambda^8&=&(x^2-a)^2\left(x^4+Ax^2+C \right), \nonu  \\
B_8(x)-2\Lambda^8&=&x^2 (x^2-b)^2\left(x^2+D \right). \nonu
\end{eqnarray}
As in previous examples, we must take into account the constraint
(\ref{Spconstrelation}). We get the equations,
\begin{eqnarray}
a^2C=4\Lambda^8,\qquad a^2A-2aC-b^2D=0, \label{ssk} \\
a^2-2aA-b^2+C+2bD=0,\qquad -2a+A+2b-D=0. \label{ssk2}
\end{eqnarray}

Although the solutions of these equations are a little bit
complicated, those are reasonable solutions. We can continuously
transit between the phases that can be obtained in semiclassical
limit. So we concentrate on surveying the semiclassical limits of
the solutions and do not obtain matrix model curves. From the
first equation of (\ref{ssk}) we can take two classical limits:

1. $\Lambda \to 0$ with $a\to 0$ and $C\to 0$,


2. $\Lambda \to 0$ with fixed $C$ and $a\to 0$.

\noindent To begin with we consider the case 1. In this case the
equations (\ref{ssk}) and (\ref{ssk2}) become
\begin{eqnarray}
b=0,\qquad A=D. \nonu
\end{eqnarray}
Thus in this classical limit, the characteristic function behaves
$P_6(x)\to x^4(x^2+A^2)$, which means that the gauge group breaks
into $Sp(6) \to Sp(4)\times U(1)$.

Next we consider the case 2. 
The equations (\ref{ssk}) and
(\ref{ssk2}) become
\begin{eqnarray}
b^2D=0,\qquad -b^2+C+2bD=0,\qquad A+2b-D=0. \nonu
\end{eqnarray}
Then we have the solutions,
\begin{eqnarray}
D=0,\qquad A=2b,\qquad C=b^2. \nonu
\end{eqnarray}
So we obtain  a relation $P_{6}(x)\to x^2(x^2-b)^2$, which means
that $Sp(6)\to Sp(2)\times U(2)$. After all, we can transit
continuously between the phases that have gauge groups
$Sp(2)\times U(2)$ and $Sp(4)\times U(1)$.

So far we have studied the multiplication maps from $Sp(2)$ to
$Sp(6)$ which is an example describing the vacua of $Sp(2KN+2K-2)$
gauge theory from $Sp(2N)$ theory. Then it is natural to consider
whether there exists other multiplication map from $Sp(2N)$ to
$Sp(2KN+2K-2)$ with different values of $N$ and $K$. For example,
when $K=3$ and $N=1$, then the vacua of $Sp(10)$ gauge theory has
the same superpotential as those of $Sp(2)$ theory. Therefore we
expect that there exists a solution that is a confining phase from
the explicit multiplication map from $Sp(2)$ to $Sp(10)$ by $K=3$
where we denote as $P^{K=3}_{Sp(2)\to Sp(10)}(x)$. Although we
have considered a couple of examples, it would be interesting to
study the possible confining vacua for each general $N$ and $K$
systematically.

We have seen that massless constraint for $SO(2N+2)$ is the same
as the one for $Sp(2N-2)$. We ask whether there exists a
multiplication map from $SO(2N)$ to $Sp(2M)$ or a map from
$Sp(2N)$ to $SO(2M)$. From the experience we considered so far,
let us consider the following characteristic function
\begin{eqnarray}
P_{K(2N-2)-2}(x)=\frac{2\eta^{2K} \Lambda^{2KN-2K}}{x^2}{\cal T}_K
\left(\frac{P_{2N}(x)}{2x^2 \eta^2 \Lambda^{2N-2}}
\right)-\frac{2\Lambda^{2KN-2K}}{x^2} \nonu
\end{eqnarray}
where $P_{2N}(x)$ satisfies the usual massless monopole constraint
of $SO(2N)$ gauge theory.
From this relation one sees that a new function $P_{K(2N-2)-2}(x)$
satisfies the massless monopole constraint of $Sp(2KN-2K-2)$ gauge
theory.
For
$Sp(6)$ case, namely $K(2N-2)-2=6$, if we choose $K=2$ we would
expect to have a map from $SO(10)\to Sp(6)$. By using the explicit
result for vacua of $SO(10)$ gauge theory that can be done by
tedious calculations, we will see that $P_{SO(10) \to
Sp(6)}^{K=2}(x)$ becomes the result for $Sp(6)$ which have already
been derived. Conversely we can construct maps from $Sp(2N)$ to
$SO(2M)$ in the same way as previous one. Of course the map with
$K=1$ gives the same as the one that we have already obtained in
terms of map from $SO(2N)$ to $Sp(2M)$ by substituting $K=1$.
Although within same gauge groups the map with
$K=1$ is trivial, the maps between the different gauge groups are
somewhat interesting one, which maps the Coulomb vacua to the
Coulomb vacua. This is naively expected because $K=1$ does not
give a common divisor.
So for the
case $SO(2N)$ to $Sp(2M)$, we can expect to have the following
relations, $2M+2 \iff K(2N- 2)$. In other words, the effect of
unoriented diagram or geometrically the effect of orientifold
plane changes under these maps.

Finally, we would like to comment on the transition between different classical limit
gauge groups $SO(N_0)\times U(N_1)$ and $SO(\widetilde{N}_0)\times
U(\widetilde{N}_1)$ or $Sp(N_0)\times U(N_1)$ and $Sp(\widetilde{N}_0)\times
U(\widetilde{N}_1)$. As in \cite{csw} these transition was interpreted as
a rearrangement of the compact one-cycles on the matrix model curve. Since the
matrix model curve for $SO/Sp$ gauge theories have ${\bf Z}_2$ symmetry,
this rearrangement should occur in the way that keeps this ${\bf Z}_2$ symmetry.


\vspace{1cm}
\centerline{\bf Acknowledgments}

This research of CA was supported by Korea Research Foundation
Grant(KRF-2002-015-CS0006). CA thanks Korea Institute for Advanced
Study (KIAS) and String Theory Journal Club where this work was
undertaken. The authors would like to thank Bo Feng and 
Hiroaki Kanno for useful suggestions and discussions.


\begin{thebibliography}{[00]}


\bibitem{vafa}
C.~Vafa,
J.\ Math.\ Phys.\  {\bf 42}, 2798 (2001), {\tt hep-th/0008142}.

\bibitem{civ}
F.~Cachazo, K.~A.~Intriligator and C.~Vafa,
Nucl.\ Phys.\ B {\bf 603}, 3 (2001), {\tt hep-th/0103067}.


\bibitem{ckv}
F.~Cachazo, S.~Katz and C.~Vafa,
{\tt hep-th/0108120}.

\bibitem{cfikv}
F.~Cachazo, B.~Fiol, K.~A.~Intriligator, S.~Katz and C.~Vafa,
Nucl.\ Phys.\ B {\bf 628}, 3 (2002), {\tt hep-th/0110028}.

\bibitem{gukov1}
S.~Gukov, C.~Vafa and E.~Witten,
Nucl.\ Phys.\ B {\bf 584}, 69 (2000) [Erratum-ibid.\ B {\bf 608},
477 (2001)], {\tt hep-th/9906070}.

\bibitem{gukov2}
S.~Gukov,
Nucl.\ Phys.\ B {\bf 574}, 169 (2000), {\tt hep-th/9911011}.

\bibitem{tv}
T.~R.~Taylor and C.~Vafa,
Phys.\ Lett.\ B {\bf 474}, 130 (2000), {\tt hep-th/9912152}.


\bibitem{eot}
J.~D.~Edelstein, K.~Oh and R.~Tatar,
JHEP {\bf 0105}, 009 (2001), {\tt hep-th/0104037}.



\bibitem{fo1}
H.~Fuji and Y.~Ookouchi,
{\tt hep-th/0205301}.

\bibitem{cv}
F.~Cachazo and C.~Vafa,
{\tt hep-th/0206017}.


\bibitem{feng1}
B.~Feng,
{\tt hep-th/0212010}.

\bibitem{ookouchi}
Y.~Ookouchi,
{\tt hep-th/0211287}.




\bibitem{dot1}
K.~Dasgupta, K.~Oh and R.~Tatar,
Nucl.\ Phys.\ B {\bf 610}, 331 (2001), {\tt hep-th/0105066}.

\bibitem{dot2}
K.~Dasgupta, K.~Oh and R.~Tatar,
JHEP {\bf 0208}, 026 (2002), {\tt hep-th/0106040}.

\bibitem{ot}
K.~Oh and R.~Tatar,
Adv.\ Theor.\ Math.\ Phys.\  {\bf 6}, 141 (2003), {\tt
hep-th/0112040}.

\bibitem{dopt}
K.~Dasgupta, K.~Oh, J.~Park and R.~Tatar,
JHEP {\bf 0201}, 031 (2002), {\tt hep-th/0110050}.



\bibitem{dv1}
R.~Dijkgraaf and C.~Vafa,
Nucl.\ Phys.\ B {\bf 644}, 3 (2002), {\tt hep-th/0206255}.

\bibitem{dv2}
R.~Dijkgraaf and C.~Vafa,
Nucl.\ Phys.\ B {\bf 644}, 21 (2002), {\tt hep-th/0207106}.

\bibitem{dv3}
R.~Dijkgraaf and C.~Vafa,
{\tt hep-th/0208048}.


\bibitem{dgkv}
R.~Dijkgraaf, S.~Gukov, V.~A.~Kazakov and C.~Vafa,
{\tt hep-th/0210238}.




\bibitem{cm}
L.~Chekhov and A.~Mironov,
Phys.\ Lett.\ B {\bf 552}, 293 (2003), {\tt hep-th/0209085}.



\bibitem{ino}
H.~Ita, H.~Nieder and Y.~Oz,
JHEP {\bf 0301}, 018 (2003), {\tt hep-th/0211261}.


\bibitem{Aganagic:2002wv}
M.~Aganagic, A.~Klemm, M.~Marino and C.~Vafa,
{\tt hep-th/0211098}.


\bibitem{dglvz}
R.~Dijkgraaf, M.~T.~Grisaru, C.~S.~Lam, C.~Vafa and D.~Zanon,
{\tt hep-th/0211017}.

\bibitem{cdsw}
F.~Cachazo, M.~R.~Douglas, N.~Seiberg and E.~Witten,
JHEP {\bf 0212}, 071 (2002), {\tt hep-th/0211170}.

\bibitem{csw}
F.~Cachazo, N.~Seiberg and E.~Witten,
{\tt hep-th/0301006}.

\bibitem{Berenstein:2002sn}
D.~Berenstein,
Phys.\ Lett.\ B {\bf 552}, 255 (2003), {\tt hep-th/0210183}.

\bibitem{Argurio:2002xv}
R.~Argurio, V.~L.~Campos, G.~Ferretti and R.~Heise,
{\tt hep-th/0210291}.

\bibitem{McGreevy:2002yg}
J.~McGreevy,
JHEP {\bf 0301}, 047 (2003), {\tt hep-th/0211009}.

\bibitem{Suzuki:2002gp}
H.~Suzuki,
{\tt hep-th/0211052}.

\bibitem{Suzuki:2002jc}
H.~Suzuki,
{\tt hep-th/0212121}.

\bibitem{Demasure:2002sc}
Y.~Demasure and R.~A.~Janik,
Phys.\ Lett.\ B {\bf 553}, 105 (2003), {\tt hep-th/0211082}.

\bibitem{Bena:2002kw}
I.~Bena and R.~Roiban,
Phys.\ Lett.\ B {\bf 555}, 117 (2003), {\tt hep-th/0211075}.

\bibitem{Tachikawa:2002wk}
Y.~Tachikawa,
{\tt hep-th/0211189}.

\bibitem{Tachikawa:2002ud}
Y.~Tachikawa,
{\tt hep-th/0211274}.

\bibitem{Feng:2002zb}
B.~Feng,
{\tt hep-th/0211202}.

\bibitem{Feng:2002yf}
B.~Feng and Y.~H.~He,
{\tt hep-th/0211234}.

\bibitem{Argurio:2002hk}
R.~Argurio, V.~L.~Campos, G.~Ferretti and R.~Heise,
Phys.\ Lett.\ B {\bf 553}, 332 (2003), {\tt hep-th/0211249}.

\bibitem{Naculich:2002hi}
S.~G.~Naculich, H.~J.~Schnitzer and N.~Wyllard,
Nucl.\ Phys.\ B {\bf 651}, 106 (2003), ~{\tt hep-th/0211123}.

\bibitem{Naculich:2002hr}
S.~G.~Naculich, H.~J.~Schnitzer and N.~Wyllard,
~JHEP ~{\bf 0301}, ~015 ~(2003), ~{\tt hep-th/0211254}.

\bibitem{Bena:2002ua}
I.~Bena, R.~Roiban and R.~Tatar,
{\tt hep-th/0211271}.

\bibitem{Ohta:2002rd}
K.~Ohta,
{\tt hep-th/0212025}.

\bibitem{Seki:2002ti}
S.~Seki,
{\tt hep-th/0212079}.

\bibitem{Bena:2002tn}
I.~Bena, S.~de Haro and R.~Roiban,
{\tt hep-th/0212083}.

\bibitem{Hofman:2002bi}
C.~Hofman,
{\tt hep-th/0212095}.

\bibitem{Demasure:2002jb}
Y.~Demasure and R.~A.~Janik,
{\tt hep-th/0212212}.

\bibitem{Seiberg:2002jq}
N.~Seiberg,
JHEP {\bf 0301}, 061 (2003), {\tt hep-th/0212225}.

\bibitem{Ahn:2002vj}
C.~Ahn and S.~Nam,
{\tt hep-th/0212231}.

\bibitem{Ahn:2003fq}
C.~Ahn,
{\tt hep-th/0301011}.

\bibitem{Feng:2002is}
B.~Feng,
{\tt hep-th/0212274}.

\bibitem{Ferrari:2002jp}
F.~Ferrari,
Nucl.\ Phys.\ B {\bf 648}, 161 (2003), {\tt hep-th/0210135}.

\bibitem{fo2}
H.~Fuji and Y.~Ookouchi,
JHEP {\bf 0212}, 067 (2002), {\tt hep-th/0210148}.

\bibitem{Gorsky:2002uk}
A.~Gorsky,
Phys.\ Lett.\ B {\bf 554}, 185 (2003), {\tt hep-th/0210281}.

\bibitem{ferrari1}
F.~Ferrari,
{\tt hep-th/0211069}.

\bibitem{Gopakumar:2002wx}
R.~Gopakumar,
{\tt hep-th/0211100}.

\bibitem{Dijkgraaf:2002wr}
R.~Dijkgraaf, A.~Neitzke and C.~Vafa,
{\tt hep-th/0211194}.

\bibitem{Klemm:2002pa}
A.~Klemm, M.~Marino and S.~Theisen,
{\tt hep-th/0211216}.

\bibitem{Dijkgraaf:2002yn}
R.~Dijkgraaf, A.~Sinkovics and M.~Temurhan,
{\tt hep-th/0211241}.

\bibitem{Itoyama:2002ys}
H.~Itoyama and A.~Morozov,
{\tt hep-th/0211245}.

\bibitem{ashoketal}
S.~K.~Ashok, R.~Corrado, N.~Halmagyi, K.~D.~Kennaway and C.~Romelsberger,
{\tt hep-th/0211291}.


\bibitem{Itoyama:2002rk}
H.~Itoyama and A.~Morozov,
{\tt hep-th/0212032}.

\bibitem{jo}
R.~A.~Janik and N.~A.~Obers,
Phys.\ Lett.\ B {\bf 553}, 309 (2003), {\tt hep-th/0212069}.

\bibitem{Balasubramanian:2002tm}
V.~Balasubramanian, J.~de Boer, B.~Feng, Y.~H.~He, M.~x.~Huang, V.~Jejjala and A.~Naqvi,
{\tt hep-th/0212082}.

\bibitem{Chekhov:2003gz}
L.~Chekhov, A.~Marshakov, A.~Mironov and D.~Vasiliev,
{\tt hep-th/0301071}.



\bibitem{Matone:2002wh}
M.~Matone,
{\tt hep-th/0212253}.

\bibitem{Dymarsky:2003qs}
A.~Dymarsky and V.~Pestun,
{\tt hep-th/0301135}.

\bibitem{Itoyama:2003qt}
H.~Itoyama and A.~Morozov,
{\tt hep-th/0301136}.

\bibitem{Bergamin:2003ub}
L.~Bergamin and P.~Minkowski,
{\tt hep-th/0301155}.



\bibitem{ferrari2}
F.~Ferrari,
{\tt hep-th/0301157}.

\bibitem{Muck:2003zf}
W.~Muck,
{\tt hep-th/0301171}.



\bibitem{Roiban:2003uq}
R.~Roiban, R.~Tatar and J.~Walcher,
{\tt hep-th/0301217}.

\bibitem{Ahn:2003ua}
C.~Ahn and S.~Nam,
{\tt hep-th/0301203}.



\bibitem{Ookouchi:2003xc}
Y.~Ookouchi and Y.~Watabiki,
{\tt hep-th/0301226}.


\bibitem{dv4}
R.~Dijkgraaf and C.~Vafa,
{\tt hep-th/0302011}.






\bibitem{ty}
S.~Terashima and S.~K.~Yang,
Phys.\ Lett.\ B {\bf 391}, 107 (1997), {\tt hep-th/9607151}.

\bibitem{kty}
T.~Kitao, S.~Terashima and S.~K.~Yang,
Phys.\ Lett.\ B {\bf 399}, 75 (1997), {\tt hep-th/9701009}.

\bibitem{as}
P.~C.~Argyres and A.~D.~Shapere,
Nucl.\ Phys.\ B {\bf 461}, 437 (1996), {\tt hep-th/9509175}.

\bibitem{aps}
P.~C.~Argyres, M.~Ronen Plesser and A.~D.~Shapere,
Nucl.\ Phys.\ B {\bf 483}, 172 (1997), {\tt hep-th/9608129}.

\bibitem{hanany}
A.~Hanany,
Nucl.\ Phys.\ B {\bf 466}, 85 (1996), {\tt hep-th/9509176}.

\bibitem{hms}
T.~Hirayama, N.~Maekawa and S.~Sugimoto,
Prog.\ Theor.\ Phys.\  {\bf 99}, 843 (1998), {\tt hep-th/9705069}.

\bibitem{dkp}
E.~D'Hoker, I.~M.~Krichever and D.~H.~Phong,
Nucl.\ Phys.\ B {\bf 489}, 211 (1997), {\tt hep-th/9609145}.


\bibitem{aot}
C.~Ahn, K.~Oh and R.~Tatar,
J.\ Geom.\ Phys.\  {\bf 28}, 163 (1998), {\tt hep-th/9712005}.

\bibitem{bl} A. ~Brandhuber and K. ~Landsteiner,  Phys. \ Lett. \ B 
{\bf 358}, 73 (1995), {\tt hep-th/9507008}. 

\bibitem{ds1} U.H. ~Danielsson and B. ~Sundborg,  Phys. \ Lett. \ B
{\bf 358}, 273 (1995), {\tt hep-th/9504102}.

\bibitem{ahn98}
C.~Ahn,
Phys.\ Lett.\ B {\bf 426}, 306 (1998), {\tt hep-th/9712149}.

\bibitem{Tera97}
S.~Terashima,
Nucl.\ Phys.\ B {\bf 526}, 163 (1998), {\tt hep-th/9712172}.



\bibitem{ds}
M.~R.~Douglas and S.~H.~Shenker, Nul. \ Phys. \ B {\bf 447}, 271
(1995), {\tt hep-th/9503163}.



\end{thebibliography}
\end{document}